\documentclass{article} 
\usepackage{preprintbox}

\usepackage{amsmath,amsfonts,bm}

\def\secref#1{section~\ref{#1}}

\def\eqref#1{equation~\ref{#1}}

\def\plaineqref#1{\ref{#1}}

\def\1{\bm{1}}

\def\eps{{\epsilon}}

\def\ve{{\bm{e}}}
\def\vf{{\bm{f}}}

\def\vm{{\bm{m}}}

\def\vs{{\bm{s}}}
\def\vt{{\bm{t}}}

\def\vw{{\bm{w}}}

\def\vz{{\bm{z}}}

\def\mW{{\bm{W}}}

\DeclareMathAlphabet{\mathsfit}{\encodingdefault}{\sfdefault}{m}{sl}
\SetMathAlphabet{\mathsfit}{bold}{\encodingdefault}{\sfdefault}{bx}{n}

\def\gC{{\mathcal{C}}}

\def\gE{{\mathcal{E}}}

\def\gG{{\mathcal{G}}}

\def\gJ{{\mathcal{J}}}
\def\gK{{\mathcal{K}}}

\def\gS{{\mathcal{S}}}
\def\gT{{\mathcal{T}}}

\newcommand{\E}{\mathbb{E}}
\newcommand{\Ls}{\mathcal{L}}
\newcommand{\R}{\mathbb{R}}

\newcommand{\softmax}{\mathrm{softmax}}

\usepackage{hyperref}
\usepackage{url}

\usepackage{algorithm}
\usepackage{algpseudocode}
\usepackage{booktabs}
\usepackage{graphicx}
\usepackage{bbm, dsfont}
\def\appref#1{Appendix~\ref{#1}}

\usepackage{nicefrac}
\usepackage{xcolor}

\usepackage[toc,page,header]{appendix}
\usepackage{titletoc}
\usepackage{tcolorbox}
\newcounter{backmatter}
\newcommand{\backmattersection}[1]{%
  \stepcounter{backmatter}%
  \section*{\large #1}%
  \pdfbookmark[1]{#1}{backmatter.\thebackmatter}%
}
\newcommand{\latfix}{\textbf{Lattice fixed.}\ }

\title{Let CSP Be Your ANCHOR: Adaptive Crystal Search over Frozen Structure Priors}

\newcommand{\eqmark}{\textsuperscript{*}}
\IfFileExists{latexml.sty}{\usepackage{latexml}}{\newif\iflatexml}

\iflatexml
  \author{Emma Lei Hovmand\eqmark \\ {Department of Energy Conversion and Storage, Technical University of Denmark}
    \and Jonas Elsborg\eqmark \\ {Department of Energy Conversion and Storage, Technical University of Denmark}
    \and Melih Kandemir \\ {Department of Mathematics and Computer Science, University of Southern Denmark}
    \and Arghya Bhowmik \\ {Department of Energy Conversion and Storage, Technical University of Denmark}
         \\ {\eqmark Equal contribution.}}
\else
  \author{Emma Lei Hovmand\textsuperscript{1,*}, Jonas Elsborg\textsuperscript{1,*}, Melih Kandemir\textsuperscript{2}, Arghya Bhowmik\textsuperscript{1}}
  \affiliation{\textsuperscript{1}Department of Energy Conversion and Storage, Technical University of Denmark\\
               \textsuperscript{2}Department of Mathematics and Computer Science, University of Southern Denmark}
  \contribution[*]{Equal contribution.}
\fi

\abstract{
De novo crystal generation (DNG) models decide where to search in composition space and how to generate structures with one set of weights. We argue that discovery is better served by separating the two. A crystal structure prediction (CSP) model is a physical prior that should be improved by likelihood training, while rewards, including novelty measured against the search's own history, should act on a search over compositions. We introduce \textsc{ANCHOR}, a GRPO composition policy trained with multi-objective rewards around a frozen CSP model, and continuous adaptive novelty (CAN), a graded novelty score against known structures and a growing discovery history. Using the frozen CSP model as a fixed ruler under one evaluator, we test where adaptation should act. Replacing DNG compositions with \textsc{ANCHOR}'s policy on the same CSP backbone raises MSUN from $11.4\%$ to $47.6\%$ and SUN from $1.1\%$ to $22.1\%$ at $99.9\%$ formula uniqueness. Fine-tuning DNG models directly on the same rewards instead moves their composition marginal without raising their on-hull fraction. We show that KL-regularized fine-tuning of a DNG model can only reweight chemistry the pretrained model already supports by a bounded factor, while unregularized DNG fine-tunes move toward known or less stable chemistry. Even a stability-only reward routed into \textsc{ANCHOR}'s CSP backbone roughly halves SUN relative to the frozen backbone, whereas likelihood training on structures found during search can improve a CSP backbone. Under MatterGen's evaluation pipeline, \textsc{ANCHOR} raises state-of-the-art MSUN from $29.2\%$ to $41.3\%$, transfers without retraining to two further CSP backbones, and reaches $47.1\%$ after distillation into Crystalite-CSP. As with any model optimised against a potential, its on-hull rate depends on that potential.}

\correspondence{EH: \email{eclh@dtu.dk}, AB: \email{arbh@dtu.dk}}

\newcommand{\ms}[2]{#1$_{\scriptscriptstyle\pm#2}$}
\begin{document}

\maketitle
\iflatexml
  {\small\eqmark Equal contribution.}\par
\fi

\begin{figure}[ht!]
    \centering
    \includegraphics[width=0.999\linewidth]{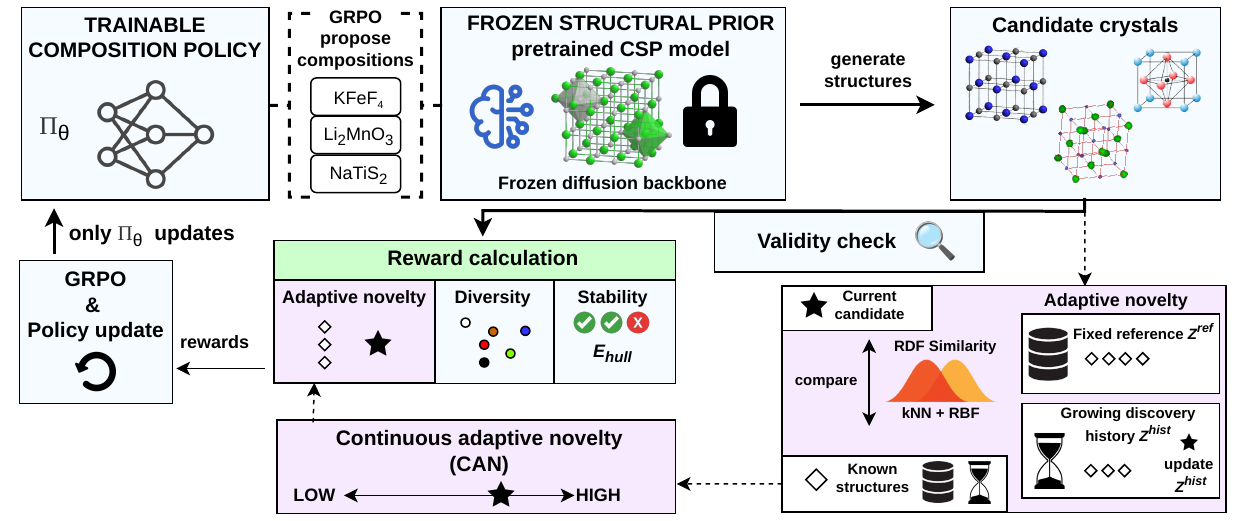}
    \caption{\textbf{\textsc{ANCHOR}.} A GRPO-trained composition policy queries a frozen crystal diffusion model to generate candidate structures. Candidates are scored by stability, diversity, and continuous adaptive novelty (CAN), a graded RDF-based score that takes the minimum novelty relative to a fixed reference index and the agent's growing discovery history. Only the composition policy is updated. 
    }
    \label{fig:placeholder}
\end{figure}

\section{Introduction \& Motivation}
\label{sec:introduction}
Crystalline materials underpin energy storage, catalysis, and electronics, but only a small fraction of possible compositions and structures are stable, and although many stable materials likely remain unexplored~\citep{merchant2023scaling}, experiments and first-principles calculations are expensive. Most diffusion-based de novo crystal generation (DNG) models~\citep{xie2021crystal,jiao2023crystal,cornet2025kinetic,zeni2025generative} learn a joint distribution
$p_0(c,x)=p_0(c)p_0(x\mid c)$ over crystal structures $x$ with chemical composition $c$. The conditional distribution $p_0(x\mid c)$ describes the crystal structure prediction (CSP) problem at fixed composition $c$, while $p_0(c)$ reflects which compositions are probable under the training distribution. The latter is a useful chemical prior, but not the distribution that should be searched, since a CSP model may remain effective for compositions that $p_0(c)$ rarely queries. The marginal answers what chemistry resembles what has been observed, while discovery asks what to investigate next. Reward-guided DNG can move beyond this learned composition marginal~\citep{park2026guiding,chen2025accelerating}, but does so by adapting the joint generator. A crystal-level reward can therefore alter both chemical exploration and the structural distribution, even when the useful adaptation is primarily compositional. Current models already work around this entanglement approximately, but use strong KL regularization that also restricts chemical movement~\citep{park2026guiding}, or a down-weighted atom-type loss because compositional memorization is hard to control independently of structural quality~\citep{veljkovic2026crystalite}. Novelty sharpens the problem. Conventional novelty is evaluated against a fixed reference collection~\citep{zeni2025generative,betala2025lemat}, so a rediscovered structure earns the same novelty reward the second time it is seen. Search history can discourage rediscovery~\citep{turk2022assessing,oschinski2026general,bellemare2016unifying}, but its value changes while the crystal itself does not, so a structural generator optimized on it is asked to adapt to campaign history rather than to physics.

We therefore give each role its own model and signal. The CSP model is treated as a physical prior, trained on structures and improved with structural signals, and a separate composition policy absorbs every discovery signal, including history-dependent novelty. A frozen CSP model also acts as a fixed ruler, since querying it with any model's compositions measures what that model's composition marginal alone contributes. With this ruler, we find that the composition marginal and not the structural model limits DNG discovery, that reward fine-tuning a DNG model mostly moves its marginal toward compositions where the reward is easiest to collect, collapsing formula uniqueness to $40$-$50\%$ on one backbone, and that even a stability-only reward routed into a CSP backbone roughly halves its SUN. Likelihood training on the structures the search found instead improves the backbone. Where a reward is applied therefore matters as much as what it measures. Rewards should guide the search over compositions, and likelihood should train the structural prior.

\paragraph{Contributions.}
We introduce \textsc{ANCHOR}, \textbf{A}daptive \textbf{N}ovelty for \textbf{C}rystals with \textbf{H}istory-aware \textbf{O}uter-loop \textbf{R}einforcement. Our contributions are:
\begin{enumerate}
\item \textbf{Composition search over a structure prior.} \textsc{ANCHOR} replaces the likelihood-trained composition marginal of DNG with a GRPO policy that queries a frozen CSP model, so discovery rewards change where to search, not how structures are generated.

\item \textbf{Continuous adaptive novelty (CAN).} A graded novelty score against known structures and the search's own history, so a crystal stops earning reward once it is found. We show that CAN alone prevents the search from collapsing onto repeated compositions, and that its graded and adaptive components together raise SUN from $8.9\%$ with a binary novelty flag to $22.1\%$.

\item \textbf{Separation of composition and structure effects.} Using a frozen CSP model as a fixed ruler under one evaluator, we show that replacing DNG compositions with \textsc{ANCHOR}'s policy raises MSUN more than 3-fold and SUN more than 10-fold on the same backbone. With our reward, fine-tuning through the denoising chain does not raise the on-hull fraction of DNG or CSP models, while distillation via likelihood training on discovered structures raises the SUN of Crystalite-CSP from $20.4\%$ to $26.0\%$ on the same compositions.

\item \textbf{State-of-the-art MatterGen results and transferability.} Under MatterGen's pipeline, \textsc{ANCHOR} reaches $41.3\%$ MSUN against $29.2\%$ for the previous best, rising to $47.1\%$ after likelihood distillation into Crystalite-CSP, and $16.0\%$ on-hull SUN against at most $2.5\%$ for any DNG model, and transfers to two further CSP backbones without retraining.
\end{enumerate}

\section{Background}
\label{sec:background}

\paragraph{Crystal generation and evaluation.}
Crystal structure prediction (CSP) generates structures for a fixed composition, while de novo crystal generation (DNG) generates composition and structure jointly~\citep{jiao2023crystal,xie2021crystal,cornet2025kinetic}. A composition $c$ specifies elements from a vocabulary $\gE$ and their stoichiometry, and a crystal structure $x=(L,R,\vt)$ adds the lattice $L\in\mathbb{R}^{3\times3}$, the fractional coordinates $R\in[0,1)^{T\times3}$ of its $T$ atoms, and their species $\vt\in\gE^{T}$, consistent with $c$.
This gives a natural decomposition of DNG into composition selection and CSP via $p(c,x)=p(c)p(x\mid c)$, where $p(x\mid c)$ represents the CSP subproblem at fixed composition. This factorization is conceptual, since CSP and DNG models are typically trained separately rather than constructing DNG from a pretrained CSP backbone.  Modern benchmarks evaluate validity, stability, novelty, uniqueness, and diversity~\citep{betala2025lemat}, commonly summarized by SUN (Stable, unique, novel) or MSUN (metastable, unique, novel). Further details on prior DNG/CSP models and benchmarks are given in Appendix~\ref{app:related_work}. 

\paragraph{Novelty and continued discovery.}
Novelty in crystal generation is conventionally evaluated against a fixed reference collection, while uniqueness measures repetition among generated samples~\citep{zeni2025generative,betala2025lemat}. Continuous structural metrics provide graded alternatives to binary structure matching~\citep{valle2010crystal,negishi2026continuous}, but are still evaluated against fixed reference data and so provide no memory of structures found during an ongoing search. 

\paragraph{Reward-guided crystal generation.}
Composition-level RL has been used to search chemical space, with structure prediction performed downstream for selected compositions~\citep{karpovich2024deep}. More recent work applies rewards directly to crystal generators, such as MatInvent, which uses RL for single- and multi-objective diffusion-based crystal design~\citep{chen2025accelerating}.
These approaches can move the generated composition distribution away from the likelihood-trained marginal, but they do so by adapting the joint crystal generator.
Chemeleon2 provides a novel formulation by treating the reverse-denoising process as an RL policy~\citep{park2026guiding}. Starting from noise $z_T$, the denoising trajectory is $\tau=z_T\rightarrow z_{T-1}\rightarrow\cdots\rightarrow z_0$, and the decoder $D_\phi$ maps the final latent state to the generated crystal $x=D_\phi(z_0)$. The RL objective maximizes the expected terminal reward
\begin{equation}
J(\theta)
=
\mathbb{E}_{\tau\sim\pi_\theta}
\left[
R(x)
\right].
\label{eq:chemeleon_rl}
\end{equation}
Because the reward is observed only for the final decoded crystal, it must be attributed across the preceding denoising decisions, creating a temporal credit-assignment problem.

However, the joint optimization also obscures whether adaptation should occur in composition or structure.  \citet{park2026guiding} identify a mismatch between likelihood-based pretraining and their novelty- and uniqueness-based "creativity" objective and use strong KL regularization toward the pretrained denoiser to restrict reward-driven drift.


More generally, reward-driven fine-tuning is known to potentially reduce diversity and sample quality in diffusion models~\citep{uehara2024fine}. The optimized model still follows a reward-tilted endpoint distribution~\citep{bergmeister2026reinforce}, which is particularly relevant for novelty objectives that can favor regions assigned lower probability by the pretrained model. This motivates separating the distribution that decides where to search from the model that generates structures.

\section{Methods}
\label{sec:method}
\paragraph{The DNG composition marginal.}
Under the factorization of Section~\ref{sec:background}, discovery effort is allocated as the product of a composition distribution and the success rate of the conditional it queries. For any conditional structural model $g(x\mid c)$, we can define
\[
s_g(c)
=
\Pr_{x\sim g(x\mid c)}\left[S(x,c)=1\right]
\]
as its discovery success rate at composition $c$, under a binary discovery criterion $S(x,c)\in\{0,1\}$. Likelihood-trained DNG models collapse this design space into one generative model. However, its two factors can be chosen independently, and \textsc{ANCHOR} takes the one combination that a jointly trained generator cannot reach: a composition policy $\pi_\theta(c)$ trained by reinforcement learning against the frozen conditional it queries. Successful candidates are then allocated as
\begin{equation}
\begin{aligned}
\text{DNG} &\quad &&p_0(c)\,s_{\mathrm{DNG}}(c),\\
\text{DNG compositions} + \text{CSP} &\quad &&p_0(c)\,s_{\mathrm{CSP}}(c),\\
\textsc{ANCHOR} &\quad &&\pi_\theta(c)\,s_{\mathrm{CSP}}(c).
\end{aligned}
\label{eq:discovery_decomposition}
\end{equation}
These substitutions give reciprocal controls: holding the composition list fixed isolates the structural generator, while holding the CSP model fixed isolates composition allocation. Replacing $p_0(c)$ with the marginal $q(c)$ of a reward fine-tuned generator gives the same control for any fine-tune, so a frozen CSP model acts as a fixed ruler for what fine-tuning did to the compositions. We expect $s_{\mathrm{CSP}}\geq s_{\mathrm{DNG}}$ because a CSP model is trained only to solve the structure problem at fixed composition whereas $p_{\mathrm{DNG}}(x\mid c)$ must also account for which compositions are probable. Separately, a composition can have high $s_{\mathrm{CSP}}(c)$ and still be rarely reached, because likelihood training optimizes $p_0(c)$ for reproducing the reference distribution rather than for discovery. \appref{app:novelty_tension} shows that the optimum is $p_0(c)$ reweighted by a per-composition partition function. Because the reward is bounded, so is this reweighting, since $q^*(c)/p_0(c)\le e^{\Delta R/\beta}$ for a reward range $\Delta R$ (\eqref{eq:bounded_tilt}), and KL-regularized fine-tuning can therefore reweight chemistry the pretrained model already proposes, but cannot reach chemistry to which it assigns negligible probability. Removing the KL term lifts this bound, but it also removes the only constraint on the conditional. Reward fine-tuning therefore moves the marginal toward compositions where the reward is easy to collect, which need not be where discovery succeeds, and it also tilts the conditional, which is the coupling \textsc{ANCHOR} removes. Section~\ref{sec:results} evaluates all three rows of~\eqref{eq:discovery_decomposition} under one evaluator.

Furthermore, reward-guided fine-tuning can move the adapted marginal $q(c)$ away from $p_0(c)$, but a crystal-level reward does not specify whether adaptation should occur in $q(c)$ or in $q(x\mid c)$. This adds a semantic credit assignment problem on top of the unavoidable temporal credit assignment across denoising steps implicit in Eq.~\ref{eq:chemeleon_rl}. A single KL coefficient does not separate them either:
\begin{equation}
D_{\mathrm{KL}}\!\left(q(c,x)\|p_0(c,x)\right)
=
D_{\mathrm{KL}}\!\left(q(c)\|p_0(c)\right)
+
\mathbb{E}_{c\sim q(c)}
D_{\mathrm{KL}}\!\left(q(x\mid c)\|p_0(x\mid c)\right).
\label{eq:kl_factorization_main}
\end{equation}
Raising the joint KL strength to restrict drift in $q(x\mid c)$ therefore also restricts the chemical adaptation needed to move beyond the pretrained marginal. Any reward that varies across structures at fixed composition can therefore change the conditional. This effect is amplified for history-dependent novelty. With $R_t(x)=R_{\mathrm{nov}}(x;Z^{\mathrm{ref}},Z^{\mathrm{hist}}_t)$ at search step $t$, the KL-regularized optimum is $q_t^*(x\mid c)\propto p_0(x\mid c)\exp(R_t(x)/\beta)$~\citep{bergmeister2026reinforce}, so a structure the agent has already found loses probability although its physical plausibility is unchanged. However, a crystal does not become less physically plausible because it has been discovered. It just becomes less useful to sample again. The straightforward solution is to freeze $p_0(x\mid c)$ while leaving the composition policy free. A novelty- or history-driven reward carries no information about how to improve the structural model. The structures the search finds do, and can serve as likelihood training data for the prior. \appref{app:novelty_tension} develops these arguments in more detail.

\paragraph{Separating discovery from structure generation.}
\textsc{ANCHOR} makes the separation architectural by replacing the pretrained marginal $p_0(c)$ with an adaptive policy and freezing the conditional:
\begin{equation}
    p_{\textsc{ANCHOR}}(c,x)=\pi_\theta(c)\,p_{\mathrm{CSP}}(x\mid c).
    \label{eq:anchor_ansatz}
\end{equation}
The policy can now concentrate search where the frozen CSP is already productive, including compositions the original DNG marginal rarely reaches. Because $p_{\textsc{ANCHOR}}(x\mid c)=p_{\mathrm{CSP}}(x\mid c)$ by construction, the structural term in Eq.~\ref{eq:kl_factorization_main} vanishes identically rather than being suppressed by a penalty that must be tuned. The reward can therefore act on $\pi_\theta$ without any route to the structural model, and can include history-dependent novelty without inducing undesirable reward tilting. We note that freezing the CSP does not freeze structural exploration, since repeated queries at the same composition can still sample different structures from $p_{\mathrm{CSP}}(x\mid c)$. Freezing is a property of the search rather than of the prior, which can be improved between searches by likelihood training on the structures they find (Section~\ref{sec:results}). 

\paragraph{Discovery as composition-level RL.} Each rollout samples a composition action
\begin{equation}
  c = \bigl(k,\; \ve,\; T,\; \rho \bigr)\sim\pi_\theta(\cdot\mid\vs),
  \label{eq:action}
\end{equation}
which fixes the number of distinct elements $k$, their identities $\ve$, the cell size $T$ and the stoichiometry $\rho$, and hence the reduced formula $f(c)$ (\appref{app:elem_vocab_action_space}). The frozen CSP model~\citep{cornet2025kinetic} samples a structure for $c$, which is relaxed over atomic positions for at most $100$ BFGS steps under MatterSim-1M~\citep{yang2024mattersim} before it is scored (Figure~\ref{fig:placeholder}). We optimize $\pi_\theta$ with GRPO, which standardizes each reward within a group of $G$ actions sampled under the same state (\appref{app:policy}).

\paragraph{Continuous Adaptive Novelty.}
Fixed-reference novelty, whether binary or graded~\citep{negishi2026continuous}, keeps rewarding a structure after it has been found. CAN instead scores each valid relaxed structure against both known structures and the search's own history. We describe a structure $x$ by an $\ell_1$-normalized radial distribution function (RDF) histogram $\vz(x)$ and compare it only with structures of the same reduced formula $f$, which separates structural novelty within a formula family from the chemical exploration done by the policy. Against a set $Z$ of such embeddings, novelty is the $k$-nearest-neighbour RBF distance
\begin{equation}
  \nu(\vz; Z)
  \;=\;
    \mathrm{clip}\Biggl(
    1 - \frac{1}{\lvert \gJ\rvert}
    \sum_{j \in \gJ}
    \exp\Bigl(-\frac{\lVert \vz - \vz_j \rVert_2^{2}}{2\sigma^{2}}\Bigr),
    \; 0,\; 1
  \Biggr),
  \qquad
  \gJ = \mathrm{kNN}_{k_{nn}}(\vz, Z).
  \label{eq:knn-novelty}
\end{equation}
The agent keeps a growing history $Z^{\mathrm{hist}}$ of the RDFs it has produced alongside the fixed MP-20 reference $Z^{\mathrm{ref}}$, and CAN takes the lower of the two scores,
\begin{equation}
  \nu(x)=\min\bigl(\nu(\vz;Z^{\mathrm{ref}}),\;\nu(\vz;Z^{\mathrm{hist}})\bigr),
  \label{eq:novelty}
\end{equation}
where an index with no structure of formula $f$ is omitted and a formula found in neither receives the bootstrap score $\nu_0$. After scoring, $\vz(x)$ joins $Z^{\mathrm{hist}}$, so regenerating a structure the policy has already found drives its novelty to zero and sustained reward requires continued discovery (\appref{app:reward/novelty}).

\paragraph{Reward and multi-objective GRPO.} Each candidate first passes a validity gate. Valid structures are scored by stability $s_{\mathrm{stab}}=1-\mathrm{clip}(E_{\mathrm{hull}},0,1)$, CAN $\nu$, and structural and compositional diversity within the group, collected as $\vm(c)=(s_{\mathrm{stab}},\nu,d_{\mathrm{struct}},d_{\mathrm{comp}})\in[0,1]^4$ (\appref{app:reward/stability}, \appref{app:reward/novelty} and \appref{app:reward/diversity}). With a count bonus that decays as a reduced formula is revisited~\citep{thiede2022curiosity}, the reward is
\begin{equation}
  r(c)
  \;=\;
  \begin{cases}
    \vm(c)^{\top}\vw \;+\; \dfrac{\lambda_c}{\sqrt{\max(n_f, 1)}},
      & \text{$c$ valid},\\[8pt]
    r_{\mathrm{inv}}, & \text{otherwise},
  \end{cases}
  \label{eq:reward}
\end{equation}
where $n_f$ counts valid generations of $f$ in the run, so the count bonus prices repeated compositions while CAN prices repeated structures. Because these terms compete and a fixed weighting would need tuning, we sample $\vw\sim\mathrm{Dirichlet}(\bm{\alpha})$ and an element sub-group $\gS\subseteq\gE$ for each rollout, hold both fixed across the group, and supply them to the policy through $\vs$ (\eqref{eq:state}). The policy is thereby conditioned on the trade-off it is asked to make, the group-relative advantage compares actions under one scalarization and one element set, and masking element draws to $\gS$ spreads exploration across the periodic table. At evaluation, $\vw$ is fixed to the Dirichlet mean (\appref{app:policy}).

\section{Experiments}
\label{sec:results}
\paragraph{Setup and evaluation.}
The main \textsc{ANCHOR} model is trained with a frozen KLDM-CSP backbone, $G=32$, $r_{\mathrm{inv}}=-0.2$ and MatterSim-1M as the training potential (\appref{app:stage1}).
Training uses the positions-only relaxation of Section~\ref{sec:method} because relaxation dominates the cost of the reward (\appref{app:relax-budget}), and evaluation is configured for fidelity instead.
Every internal number is produced after training by sampling $N_{\text{eval}}=1000$ post-training generations across three seeds. Structures are relaxed with MatterSim-5M under BFGS for at most 400 steps with positions and lattice both free, and scored against one hull. \appref{app:eval-cellrelax} repeats every internal table with the lattice fixed, with the same conclusions. We report \textbf{Valid} (the validity gate, see \appref{app:reward/validity}), \textbf{Unique} (distinct reduced formulas), \textbf{Novel}, and \textbf{SUN}\,/\,\textbf{MSUN} (valid $\wedge$ unique $\wedge$ novel $\wedge\ E_{\mathrm{hull}}\le0$ / $\le0.1$~eV/atom). We keep these names for the MatterGen and LeMat-GenBench pipelines, and write Stable and Meta for the stability rates at $E_{\mathrm{hull}}\le0$ and $\le0.1$~eV/atom without the uniqueness and novelty conditions.  We selected the main configuration from eight runs (\appref{sec:app-setup}, \appref{app:stage1} and \appref{app:stage2-invalid_penalty}). Table~\ref{tab:explore-ablation} ablates the exploration mechanisms, Table~\ref{tab:internal} varies where adaptation acts under our internal evaluator, and Table~\ref{tab:mattergen_dng} scores the same rows with MatterGen's pipeline next to published DNG results, with LeMat-GenBench in \appref{app:eval-lemat}. In Table~\ref{tab:internal}, paired comparisons use matched compositions wherever the structural generator is being isolated: the $\to$ CSP rows hold the preceding model's compositions fixed while changing only the structure generator, and the likelihood-tuned rows query each CSP model with the same \textsc{ANCHOR} compositions. 

\begin{table}[t]
\centering
\scriptsize
\renewcommand{\arraystretch}{0.95}
\setlength{\tabcolsep}{3pt}
\caption{Exploration-mechanism ablation ($N_{\text{eval}}=1000$, MatterSim-5M, BFGS-400, positions and lattice free; \appref{app:eval_protocol}). Rows give the full method, each mechanism removed (--), each mechanism alone, and the full method with CAN replaced by the binary novelty flag of Chemeleon2 (\appref{app:explore_ablation}). Removing adaptive novelty scores novelty against the frozen MP-20 index alone. All columns are percentages except \emph{MP-20}, which counts valid samples whose reduced formula occurs in MP-20. Best in MSUN and SUN is bold, second-best underlined. Runs were matched on wall clock. Entries are mean $\pm$ std over three sampling seeds.}
\label{tab:explore-ablation}
\begin{tabular*}{\textwidth}{@{\extracolsep{\fill}}lcccccccc@{}}
\toprule
Configuration & Valid & Unique & Novel & Meta & Stable & MSUN & SUN & MP-20 \\
\midrule
\textbf{Full method}     & \ms{97.5}{0.4} & \ms{99.9}{0.1} & \ms{97.3}{0.5} & \ms{47.7}{2.0} & \ms{22.2}{1.2} & \underline{\ms{47.6}{2.1}} & \textbf{\ms{22.1}{1.2}} & \ms{0}{0} \\
\midrule
\quad-- count bonus      & \ms{96.4}{0.2} & \ms{100.0}{0.0} & \ms{96.4}{0.3} & \ms{32.8}{1.5} & \ms{12.6}{0.4} & \ms{32.8}{1.5} & \ms{12.6}{0.4} & \ms{0}{0} \\
\quad-- sub-group mask   & \ms{96.6}{0.3} & \ms{99.9}{0.1} & \ms{96.5}{0.1} & \ms{43.5}{2.5} & \ms{11.6}{0.8} & \ms{43.4}{2.4} & \ms{11.6}{0.9} & \ms{0}{0} \\
\quad-- adaptive nov.    & \ms{97.0}{0.1} & \ms{99.8}{0.1} & \ms{96.9}{0.0} & \ms{39.7}{2.1} & \ms{14.7}{2.0} & \ms{39.7}{2.1} & \ms{14.7}{2.0} & \ms{0}{0} \\
\midrule
adaptive nov.\ only      & \ms{97.7}{0.6} & \ms{100.0}{0.0} & \ms{97.6}{0.6} & \ms{52.0}{0.4} & \ms{16.3}{2.0} & \textbf{\ms{52.0}{0.5}} & \underline{\ms{16.3}{2.0}} & \ms{0}{0} \\
count bonus only         & \ms{96.0}{0.6} & \ms{100.0}{0.0} & \ms{96.0}{0.6} & \ms{41.3}{1.4} & \ms{12.4}{0.8} & \ms{41.3}{1.4} & \ms{12.4}{0.8} & \ms{0}{0} \\
sub-group mask only      & \ms{99.7}{0.2} & \ms{6.2}{0.1}   & \ms{99.6}{0.2} & \ms{66.8}{0.6} & \ms{21.6}{0.8} & \ms{3.4}{0.2}  & \ms{1.0}{0.1}  & \ms{358}{1} \\
none                     & \ms{99.9}{0.1} & \ms{0.7}{0.5}   & \ms{99.9}{0.1} & \ms{96.3}{0.8} & \ms{69.0}{0.4} & \ms{0.4}{0.3}  & \ms{0.3}{0.4}  & \ms{997}{1} \\
\midrule
binary novelty flag      & \ms{97.2}{0.0} & \ms{100.0}{0.1} & \ms{97.2}{0.1} & \ms{25.1}{2.3} & \ms{8.9}{0.8}  & \ms{25.0}{2.3} & \ms{8.9}{0.8}  & \ms{0}{0} \\
\bottomrule
\end{tabular*}
\end{table}

\paragraph{Adaptive novelty sustains discovery.}
Table~\ref{tab:explore-ablation} ablates the three mechanisms that penalize sampling the same compositions: adaptive novelty, count bonus, and element sub-group mask, as a full factorial. With none of them enabled, the policy collapses to $0.7\%$ unique compositions, so although $96.3\%$ of structures are metastable, MSUN is $0.4\%$. Adaptive novelty alone restores $100\%$ uniqueness and reaches the highest MSUN ($52.0\%$, against $41.3\%$ for the count bonus alone), while sub-group masking alone does not prevent collapse. The full method has the highest SUN ($22.1\%$ against $16.3\%$ for adaptive novelty alone) at $4.4$ points lower MSUN. Replacing the growing history with the frozen MP-20 index lowers SUN to $14.7\%$ and MSUN to $39.7\%$, and replacing the graded score with binary novelty lowers them to $8.9\%$ and $25.0\%$, so both components of the adaptive score matter. The graded score matters because a binary flag is constant on formulas outside MP-20, so it cancels in the group-relative advantage once the policy leaves the reference set (\appref{app:explore_ablation}). Over a full training run, the policy proposes $97{,}599$ distinct reduced formulas among $101{,}237$ valid structures, and its formula collisions with MP-20 fall from $346$ in the first fifth of the run to $3$ in the last. 

\begin{table*}[t]
\centering
\scriptsize
\renewcommand{\arraystretch}{0.95}
\setlength{\tabcolsep}{3pt}
\caption{\textbf{Separating composition and structure effects.} One validity gate, one relaxer (MatterSim-5M, BFGS-400, positions and lattice free), one reference hull, and one fixed MP-20 novelty index for every row; $N_{\mathrm{eval}}=1000$. All columns are percentages except \emph{MP-20}, which counts valid samples whose reduced formula occurs in MP-20. $\to$ CSP queries the corresponding frozen CSP model with the preceding model's compositions; \emph{FT} is DDPO reward fine-tuning at Dirichlet-mean (\emph{mean}) or equal stability and novelty (\emph{bal.}) objective weights. \emph{Untrained composition sources} are defined in \appref{app:composition_ladder}. In the \emph{likelihood-tuned CSP} block, each backbone is fine-tuned by likelihood on MP-20 alone (step-matched control) or with the structures found by \textsc{ANCHOR} added (+ \textsc{ANCHOR} data), and queried with the \textsc{ANCHOR} compositions. Entries are mean$_{\pm\mathrm{std}}$ over three sampling seeds. Best in MSUN and SUN is bold, second-best underlined. Row definitions in \appref{sec:app-eval-rows}.}
\label{tab:internal}

\begin{tabular*}{\textwidth}{@{\extracolsep{\fill}}lcccccccc@{}}
\toprule
Configuration & Valid & Unique & Novel & Meta & Stable & MSUN & SUN & MP-20 \\
\midrule

\multicolumn{9}{@{}l}{\textbf{Untrained composition sources (frozen KLDM-CSP)}} \\[2pt]
Random formulas
& \ms{9.8}{0.4} & \ms{100.0}{0.0} & \ms{8.3}{0.6} & \ms{2.4}{0.5}
& \ms{0.4}{0.1} & \ms{1.5}{0.4} & \ms{0.2}{0.0} & \ms{17.5}{3.5} \\

Enumerated charge-neutral
& \ms{68.9}{1.6} & \ms{100.0}{0.0} & \ms{68.2}{1.3} & \ms{9.4}{0.3}
& \ms{1.2}{0.1} & \ms{8.8}{0.0} & \ms{1.2}{0.1} & \ms{9.5}{4.9} \\

\midrule
\multicolumn{9}{@{}l}{\textbf{DNG and reward fine-tuned DNG (ours)}} \\[2pt]

Crystalite-DNG
& \ms{47.0}{2.7} & \ms{98.8}{0.1} & \ms{31.9}{1.8} & \ms{26.2}{2.1}
& \ms{2.8}{0.3} & \ms{14.0}{0.7} & \ms{1.5}{0.4} & \ms{165.7}{17.2} \\

\quad$\to$ CSP
& \ms{48.9}{2.6} & \ms{98.8}{0.1} & \ms{33.3}{1.4} & \ms{28.0}{3.0}
& \ms{2.9}{0.6} & \ms{15.0}{1.0} & \ms{1.7}{0.3} & \ms{167.3}{16.6} \\

\quad+ FT (mean)
& \ms{96.7}{0.4} & \ms{40.5}{0.5} & \ms{50.4}{1.1} & \ms{54.0}{0.7}
& \ms{1.9}{0.5} & \ms{20.2}{2.6} & \ms{0.9}{0.5} & \ms{631.5}{20.5} \\

\quad+ FT (bal.)
& \ms{97.0}{0.8} & \ms{49.5}{0.7} & \ms{61.1}{1.0} & \ms{51.4}{0.4}
& \ms{2.6}{0.9} & \ms{22.9}{0.8} & \ms{1.5}{0.7} & \ms{612.0}{25.5} \\

\quad+ FT (mean) $\to$ CSP
& \ms{97.2}{0.6} & \ms{40.5}{0.5} & \ms{50.0}{2.0} & \ms{47.0}{1.6}
& \ms{0.9}{0.1} & \ms{18.2}{2.3} & \ms{0.7}{0.2} & \ms{638.0}{17.0} \\

\quad+ FT (bal.) $\to$ CSP
& \ms{97.3}{0.5} & \ms{49.5}{0.7} & \ms{60.1}{1.8} & \ms{47.9}{0.7}
& \ms{1.6}{0.5} & \ms{21.0}{2.0} & \ms{1.0}{0.5} & \ms{614.0}{22.6} \\

KLDM-DNG
& \ms{44.6}{0.7} & \ms{99.2}{0.2} & \ms{35.6}{1.3} & \ms{18.0}{0.6}
& \ms{2.4}{0.7} & \ms{11.1}{0.9} & \ms{1.4}{0.4} & \ms{113.3}{9.2} \\

\quad$\to$ CSP
& \ms{43.8}{0.2} & \ms{99.2}{0.2} & \ms{34.4}{1.5} & \ms{19.0}{0.7}
& \ms{2.2}{0.6} & \ms{11.4}{0.8} & \ms{1.1}{0.2} & \ms{112.3}{9.3} \\

\quad+ FT (mean)
& \ms{53.4}{0.8} & \ms{99.2}{0.4} & \ms{44.0}{1.1} & \ms{19.3}{0.5}
& \ms{2.0}{0.4} & \ms{12.0}{0.6} & \ms{1.3}{0.3} & \ms{115.0}{13.0} \\

\quad+ FT (bal.)
& \ms{52.8}{2.4} & \ms{99.0}{0.5} & \ms{43.0}{1.9} & \ms{19.1}{0.9}
& \ms{2.1}{0.4} & \ms{11.4}{1.7} & \ms{1.1}{0.4} & \ms{114.7}{18.2} \\

\quad+ FT (mean) $\to$ CSP
& \ms{52.5}{0.7} & \ms{99.5}{0.1} & \ms{44.6}{0.9} & \ms{19.1}{0.0}
& \ms{2.1}{0.4} & \ms{13.3}{0.3} & \ms{1.1}{0.5} & \ms{109.5}{3.5} \\

\quad+ FT (bal.) $\to$ CSP
& \ms{50.5}{2.5} & \ms{98.8}{0.6} & \ms{41.7}{0.4} & \ms{18.1}{0.9}
& \ms{1.8}{0.4} & \ms{11.1}{0.6} & \ms{0.9}{0.4} & \ms{116.0}{24.0} \\

\midrule
\multicolumn{9}{@{}l}{\textbf{ANCHOR + frozen CSP}} \\[2pt]

Trained on KLDM
& \ms{97.5}{0.4} & \ms{99.9}{0.1} & \ms{97.3}{0.5} & \ms{47.7}{2.0}
& \ms{22.2}{1.2} & \ms{47.6}{2.1} & \ms{22.1}{1.2} & \ms{0.0}{0.0} \\

\quad$\to$ Crystalite
& \ms{93.1}{0.9} & \ms{99.9}{0.1} & \ms{93.0}{1.0} & \ms{46.8}{2.3}
& \ms{19.5}{1.9} & \ms{46.7}{2.3} & \ms{19.5}{1.9} & \ms{0.0}{0.0} \\

\quad$\to$ OMatG
& \ms{98.1}{0.2} & \ms{99.9}{0.1} & \ms{98.0}{0.3} & \ms{48.2}{2.3}
& \ms{21.8}{0.6} & \underline{\ms{48.1}{2.3}} & \ms{21.7}{0.6} & \ms{0.0}{0.0} \\

\midrule
\multicolumn{9}{@{}l}{\textbf{ANCHOR + reward-trained KLDM-CSP (GRPO)}} \\[2pt]

Shared reward
& \ms{97.6}{0.8} & \ms{99.9}{0.1} & \ms{97.5}{0.9} & \ms{22.7}{1.6}
& \ms{6.3}{1.3} & \ms{22.6}{1.7} & \ms{6.3}{1.3} & \ms{0.0}{0.0} \\

Stability-only CSP reward
& \ms{97.9}{0.5} & \ms{100.0}{0.0} & \ms{97.9}{0.5} & \ms{29.5}{0.7}
& \ms{10.7}{0.2} & \ms{29.5}{0.7} & \ms{10.7}{0.2} & \ms{0.0}{0.0} \\

\midrule
\multicolumn{9}{@{}l}{\textbf{ANCHOR + likelihood-tuned CSP}} \\[2pt]

KLDM-CSP
& \ms{96.8}{0.1} & \ms{99.9}{0.1} & \ms{96.7}{0.3} & \ms{48.0}{1.2}
& \ms{22.1}{0.9} & \ms{48.0}{1.3} & \ms{22.1}{1.0} & \ms{0.0}{0.0} \\

\quad+ \textsc{ANCHOR} data
& \ms{97.5}{0.4} & \ms{99.9}{0.1} & \ms{97.3}{0.5} & \ms{48.2}{2.4}
& \ms{22.8}{1.1} & \underline{\ms{48.1}{2.5}} & \underline{\ms{22.7}{1.1}} & \ms{0.0}{0.0} \\

Crystalite-CSP
& \ms{93.6}{1.2} & \ms{99.9}{0.1} & \ms{93.5}{1.1} & \ms{46.8}{1.1}
& \ms{20.5}{0.3} & \ms{46.7}{1.3} & \ms{20.4}{0.4} & \ms{0.0}{0.0} \\

\quad+ \textsc{ANCHOR} data
& \ms{98.8}{0.1} & \ms{99.9}{0.1} & \ms{98.7}{0.3} & \ms{52.9}{3.5}
& \ms{26.1}{1.3} & \textbf{\ms{52.8}{3.7}} & \textbf{\ms{26.0}{1.4}} & \ms{0.0}{0.0} \\

\bottomrule
\end{tabular*}
\end{table*}

\begin{table*}[t]
\centering
\scriptsize
\renewcommand{\arraystretch}{0.95}
\setlength{\tabcolsep}{3pt}
\caption{MP-20 under MatterGen's evaluation pipeline. MSUN is the quantity MatterGen and the cited works report as SUN; SUN applies the same conjunction at $E_{\mathrm{hull}}\le0$. All columns are percentages except $\overline{E}_h$ (eV/atom) and RMSD (\AA). Upper block: the strongest published variant of each model family, rounded to one decimal. All published results are in Table~\ref{tab:app-mattergen-baselines}. Middle blocks: our own sampling of the released Crystalite-DNG checkpoint and of a KLDM-DNG model we retrained, with their DDPO fine-tunes, which reproduce the published MSUN of their families ($23.6\%$ and $18.0\%$ against $24.3\%$ and $18.5\%$). Our rows report mean$_{\pm\mathrm{std}}$ over three sampling seeds, with row blocks as in Table~\ref{tab:internal}. Best in each column is bold, second-best underlined. \\
\textsuperscript{a}\cite{cornet2025kinetic}, \textsuperscript{b}\cite{veljkovic2026crystalite}, \textsuperscript{c}\cite{zhang2026crystalrepa}.}
\label{tab:mattergen_dng}
\begin{tabular*}{\textwidth}{@{\extracolsep{\fill}}lcccccccc@{}}
\toprule
Model & Unique & Novel & Meta & Stable & $\overline{E}_h$ & RMSD & MSUN & SUN \\
\midrule
\multicolumn{9}{@{}l}{\textbf{DNG baselines}} \\[2pt]
MatterGen-MP\textsuperscript{a}     & --   & --   & 47.1 & -- & 0.20 & \underline{0.15} & 25.8 & -- \\
KLDM-\(x_0\)(D)\textsuperscript{a}  & --   & --   & 59.2 & -- & 0.16 & 0.28 & 18.5 & -- \\
Crystalite\textsuperscript{b}       & 94.7 & 56.6 & \underline{64.5} & -- & 0.15 & 0.27 & 24.3 & -- \\
MatterGen+REPA\textsuperscript{c}   & 99.7 & 63.5 & \textbf{66.2} & -- & \underline{0.12} & \textbf{0.11} & 29.2 & -- \\
\midrule
\multicolumn{9}{@{}l}{\textbf{Untrained composition sources (frozen KLDM-CSP)}} \\[2pt]
Random formulas                & \ms{100.0}{0.0} & \ms{94.8}{0.3} & \ms{14.2}{2.2} & \ms{1.5}{0.5} & \ms{0.329}{0.012} & \ms{0.85}{0.02} & \ms{10.3}{1.7} & \ms{0.8}{0.4} \\
Enumerated charge-neutral      & \ms{100.0}{0.0} & \ms{99.2}{0.1} & \ms{10.7}{0.6} & \ms{1.8}{0.0} & \ms{0.325}{0.008} & \ms{1.08}{0.01} & \ms{10.1}{0.6} & \ms{1.8}{0.1} \\
\midrule
\multicolumn{9}{@{}l}{\textbf{DNG and reward fine-tuned DNG (ours)}} \\[2pt]
Crystalite-DNG           & \ms{99.3}{0.3} & \ms{63.1}{0.9} & \ms{57.0}{1.9} & \ms{7.2}{1.1} & \ms{0.176}{0.017} & \ms{0.32}{0.01} & \ms{23.6}{0.6} & \ms{2.3}{0.4} \\
\quad$\to$ CSP           & \ms{99.2}{0.4} & \ms{61.5}{1.3} & \ms{59.1}{1.7} & \ms{7.9}{0.7} & \ms{0.169}{0.017} & \ms{0.26}{0.01} & \ms{23.9}{0.5} & \ms{2.5}{0.4} \\
\quad+ FT (mean)         & \ms{61.9}{1.1} & \ms{64.8}{0.7} & \ms{51.9}{0.9} & \ms{4.4}{0.4} & \ms{0.154}{0.012} & \ms{0.46}{0.01} & \ms{17.2}{0.3} & \ms{1.5}{0.3} \\
\quad+ FT (bal.)         & \ms{68.8}{0.9} & \ms{64.8}{0.8} & \ms{51.9}{1.0} & \ms{4.3}{0.4} & \ms{0.175}{0.008} & \ms{0.48}{0.01} & \ms{19.4}{0.8} & \ms{1.6}{0.3} \\
\quad+ FT (mean)\ $\to$ CSP & \ms{43.0}{0.8} & \ms{51.4}{1.6} & \ms{49.5}{0.1} & \ms{1.9}{0.2} & \ms{0.250}{0.008} & \ms{0.22}{0.02} & \ms{10.5}{0.8} & \ms{0.8}{0.1} \\
\quad\rlap{+ FT (bal.)}\hphantom{+ FT (mean)}\ $\to$ CSP & \ms{52.2}{1.2} & \ms{51.5}{3.9} & \ms{47.7}{0.8} & \ms{2.2}{0.4} & \ms{0.262}{0.011} & \ms{0.22}{0.01} & \ms{13.4}{0.3} & \ms{1.1}{0.3} \\
KLDM-DNG                 & \ms{99.8}{0.1} & \ms{76.2}{0.7} & \ms{38.3}{1.4} & \ms{6.2}{0.6} & \ms{0.212}{0.009} & \ms{0.58}{0.02} & \ms{18.0}{0.9} & \ms{2.3}{0.4} \\
\quad$\to$ CSP           & \ms{99.8}{0.1} & \ms{75.1}{1.3} & \ms{40.6}{1.5} & \ms{6.1}{0.4} & \ms{0.200}{0.006} & \ms{0.59}{0.01} & \ms{18.8}{2.0} & \ms{2.1}{0.4} \\
\quad+ FT (mean)         & \ms{99.6}{0.3} & \ms{78.7}{1.0} & \ms{34.5}{1.2} & \ms{3.9}{0.4} & \ms{0.234}{0.007} & \ms{0.72}{0.02} & \ms{16.2}{0.8} & \ms{1.5}{0.5} \\
\quad+ FT (bal.)         & \ms{99.5}{0.4} & \ms{78.5}{1.6} & \ms{34.7}{2.4} & \ms{4.8}{0.6} & \ms{0.222}{0.013} & \ms{0.69}{0.02} & \ms{16.2}{1.1} & \ms{1.9}{0.2} \\
\quad+ FT (mean)\ $\to$ CSP & \ms{99.5}{0.1} & \ms{77.5}{1.3} & \ms{36.0}{1.2} & \ms{3.9}{0.9} & \ms{0.225}{0.015} & \ms{0.65}{0.01} & \ms{17.0}{2.0} & \ms{1.7}{0.6} \\
\quad\rlap{+ FT (bal.)}\hphantom{+ FT (mean)}\ $\to$ CSP & \ms{99.0}{0.6} & \ms{77.2}{1.2} & \ms{34.7}{2.5} & \ms{4.2}{1.1} & \ms{0.228}{0.006} & \ms{0.61}{0.01} & \ms{16.3}{0.6} & \ms{1.6}{0.1} \\
\midrule
\multicolumn{9}{@{}l}{\textbf{ANCHOR + frozen CSP (ours)}} \\[2pt]
Trained on KLDM          & \ms{100.0}{0.0} & \ms{100.0}{0.0} & \ms{41.3}{1.0} & \ms{16.0}{1.0} & \ms{0.138}{0.006} & \ms{1.39}{0.01} & \ms{41.3}{1.0} & \ms{16.0}{1.0} \\
\quad$\to$ Crystalite    & \ms{100.0}{0.0} & \ms{100.0}{0.0} & \ms{41.1}{2.1} & \ms{14.4}{0.7} & \ms{0.142}{0.006} & \ms{1.27}{0.02} & \ms{41.1}{2.1} & \ms{14.4}{0.7} \\
\quad$\to$ OMatG         & \ms{100.0}{0.0} & \ms{100.0}{0.0} & \ms{43.1}{0.9} & \underline{\ms{16.7}{0.6}} & \ms{0.136}{0.005} & \ms{1.27}{0.01} & \underline{\ms{43.1}{0.9}} & \underline{\ms{16.7}{0.6}} \\
\midrule
\multicolumn{9}{@{}l}{\textbf{ANCHOR + reward-trained KLDM-CSP (GRPO)}} \\[2pt]
Shared reward                 & \ms{100.0}{0.0} & \ms{100.0}{0.0} & \ms{17.0}{1.9} & \ms{4.7}{0.4} & \ms{0.211}{0.008} & \ms{1.37}{0.01} & \ms{17.0}{1.9} & \ms{4.7}{0.4} \\
Stability-only CSP reward     & \ms{100.0}{0.0} & \ms{100.0}{0.1} & \ms{22.6}{1.9} & \ms{8.1}{0.6} & \ms{0.187}{0.006} & \ms{1.27}{0.02} & \ms{22.6}{1.8} & \ms{8.1}{0.6} \\
\midrule
\multicolumn{9}{@{}l}{\textbf{ANCHOR + likelihood-tuned CSP (ours)}} \\[2pt]
KLDM-CSP                               & \ms{100.0}{0.0} & \ms{100.0}{0.0} & \ms{41.2}{2.1} & \ms{16.3}{0.7} & \ms{0.138}{0.008} & \ms{1.40}{0.01} & \ms{41.2}{2.1} & \ms{16.3}{0.7} \\
\quad+ \textsc{ANCHOR} data             & \ms{100.0}{0.0} & \ms{100.0}{0.0} & \ms{40.6}{1.4} & \ms{15.8}{1.0} & \ms{0.142}{0.006} & \ms{1.24}{0.02} & \ms{40.6}{1.4} & \ms{15.8}{1.0} \\
Crystalite-CSP                          & \ms{100.0}{0.0} & \ms{100.0}{0.0} & \ms{39.5}{2.7} & \ms{14.1}{0.3} & \ms{0.148}{0.011} & \ms{1.27}{0.02} & \ms{39.5}{2.7} & \ms{14.1}{0.3} \\
\quad+ \textsc{ANCHOR} data             & \ms{100.0}{0.0} & \ms{100.0}{0.0} & \ms{47.1}{2.4} & \textbf{\ms{22.0}{2.2}} & \textbf{\ms{0.114}{0.007}} & \ms{0.86}{0.02} & \textbf{\ms{47.1}{2.4}} & \textbf{\ms{22.0}{2.2}} \\
\bottomrule
\end{tabular*}
\end{table*}

\paragraph{The composition marginal limits DNG discovery.}
Replacing only the DNG generator with the corresponding CSP model at fixed DNG compositions changes MSUN and SUN by at most $1.0$ point under either pipeline, whereas replacing the composition source with \textsc{ANCHOR}'s policy at a fixed CSP backbone raises MSUN by $32$-$36$ points and SUN by a factor of $11$-$20$. Uniformly sampled charge-neutral formulas reach only $8.8\%$ MSUN, against $1.5\%$ for random formulas, so the policy learns more than charge neutrality. Under MatterGen, \textsc{ANCHOR} reaches $41.3\%$ MSUN against $29.2\%$ for the previous state-of-the-art and $18.5\%$ for the strongest published KLDM variant, and $14.4$-$16.7\%$ on-hull SUN, while every row sampling DNG compositions, including the fine-tuned ones, stays at or below $2.5\%$. Substituting Crystalite-CSP or OMatG-CSP under the KLDM-trained policy preserves the same high discovery rate, so the learned search transfers across structural priors without retraining. The pipelines differ in potential, optimiser, and novelty definition (\appref{app:eval-mattergen}), but rank \textsc{ANCHOR} against the DNG models identically.

\paragraph{Reward fine-tuning mostly moves compositions.}
We fine-tune both DNG models on the same reward as \textsc{ANCHOR} using DDPO~\citep{black2024training}, the standard policy gradient method for diffusion models. We do not use KL regularization so that the composition marginal is not constrained by the bounded reweighting in \eqref{eq:bounded_tilt} (\appref{app:reward-ft}). This doubles the validity and Meta of Crystalite-DNG, but does not improve the metastable fraction of valid structures and lowers the on-hull fraction, so validity drives the apparent MSUN gain. The fine-tune passes the gate by moving toward known chemistry: MP-20 formula overlap rises sharply,    uniqueness falls to roughly half, so the model increasingly repeats its own formulas. This is the compositional memorization that \citet{veljkovic2026crystalite} limit during pretraining by down-weighting the atom-type loss, a constraint the reward gradient does not have. Novel rises only with validity, as the novel fraction of valid samples (Novel/Valid in Table~\ref{tab:internal}) falls from $68\%$ to $52$-$63\%$, so the fine-tune optimizes around the history-dependent term rather than following it, as expected for a target that changes with the search history (Section~\ref{sec:method}). MatterGen, which does not gate stability on validity, confirms that both fine-tunes lower Meta ($51.9\%$ against $57.0\%$) and Stable ($4.3$-$4.4\%$ against $7.2\%$). When realised by the frozen Crystalite-CSP, the fine-tuned compositions alone lower stability, while the fine-tuned structural model recovers only about one point, showing that the loss comes primarily from the compositions rather than structural drift. KLDM-DNG retains high uniqueness, but its fine-tuned structural model is likewise no better than the frozen CSP at the same compositions. When realised by the frozen CSP, its compositions become slightly more novel but less stable under MatterGen. Dirichlet-mean weights give an on-hull fraction of $1.9$-$2.0\%$, and raising the stability weight to equal novelty lifts it only to $2.1$-$2.6\%$, against $2.8\%$ and $2.4\%$ for the pretrained models, consistent with the reward having to be assigned across the denoising steps (Section~\ref{sec:background}). Even without a KL term, both fine-tunes behave as the reward tilting of Section~\ref{sec:method} describes (\appref{app:novelty_tension}), and concentrate on compositions where the reward is easiest to collect, which for Crystalite-DNG lie within the chemistry the pretrained model already proposes. \citet{park2026guiding} report larger gains for a latent DNG model trained with GRPO and strong KL regularization under a fixed-reference novelty reward, raising metastability from $51.2\%$ to $72.1\%$ under the potential used for training. Much of this change is compositional, as their fine-tuned samples reach novelty through new element combinations on familiar prototype structures, and their policy collapses onto a single composition when diversity reward is removed~\citep{park2026guiding}. Our fine-tunes differ in model, reward, and the KL penalty (\appref{app:reward-ft}), but both results are consistent with reward fine-tuning acting mainly through the composition marginal, which querying a frozen CSP model with the fine-tuned compositions would separate. On the same reward, \textsc{ANCHOR} steers completely away from these rediscoveries, with no MP-20 formulas against $612$-$632$ per $1000$ for the Crystalite-DNG fine-tunes.

\paragraph{Structure improves through structural signals.}
Since MSUN $\approx$ Meta and SUN $\approx$ Stable\ for \textsc{ANCHOR}, its discovery rate is bounded by how often the CSP model generates stable structures, and its structures move further during relaxation than those of any DNG model (RMSD $1.27$-$1.39$\,\AA{} against at most $0.72$\,\AA{}), so the relaxer still does much of the structural work. Letting GRPO update the backbone with the full reward exposes it to the same tilting and reaches only $22.6\%$ MSUN and $6.3\%$ SUN. Because \textsc{ANCHOR} separates the two factors, we can instead route only the stability term to the backbone while the policy still receives the full reward. This keeps history-dependent novelty away from the structural model and raises MSUN to $29.5\%$ and SUN to $10.7\%$, with the same ordering under MatterGen, but still roughly halves SUN against the frozen backbone ($22.1\%$). Part of this gap may come from co-training against a moving backbone, but a terminal reward still does little to improve structures through the denoising chain, even when it measures only stability. To test whether the discovered structures have any useful structural signal, we fine-tune the backbone by likelihood on MP-20 together with $27{,}136$ randomly drawn valid structures from one full \textsc{ANCHOR} run, with no stability or novelty threshold, thereby doubling the training data. None of the structures used for the three-seed evaluation is part of this fine-tuning corpus. To isolate the structural model, we compare against a step-matched control on MP-20 alone and query both with the same \textsc{ANCHOR} compositions as the frozen backbone. On Crystalite-CSP, this raises internal MSUN/SUN from $46.7/20.4\%$ to $52.8/26.0\%$, the highest values in Table~\ref{tab:internal}, and MatterGen MSUN/SUN from $39.5/14.1\%$ to $47.1/22.0\%$. KLDM-CSP is unchanged in MSUN and SUN despite the same doubled training set, so data volume alone does not explain the Crystalite gain. Distillation does lower the relaxation RMSD of both backbones ($0.86$ and $1.24$ against $1.27$ and $1.40$\,\AA{}, \appref{app:csp-finetune}), so the prior takes back structural work from the relaxer. The reward reaches these structures only through the validity gate and the compositions the policy queries, so the same search improves the structural model as likelihood data but not as a reward gradient through the denoising chain. The search can therefore improve where it queries the prior while also generating useful training data for the prior, although the benefit of distillation is backbone-dependent.

\paragraph{Strong search exploits the scorer.}
Our pipeline, MatterGen, and the training reward all use MatterSim potentials. Following \citet{veljkovic2026crystalite}, we also test pre-relaxing structures with NequIP and submitting them to LeMat-GenBench, which computes $E_{\mathrm{hull}}$ as the mean of ORB-v3, MACE-MP, and UMA. With the LeMat validity gate held fixed, \textsc{ANCHOR}$\to$KLDM-CSP falls from $13.6\%$ SUN in our pipeline to $0.4\%$ under LeMat-GenBench, while Crystalite-DNG rises from $1.7\%$ to $9.8\%$ (Table~\ref{tab:gates}). The on-hull ranking therefore reverses when the potential, relaxation, and hull change, and each model is strongest under the potential it was optimised against, since the released Crystalite-DNG checkpoint was selected for SUN under this NequIP relaxation~\citep{veljkovic2026crystalite}. \textsc{ANCHOR}'s MSUN also falls ($34.1\%$ to $7.8\%$) while that of Crystalite-DNG does not ($27.1\%$ to $30.0\%$), so a policy trained against a potential exploits it more strongly than checkpoint selection does. The policy exploits a validity gate in the same way: when trained against LeMat-GenBench's own gate, \textsc{ANCHOR} moves to roughly $80\%$ all-metal formulas and reaches $32.5\%$ LeMat-GenBench MSUN, above every DNG model we evaluate, but has almost no on-hull structures (\appref{app:gates}, Figure~\ref{fig:element_sampling}). We therefore treat on-hull SUN as the headline metric. The gate also changes what an evaluator rewards, since under MatterGen, which does not gate MSUN on validity, even random formulas reach $10.3\%$ MSUN although only $9.8\%$ pass our gate. 

\section{Discussion}
\label{sec:discussion}
Our results suggest separating search from structural learning in reward-driven crystal discovery, with rewards guiding the search over compositions and likelihood training the structural prior. On the search side, discovery rewards, including history-dependent novelty, belong in the composition policy, where \textsc{ANCHOR} keeps $99.9\%$ uniqueness and proposes no MP-20 formulas, whereas the same signals propagated through the denoising chain of a DNG model mostly move its composition marginal toward what is easiest to reward. CAN is therefore a search objective rather than an evaluation metric, and fixed-reference novelty and on-hull SUN remain the benchmark quantities. On the structural side, the structures the search finds carry useful signal, which in our experiments improves the prior when distilled by likelihood but not when delivered as reward gradients, even when gradients measure only stability. Search over a frozen CSP model can therefore be alternated with distillation of what it finds. Querying the frozen CSP model with a fine-tuned model's compositions tells the two kinds of change apart. The same separation could be applied beyond crystals, for example to searching protein sequences over a frozen structure predictor. The strength that makes \textsc{ANCHOR} effective also makes it exploit whatever it is scored with. This includes the validity gate and potential, so its gains are gains under the reward it was trained against, consistent with reward hacking observed in other reward-fine-tuned crystal generators~\citep{park2026guiding}. This is Goodhart's law applied to materials search~\citep{gao2023scaling}, i.e., a limit of the reward rather than of the search. Our stability estimates rely on MatterSim potentials, and scoring under the LeMat-GenBench potential ensemble and validity gate, neither used in training, is only a partial check. Training against potential ensembles and DFT validation are natural extensions. However, because \textsc{ANCHOR} keeps all structural knowledge in transferable and replaceable components, its gains should translate more directly into discoveries as potentials and CSP models improve. Better structural priors can be substituted directly rather than rediscovered through reward optimization. In the limit of a DFT-accurate potential and a physically faithful CSP model, discovery reduces to learning where to query the prior, and any modification of the prior can only make it less faithful.

\newpage

\
\backmattersection{Software and Data}
All backbones, reference datasets, and interatomic potentials used in this paper are publicly available and cited. The code for \textsc{ANCHOR} and its associated experiments will be released with the article when accepted for publication.

\backmattersection{Acknowledgments}
We received funding from DNRF for the Pioneer Center for Accelerating P2X Materials Discovery (CAPeX) grant number P3, from the Det Frie Forskningsråd for Project “Data-driven quest for TWh scalable Na-ion battery (TeraBatt)” (2035-00232B) and “Autonomous agents of Discovery for earth-Abundant Na-ion battery cathodes (ADANA)” (3164-00297B). We also acknowledge the Novo Nordisk Foundation grant number NNF25OC0101622 (AutoMLP) and the Data Science Research Infrastructure 2022 Grant: A high-performance computing infrastructure for data-driven research on sustainable energy materials, Grant no. NNF22OC0078009.

\bibliography{references}

@article{merchant2023scaling,
  title={Scaling deep learning for materials discovery},
  author={Merchant, Amil and Batzner, Simon and Schoenholz, Samuel S and Aykol, Muratahan and Cheon, Gowoon and Cubuk, Ekin Dogus},
  journal={Nature},
  volume={624},
  number={7990},
  pages={80--85},
  year={2023},
  publisher={Nature Publishing Group UK London}
}

@article{xie2021crystal,
  title={Crystal diffusion variational autoencoder for periodic material generation},
  author={Xie, Tian and Fu, Xiang and Ganea, Octavian-Eugen and Barzilay, Regina and Jaakkola, Tommi},
  journal={arXiv preprint arXiv:2110.06197},
  year={2021}
}

@article{miller2024flowmm,
  title={Flowmm: Generating materials with riemannian flow matching},
  author={Miller, Benjamin Kurt and Chen, Ricky TQ and Sriram, Anuroop and Wood, Brandon M},
  journal={arXiv preprint arXiv:2406.04713},
  year={2024}
}

@article{cornet2025kinetic,
  title={Kinetic langevin diffusion for crystalline materials generation},
  author={Cornet, Fran{\c{c}}ois and Bergamin, Federico and Bhowmik, Arghya and Lastra, Juan Maria Garcia and Frellsen, Jes and Schmidt, Mikkel N},
  journal={arXiv preprint arXiv:2507.03602},
  year={2025}
}

@article{zeni2025generative,
  title={A generative model for inorganic materials design},
  author={Zeni, Claudio and Pinsler, Robert and Z{\"u}gner, Daniel and Fowler, Andrew and Horton, Matthew and Fu, Xiang and Wang, Zilong and Shysheya, Aliaksandra and Crabb{\'e}, Jonathan and Ueda, Shoko and others},
  journal={Nature},
  volume={639},
  number={8055},
  pages={624--632},
  year={2025},
  publisher={Nature Publishing Group UK London}
}

@article{park2026guiding,
  title={Guiding generative models to uncover diverse and novel crystals via reinforcement learning},
  author={Park, Hyunsoo and Walsh, Aron},
  journal={Nature Machine Intelligence},
  pages={1--13},
  year={2026},
  publisher={Nature Publishing Group UK London}
}

@article{chen2025accelerating,
  title={Accelerating inverse materials design using generative diffusion models with reinforcement learning},
  author={Chen, Junwu and Guo, Jeff and Fako, Edvin and Schwaller, Philippe},
  journal={arXiv preprint arXiv:2511.03112},
  year={2025}
}

@article{betala2025lemat,
  title={Lemat-genbench: A unified evaluation framework for crystal generative models},
  author={Betala, Siddharth and Gleason, Samuel P and Ramlaoui, Ali and Xu, Andy and Channing, Georgia and Levy, Daniel and Fourrier, Cl{\'e}mentine and Kazeev, Nikita and Joshi, Chaitanya K and Kaba, S{\'e}kou-Oumar and others},
  journal={arXiv preprint arXiv:2512.04562},
  year={2025}
}

@article{jain2013commentary,
  title={Commentary: The Materials Project: A materials genome approach to accelerating materials innovation},
  author={Jain, Anubhav and Ong, Shyue Ping and Hautier, Geoffroy and Chen, Wei and Richards, William Davidson and Dacek, Stephen and Cholia, Shreyas and Gunter, Dan and Skinner, David and Ceder, Gerbrand and others},
  journal={APL materials},
  volume={1},
  number={1},
  year={2013},
  publisher={AIP Publishing}
}

@article{ong2013python,
  title={Python Materials Genomics (pymatgen): A robust, open-source python library for materials analysis},
  author={Ong, Shyue Ping and Richards, William Davidson and Jain, Anubhav and Hautier, Geoffroy and Kocher, Michael and Cholia, Shreyas and Gunter, Dan and Chevrier, Vincent L and Persson, Kristin A and Ceder, Gerbrand},
  journal={Computational Materials Science},
  volume={68},
  pages={314--319},
  year={2013},
  publisher={Elsevier}
}

@article{schmidt2024improving,
  title={Improving machine-learning models in materials science through large datasets},
  author={Schmidt, Jonathan and Cerqueira, Tiago FT and Romero, Aldo H and Loew, Antoine and J{\"a}ger, Fabian and Wang, Hai-Chen and Botti, Silvana and Marques, Miguel AL},
  journal={Materials Today Physics},
  volume={48},
  pages={101560},
  year={2024},
  publisher={Elsevier}
}

@article{siron2025lemat,
  title={Lemat-bulk: aggregating, and de-duplicating quantum chemistry materials databases},
  author={Siron, Martin and Djafar, Inel and Ramlaoui, Ali and Fayette, Etienne du and Rossello, Amandine and Fako, Edvin and McDermott, Matthew and Therrien, Felix and Barroso-Luque, Luis and Cipcigan, Flaviu and others},
  journal={arXiv preprint arXiv:2511.05178},
  year={2025}
}

@article{negishi2026continuous,
  title={Continuous SUN (stable, unique, and novel) metric for generative modeling of inorganic crystals},
  author={Negishi, Masahiro and Park, Hyunsoo and Mastej, Kinga Oliwia and Walsh, Aron},
  journal={Machine Learning: Science and Technology},
  volume={7},
  number={3},
  pages={035064},
  year={2026},
  publisher={IOP Publishing}
}

@article{noh2019inverse,
  title={Inverse design of solid-state materials via a continuous representation},
  author={Noh, Juhwan and Kim, Jaehoon and Stein, Helge S and Sanchez-Lengeling, Benjamin and Gregoire, John M and Aspuru-Guzik, Alan and Jung, Yousung},
  journal={Matter},
  volume={1},
  number={5},
  pages={1370--1384},
  year={2019},
  publisher={Elsevier}
}

@article{dan2020generative,
  title={Generative adversarial networks (GAN) based efficient sampling of chemical composition space for inverse design of inorganic materials},
  author={Dan, Yabo and Zhao, Yong and Li, Xiang and Li, Shaobo and Hu, Ming and Hu, Jianjun},
  journal={npj Computational Materials},
  volume={6},
  number={1},
  pages={84},
  year={2020},
  publisher={Nature Publishing Group UK London}
}

@article{ren2022invertible,
  title={An invertible crystallographic representation for general inverse design of inorganic crystals with targeted properties},
  author={Ren, Zekun and Tian, Siyu Isaac Parker and Noh, Juhwan and Oviedo, Felipe and Xing, Guangzong and Li, Jiali and Liang, Qiaohao and Zhu, Ruiming and Aberle, Armin G and Sun, Shijing and others},
  journal={Matter},
  volume={5},
  number={1},
  pages={314--335},
  year={2022},
  publisher={Elsevier}
}

@article{jiao2023crystal,
  title={Crystal structure prediction by joint equivariant diffusion},
  author={Jiao, Rui and Huang, Wenbing and Lin, Peijia and Han, Jiaqi and Chen, Pin and Lu, Yutong and Liu, Yang},
  journal={Advances in Neural Information Processing Systems},
  volume={36},
  pages={17464--17497},
  year={2023}
}

@article{park2025exploration,
  title={Exploration of crystal chemical space using text-guided generative artificial intelligence},
  author={Park, Hyunsoo and Onwuli, Anthony and Walsh, Aron},
  journal={Nature Communications},
  volume={16},
  number={1},
  pages={4379},
  year={2025},
  publisher={Nature Publishing Group UK London}
}

@article{veljkovic2026crystalite,
  title={Crystalite: A lightweight transformer for efficient crystal modeling},
  author={Veljkovi{\'c}, Tin Had{\v{z}}i and Rosenthal, Joshua and Lon{\v{c}}ari{\'c}, Ivor and van de Meent, Jan-Willem},
  journal={arXiv preprint arXiv:2604.02270},
  year={2026}
}

@article{zhao2021high,
  title={High-throughput discovery of novel cubic crystal materials using deep generative neural networks},
  author={Zhao, Yong and Al-Fahdi, Mohammed and Hu, Ming and Siriwardane, Edirisuriya MD and Song, Yuqi and Nasiri, Alireza and Hu, Jianjun},
  journal={Advanced Science},
  volume={8},
  number={20},
  pages={2100566},
  year={2021},
  publisher={Wiley Online Library}
}

@article{luo2025crystalflow,
  title={CrystalFlow: a flow-based generative model for crystalline materials},
  author={Luo, Xiaoshan and Wang, Zhenyu and Wang, Qingchang and Shao, Xuechen and Lv, Jian and Wang, Lei and Wang, Yanchao and Ma, Yanming},
  journal={Nature Communications},
  volume={16},
  number={1},
  pages={9267},
  year={2025},
  publisher={Nature Publishing Group UK London}
}

@inproceedings{simm2020reinforcement,
  title={Reinforcement learning for molecular design guided by quantum mechanics},
  author={Simm, Gregor and Pinsler, Robert and Hern{\'a}ndez-Lobato, Jos{\'e} Miguel},
  booktitle={International Conference on Machine Learning},
  pages={8959--8969},
  year={2020},
  organization={PMLR}
}

@article{turk2022assessing,
  title={Assessing deep generative models in chemical composition space},
  author={Turk, Hanna and Landini, Elisabetta and Kunkel, Christian and Margraf, Johannes T and Reuter, Karsten},
  journal={Chemistry of Materials},
  volume={34},
  number={21},
  pages={9455--9467},
  year={2022},
  publisher={ACS Publications}
}

@article{elsborg2023equivariant,
  title={Equivariant graph-representation-based actor--critic reinforcement learning for nanoparticle design},
  author={Elsborg, Jonas and Bhowmik, Arghya},
  journal={Journal of chemical information and modeling},
  volume={63},
  number={12},
  pages={3731--3741},
  year={2023},
  publisher={ACS Publications}
}

@article{elsborg2024artisan,
  title={ArtiSAN: navigating the complexity of material structures with deep reinforcement learning},
  author={Elsborg, Jonas and Bhowmik, Arghya},
  journal={Machine Learning: Science and Technology},
  volume={5},
  number={3},
  pages={035043},
  year={2024},
  publisher={IOP Publishing}
}

@article{elsborg2026reinforcement,
  title={Reinforcement Learning for Chemical Ordering in Alloy Nanoparticles},
  author={Elsborg, Jonas and Hovmand, Emma Lei and Bhowmik, Arghya},
  journal={ACS Materials Au},
  volume={6},
  number={4},
  pages={696},
  year={2026}
}

@article{shao2024deepseekmath,
  title={Deepseekmath: Pushing the limits of mathematical reasoning in open language models},
  author={Shao, Zhihong and Wang, Peiyi and Zhu, Qihao and Xu, Runxin and Song, Junxiao and Bi, Xiao and Zhang, Haowei and Zhang, Mingchuan and Li, YK and Wu, Yang and others},
  journal={arXiv preprint arXiv:2402.03300},
  year={2024}
}

@article{hollmer2026open,
  title={Open materials generation with inference-time reinforcement learning},
  author={H{\"o}llmer, Philipp and Martiniani, Stefano},
  journal={arXiv preprint arXiv:2602.00424},
  year={2026}
}

@article{valle2010crystal,
  title={Crystal fingerprint space--a novel paradigm for studying crystal-structure sets},
  author={Valle, Mario and Oganov, Artem R},
  journal={Foundations of Crystallography},
  volume={66},
  number={5},
  pages={507--517},
  year={2010},
  publisher={International Union of Crystallography}
}

@inproceedings{pathak2017curiosity,
  title={Curiosity-driven exploration by self-supervised prediction},
  author={Pathak, Deepak and Agrawal, Pulkit and Efros, Alexei A and Darrell, Trevor},
  booktitle={International conference on machine learning},
  pages={2778--2787},
  year={2017},
  organization={PMLR}
}

@article{burda2018exploration,
  title={Exploration by random network distillation},
  author={Burda, Yuri and Edwards, Harrison and Storkey, Amos and Klimov, Oleg},
  journal={arXiv preprint arXiv:1810.12894},
  year={2018}
}

@article{thiede2022curiosity,
  title={Curiosity in exploring chemical spaces: intrinsic rewards for molecular reinforcement learning},
  author={Thiede, Luca A and Krenn, Mario and Nigam, AkshatKumar and Aspuru-Guzik, Al{\'a}n},
  journal={Machine Learning: Science and Technology},
  volume={3},
  number={3},
  pages={035008},
  year={2022},
  publisher={IOP Publishing}
}

@article{oschinski2026general,
  title={General-purpose LLMs as Constrained Crystal Composition Generators},
  author={Oschinski, Hedda and Ach, Maximilian L and Jakob, Konstantin S and Carbogno, Christian and Reuter, Karsten},
  journal={arXiv preprint arXiv:2605.31495},
  year={2026}
}

@article{you2018graph,
  title={Graph convolutional policy network for goal-directed molecular graph generation},
  author={You, Jiaxuan and Liu, Bowen and Ying, Zhitao and Pande, Vijay and Leskovec, Jure},
  journal={Advances in neural information processing systems},
  volume={31},
  year={2018}
}

@article{jorgensen2019atomistic,
  title={Atomistic structure learning},
  author={J{\o}rgensen, Mathias S and Mortensen, Henrik L and Meldgaard, S{\o}ren A and Kolsbjerg, Esben L and Jacobsen, Thomas L and S{\o}rensen, Knud H and Hammer, Bj{\o}rk},
  journal={The Journal of Chemical Physics},
  volume={151},
  number={5},
  year={2019},
  publisher={AIP Publishing}
}

@article{karpovich2024deep,
  title={Deep reinforcement learning for inverse inorganic materials design},
  author={Karpovich, Christopher and Pan, Elton and Olivetti, Elsa A},
  journal={npj Computational Materials},
  volume={10},
  number={1},
  pages={287},
  year={2024},
  publisher={Nature Publishing Group UK London}
}

@article{bergmeister2026reinforce,
  title={Reinforce Adjoint Matching: Scaling RL Post-Training of Diffusion and Flow-Matching Models},
  author={Bergmeister, Andreas and Jegelka, Stefanie and N{\"u}sken, Nikolas and Domingo-Enrich, Carles and Pidstrigach, Jakiw},
  journal={arXiv preprint arXiv:2605.10759},
  year={2026}
}

@article{song2020score,
  title={Score-based generative modeling through stochastic differential equations},
  author={Song, Yang and Sohl-Dickstein, Jascha and Kingma, Diederik P and Kumar, Abhishek and Ermon, Stefano and Poole, Ben},
  journal={arXiv preprint arXiv:2011.13456},
  year={2020}
}

@article{uehara2024fine,
  title={Fine-tuning of continuous-time diffusion models as entropy-regularized control},
  author={Uehara, Masatoshi and Zhao, Yulai and Black, Kevin and Hajiramezanali, Ehsan and Scalia, Gabriele and Diamant, Nathaniel Lee and Tseng, Alex M and Biancalani, Tommaso and Levine, Sergey},
  journal={arXiv preprint arXiv:2402.15194},
  year={2024}
}

@article{bellemare2016unifying,
  title={Unifying count-based exploration and intrinsic motivation},
  author={Bellemare, Marc and Srinivasan, Sriram and Ostrovski, Georg and Schaul, Tom and Saxton, David and Munos, Remi},
  journal={Advances in neural information processing systems},
  volume={29},
  year={2016}
}

@article{savinov2018episodic,
  title={Episodic curiosity through reachability},
  author={Savinov, Nikolay and Raichuk, Anton and Marinier, Rapha{\"e}l and Vincent, Damien and Pollefeys, Marc and Lillicrap, Timothy and Gelly, Sylvain},
  journal={arXiv preprint arXiv:1810.02274},
  year={2018}
}

@article{hollmer2025open,
  title={Open materials generation with stochastic interpolants},
  author={H{\"o}llmer, Philipp and Egg, Thomas and Martirossyan, Maya M and Fuemmeler, Eric and Shui, Zeren and Gupta, Amit and Prakash, Pawan and Roitberg, Adrian and Liu, Mingjie and Karypis, George and others},
  journal={arXiv preprint arXiv:2502.02582},
  year={2025}
}

@article{zhang2026crystalrepa,
  title={CrystalREPA: Transferring Physical Priors from Universal MLIPs to Crystal Generative Models},
  author={Zhang, Chengqian and Jin, Yucheng and Zhang, Duo and Li, Tiejun and Wang, Han},
  journal={arXiv preprint arXiv:2605.08960},
  year={2026}
}

@article{yang2024mattersim,
  title={Mattersim: A deep learning atomistic model across elements, temperatures and pressures},
  author={Yang, Han and Hu, Chenxi and Zhou, Yichi and Liu, Xixian and Shi, Yu and Li, Jielan and Li, Guanzhi and Chen, Zekun and Chen, Shuizhou and Zeni, Claudio and others},
  journal={arXiv preprint arXiv:2405.04967},
  year={2024}
}

@article{schulman2017proximal,
  title={Proximal policy optimization algorithms},
  author={Schulman, John and Wolski, Filip and Dhariwal, Prafulla and Radford, Alec and Klimov, Oleg},
  journal={arXiv preprint arXiv:1707.06347},
  year={2017}
}

@article{ase-paper,
  author={Ask Hjorth Larsen and Jens Jørgen Mortensen and Jakob Blomqvist and Ivano E Castelli and Rune Christensen and Marcin Dułak and Jesper Friis and Michael N Groves and Bjørk Hammer and Cory Hargus and Eric D Hermes and Paul C Jennings and Peter Bjerre Jensen and James Kermode and John R Kitchin and Esben Leonhard Kolsbjerg and Joseph Kubal and Kristen Kaasbjerg and Steen Lysgaard and Jón Bergmann Maronsson and Tristan Maxson and Thomas Olsen and Lars Pastewka and Andrew Peterson and Carsten Rostgaard and Jakob Schiøtz and Ole Schütt and Mikkel Strange and Kristian S Thygesen and Tejs Vegge and Lasse Vilhelmsen and Michael Walter and Zhenhua Zeng and Karsten W Jacobsen},
  title={The atomic simulation environment—a Python library for working with atoms},
  journal={Journal of Physics: Condensed Matter},
  volume={29},
  number={27},
  pages={273002},
  url={http://stacks.iop.org/0953-8984/29/i=27/a=273002},
  year={2017}
}

@article{davies2019smact,
  title={{SMACT}: Semiconducting Materials by Analogy and Chemical Theory},
  author={Davies, Daniel W. and Butler, Keith T. and Jackson, Adam J. and Skelton, Jonathan M. and Morita, Kazuki and Walsh, Aron},
  journal={Journal of Open Source Software},
  volume={4},
  number={38},
  pages={1361},
  year={2019},
  publisher={The Open Journal},
  doi={10.21105/joss.01361}
}

@inproceedings{black2024training,
  title={Training diffusion models with reinforcement learning},
  author={Black, Kevin and Janner, Michael and Du, Yilun and Kostrikov, Ilya and Levine, Sergey},
  booktitle={International Conference on Learning Representations},
  volume={2024},
  pages={4965--4987},
  year={2024}
}

@article{liu2026composable,
  title={Composable Crystals: Controllable Materials Discovery via Concept Learning},
  author={Liu, Nian and Zeng, Yuwei and Kubo, Ryoji and Kazeev, Nikita and Dale, Stephen Gregory and Maevskiy, Artem and Huang, Pengru and Laurent, Thomas and Novoselov, Kostya S and Bresson, Xavier},
  journal={arXiv preprint arXiv:2605.14769},
  year={2026}
}

@article{widdowson2022average,
  title={Average Minimum Distances of periodic point sets are fundamental invariants for mapping all periodic crystals.},
  author={Widdowson, Daniel and Mosca, Marco and Pulido, Angeles and Kurlin, Vitaliy and Cooper, Andrew I},
  journal={MATCH Communications in Mathematical and in Computer Chemistry},
  volume={87},
  number={3},
  pages={529--559},
  year={2022},
  publisher={University Library in Kragujevac}
}

@inproceedings{gao2023scaling,
  title={Scaling laws for reward model overoptimization},
  author={Gao, Leo and Schulman, John and Hilton, Jacob},
  booktitle={International conference on machine learning},
  pages={10835--10866},
  year={2023},
  organization={PMLR}
}
\bibliographystyle{iclr2027_conference}


\newpage
\appendix
\startcontents[appendix]

\section*{Appendix}
\subsection*{Table of Contents}

\begin{small}
\printcontents[appendix]{}{1}{\setcounter{tocdepth}{2}}
\end{small}
\newpage
%

\section{Notation}
\label{app:notation}

\begin{table}[h!]
\centering
\caption{Symbols for the composition policy, reward, and evaluation. Rates are fractions of an evaluation batch of $N_{\mathrm{eval}}=1000$ structures, unless stated. Generative-model notation is defined in \secref{sec:background}.}
\label{tab:notation}
\footnotesize
\setlength{\tabcolsep}{4pt}
\renewcommand{\arraystretch}{0.92}
\begin{tabular}{llp{0.50\linewidth}}
\toprule
\textbf{Symbol} & \textbf{Type} & \textbf{Meaning} \\
\midrule
\multicolumn{3}{l}{\emph{Composition action space}} \\
$\gC$                     & set          & discrete chemical space searched by the policy \\
$\gE$                     & set          & element vocabulary; $E = |\gE| = 84$ \\
$\gK$                     & set          & admissible element counts, $\gK = \{2,3,4\}$; $k_{\max} = \max\gK = 4$ \\
$k$                       & integer      & number of distinct elements in an action, $k \in \gK$ \\
$\ve$                     & $\gE^{k}$    & the $k$ distinct chosen elements \\
$\rho$                    & $\mathbb{Z}_{>0}^{k}$ & atom stoichiometry \\
$T$                       & integer      & total atoms in the cell, $T = \sum_i \rho_i$, $k \le T \le N$ \\
$N$                       & integer      & maximum atoms per unit cell, $N = 20$ \\
$c = (k, \ve, T, \rho)$   & action       & one composition action \\
$f(c)$                    & string       & reduced chemical formula of action $c$ \\
$\gS \subseteq \gE$       & set          & element sub-group active for a rollout (mask) \\
$n_{\gS} = |\gS|$         & integer      & sub-group size, $n_{\gS} \sim \mathrm{Uniform}\{k_{\max},\dots,E\}$ \\
\addlinespace[2pt]
\multicolumn{3}{l}{\emph{Structures and novelty}} \\
$x$                       & structure    & crystal returned by the frozen backbone \\
$\vz(x)$                  & $\R^{n_{\text{bins}}}$ & $\ell_1$-normalized RDF histogram of $x$ over $n_{\text{bins}}$ radial bins \\
$Z^{\mathrm{ref}}$        & set          & frozen MP-20 reference embeddings for formula $f(c)$ \\
$Z^{\mathrm{hist}}$       & set          & growing agent-history embeddings for formula $f(c)$ \\
$\mathcal{F}_{\mathrm{ref}}$ & set       & reduced formulas present in MP-20 \\
$\nu(\vz; Z)$             & $[0,1]$      & kNN--RBF novelty of $\vz$ against reference set $Z$ \\
$\nu(x)$                  & $[0,1]$      & adaptive novelty (min over reference and history) \\
$\nu_0$                   & scalar       & bootstrap novelty for an unseen off-MP-20 formula \\
$k_{\mathrm{nn}}$         & integer      & kNN neighbourhood size used by $\nu(\vz; Z)$ \\
$\sigma$                  & scalar       & RBF bandwidth (per formula, with global floor) \\
\addlinespace[2pt]
\multicolumn{3}{l}{\emph{Reward}} \\
$s_{\mathrm{stab}}$       & $[0,1]$      & stability score from $E_{\mathrm{hull}}$ \\
$E_{\mathrm{p}}$          & eV           & relaxed potential energy under the interatomic potential \\
$E_{\mathrm{hull}}$       & eV/atom      & relaxed energy above the convex hull \\
$d_{\mathrm{struct}}, d_{\mathrm{comp}}$ & $[0,1]$ & within-group structural / compositional diversity \\
$\vm(c)$                  & $[0,1]^4$    & objective vector $(s_{\mathrm{stab}}, \nu, d_{\mathrm{struct}}, d_{\mathrm{comp}})$ \\
$\vw$                     & simplex      & objective weights, $\vw \sim \mathrm{Dirichlet}(\bm{\alpha})$ \\
$\bm{\alpha}$             & $\R_{>0}^{4}$ & Dirichlet concentration over the four objectives \\ 
$n_f$                     & integer      & times formula $f$ has been generated in the run \\
$\lambda_c$               & scalar       & count-bonus scale \\
$r_{\mathrm{inv}}$        & scalar $<0$  & fixed penalty for invalid structures \\
$r(c)$                    & scalar       & scalar reward for action $c$ \\
\addlinespace[2pt]
\multicolumn{3}{l}{\emph{Policy and optimization}} \\
$\pi_\theta$              & policy       & autoregressive composition policy \\
$\vs$                     & vector       & policy state, $[\,\mW_{\!f}\bar{\vf}_{\gS}\,;\,\vw\,]$ \\
$\vf_e$                   & vector       & fixed chemistry-aware feature vector of element $e$ \\
$\bar{\vf}_{\gS}$         & vector       & mean of $\vf_e$ over the active sub-group $\gS$ \\
$\mW_{\!f}$               & matrix       & learned projection of $\bar{\vf}_{\gS}$ into the state \\
$G$                       & integer      & group size (samples per rollout) \\
$A_i$                     & scalar       & group-relative advantage of sample $i$ \\
$\eta_i$                  & scalar       & importance ratio $\pi_\theta / \pi_{\theta_{\mathrm{old}}}$ \\
$\eps$                    & scalar       & PPO clipping parameter \\
$\eps_A$                  & scalar       & floor on the group reward standard deviation \\
$\beta_H$                 & scalar       & entropy-bonus coefficient (scheduled) \\
\addlinespace[2pt]
\multicolumn{3}{l}{\emph{Evaluation}} \\
Valid                     & rate         & fraction passing the validity gate of \appref{app:eval_protocol} \\
Unique                    & rate         & distinct reduced formulas, as a fraction of the batch \\
Stable                    & rate         & valid $\wedge$ $E_{\mathrm{hull}} \le 0$ \\
Metastable                & rate         & valid $\wedge$ $E_{\mathrm{hull}} \le 0.1$~eV/atom \\
$\mathrm{SUN}$            & rate         & valid $\wedge$ unique $\wedge$ novel $\wedge$ $E_{\mathrm{hull}} \le 0$ \\
$\mathrm{MSUN}$           & rate         & valid $\wedge$ unique $\wedge$ novel $\wedge$ $E_{\mathrm{hull}} \le 0.1$~eV/atom \\
\bottomrule
\end{tabular}
\end{table}

\newpage
\section{Additional Related Work}
\label{app:related_work}

\paragraph{Crystal-generation tasks and benchmark datasets.}
A crystalline material is specified by its chemical composition and periodically repeated unit cell, typically represented by atomic species, fractional coordinates, and lattice vectors. This gives rise to two closely related generative tasks. Crystal structure prediction (CSP) conditions on a fixed composition and searches over compatible structures, while de novo crystal generation explores composition and structure jointly~\citep{jiao2023crystal,xie2021crystal,cornet2025kinetic}.

Early crystal-generation benchmarks commonly used datasets derived from the Materials Project, including MP-20~\citep{jain2013commentary,xie2021crystal}. Evaluation has since expanded toward larger reference collections such as Alexandria, Alex-MP, and LeMat-Bulk~\citep{schmidt2024improving,zeni2025generative,siron2025lemat}. This expansion matters directly for novelty evaluation because the reference corpus forms part of the definition of whether a generated candidate is considered known. A structure judged novel with respect to a smaller reference may no longer be novel when evaluated against a larger collection.

LeMat-GenBench provides a broader evaluation framework spanning validity, stability, novelty, uniqueness, diversity, and computational efficiency~\citep{betala2025lemat}. Its stable, unique, and novel metric can be written as
\begin{equation}
\mathrm{SUN}(x)=
\mathbbm{1}\!\left[
\mathrm{valid}(x)
\wedge
\mathrm{unique}(x)
\wedge
\mathrm{novel}(x)
\wedge
E_{\mathrm{hull}}(x)\leq0
\right],
\end{equation}
with mSUN (MSUN in this paper) relaxing the stability threshold to
$E_{\mathrm{hull}}\leq0.1\,\mathrm{eV/atom}$. MatterGen~\citep{zeni2025generative} instead reports the $0.1$ quantity as SUN.
Benchmark results expose a persistent interaction between stability, novelty, and diversity, motivating evaluation of all three rather than stability in isolation.


\paragraph{Generative models for crystalline materials.}
Early latent-variable and adversarial approaches established learned generative design for inorganic crystals~\citep{noh2019inverse,ren2022invertible,dan2020generative,zhao2021high}. Subsequent models increasingly incorporated the periodic geometry of crystals directly into their generative parameterization. CDVAE introduced diffusion-based decoding for periodic crystal structures, while DiffCSP jointly models lattice parameters and fractional coordinates for composition-conditioned crystal structure prediction~\citep{xie2021crystal,jiao2023crystal}. KLDM models fractional coordinates on the hypertorus through kinetic Langevin dynamics~\citep{cornet2025kinetic} and is used as the primary structural backbone in this work.

More recent models have expanded conditioning and generative scale. MatterGen targets property-conditioned inorganic materials generation, while Chemeleon extends diffusion-based crystal generation toward broader controllable exploration~\citep{zeni2025generative,park2025exploration}. Flow- and Transformer-based approaches have also been proposed to simplify or accelerate generative sampling~\citep{miller2024flowmm,luo2025crystalflow,veljkovic2026crystalite}. These developments motivate treating a pretrained crystal generator as a reusable source of structural knowledge rather than necessarily retraining it for each downstream discovery objective.

\paragraph{Novelty and structural similarity.}
Novelty and uniqueness answer different questions in crystal-generation evaluation. Novelty conventionally measures whether a generated structure is absent from a fixed external reference collection, while uniqueness measures repetition among generated samples~\citep{zeni2025generative,betala2025lemat}. Stability is evaluated separately using formation energy or energy above the convex hull~\citep{ong2013python}. This separation makes fixed-reference novelty well suited to reproducible comparison between generative models, but it does not encode what has already been discovered during a continuing search.

Binary structure matching further discards information about the degree of structural similarity. Continuous structural metrics provide a graded alternative~\citep{valle2010crystal,negishi2026continuous}. In particular, continuous formulations can distinguish near-repeats from structures that are substantially different even when both would receive the same binary novel/known label. The continuous component of CAN builds on this general idea, while its adaptive component introduces persistent search history.

A concurrent compositional approach makes the same trade visible from a different direction. \citet{liu2026composable} learn a discrete codebook of local atomic environments with a VQ-VAE and condition generation on recombined codes, reporting large novelty gains (78.0\% to 90.0\% on MP-20, 61.8\% to 85.9\% on Alex-MP-20) alongside a drop in stability (48.4\% to 46.9\% and 69.0\% to 59.1\%). Their intervention is a structural conditioning signal rather than a reward, but the signature is the one \appref{app:novelty_tension} predicts for any pressure that moves the conditional distribution toward novelty: candidates further from the reference set are also further from the region the pretrained model assigns high probability. Their headline scores are computed against MP-2023 for both novelty and the hull, so they are not directly comparable to the MatterGen-pipeline numbers in Table~\ref{tab:mattergen_dng}.

Repeated proposals are not unique to crystal-generation benchmarks. Prior materials-search work has observed repeated sampling during prolonged exploration and has used search history to discourage revisiting previously explored candidates~\citep{turk2022assessing,oschinski2026general}. More broadly, reinforcement learning has a long history of history-dependent exploration mechanisms, including count-based bonuses, curiosity, and related intrinsic rewards~\citep{bellemare2016unifying,pathak2017curiosity,burda2018exploration,thiede2022curiosity}. These approaches establish precedent for allowing an exploration objective to depend on the state of the search rather than only on a fixed external dataset.

\paragraph{Reinforcement learning for molecular and materials discovery.}
Reinforcement learning has been applied across several levels of molecular and materials design. Molecular graph-generation methods use sequential actions to construct molecules optimized for target properties~\citep{you2018graph}. Other work applies RL directly to atomistic or molecular construction and assembly~\citep{jorgensen2019atomistic,simm2020reinforcement}. Materials-focused approaches have used reinforcement learning for composition search, atomic ordering, and experimentally or computationally evaluated design problems~\citep{turk2022assessing,elsborg2023equivariant,elsborg2024artisan,karpovich2024deep,elsborg2026reinforcement}.

These studies are important precedent for the composition-level policy used in \textsc{ANCHOR}. Composition-level reinforcement learning itself is therefore not the central distinction of the present work. The distinction is the role assigned to that policy: it acts as an adaptive outer search over a separately pretrained conditional crystal generator, which remains frozen throughout discovery.

\paragraph{Reward-guided crystal generation.}
A separate line of work applies optimization directly to generative crystal models. MatInvent uses reinforcement learning for single- and multi-objective diffusion-based crystal design~\citep{chen2025accelerating}. OMatG-IRL applies policy-gradient optimization to continuous-time crystal generators at inference~\citep{hollmer2026open}. These methods use downstream rewards to modify the behavior of the model responsible for generating crystal structures.

Chemeleon2 gives an explicit reinforcement-learning formulation of latent diffusion~\citep{park2026guiding}. Starting from noise $z_T$, generation follows the reverse-denoising trajectory
\[
\tau=z_T\rightarrow z_{T-1}\rightarrow\cdots\rightarrow z_0.
\]
The decoder $D_\phi$ maps the final latent state to a generated crystal $x=D_\phi(z_0)$, and the RL objective maximizes the expected terminal crystal-level reward,
\begin{equation}
J(\theta)
=
\mathbb{E}_{\tau\sim\pi_\theta}
\left[
R(x)
\right].
\end{equation}
A terminal crystal reward must therefore be attributed across the denoising trajectory.


Chemeleon2 explicitly discusses an objective mismatch between likelihood-based pretraining and discovery. Pretraining concentrates probability in regions represented by the training distribution, while its ``creativity'' reward promotes novelty and uniqueness and therefore encourages exploration of less represented regions~\citep{park2026guiding}. The resulting GRPO objective uses strong KL regularization to the pretrained denoiser to restrict drift from the learned prior.


This setup differs from \textsc{ANCHOR} in where reinforcement learning acts. Reward-guided DNG adapts the joint generator and can therefore change both $p(c)$ and $p(x\mid c)$, whereas \textsc{ANCHOR} adapts only the composition-level search while keeping the conditional structural distribution fixed. The novelty-prior consequences of modifying $p(x\mid c)$ are developed separately in Appendix~\ref{app:novelty_tension}.
\newpage
\section{Composition Search, Novelty, and the Structural Prior}
\label{app:novelty_tension}

The main text separates two questions that are combined in conventional de novo crystal generation. The first is where in chemical space to search. The second is which structures are plausible for a queried composition. We develop this distinction further here and examine how it interacts with reward-guided generator fine-tuning and history-dependent novelty.

\paragraph{DNG, CSP, and the composition marginal.}
De novo crystal generation concerns a joint distribution over composition $c$ and structure $x$, which admits the conceptual factorization
\begin{equation}
p(c,x)
=
p(c)\,p(x\mid c).
\label{eq:joint_factorization_app}
\end{equation}
The conditional factor corresponds to the CSP problem at fixed composition. This factorization does not require the underlying DNG model to contain separately parameterized composition and CSP components. In practice, DNG and CSP models are typically trained separately.

The factorization nevertheless exposes two distinct roles during discovery. A likelihood-trained DNG model learns a composition marginal that reflects which chemistries are probable under its training distribution. A CSP model instead answers which structures are plausible after a composition has already been specified. These need not have the same domain of usefulness. In particular, a pretrained CSP model may generate useful structures for compositions that receive little probability under the composition marginal learned by DNG.

For a frozen CSP model $p_0(x\mid c)$, define the probability that querying composition $c$ produces a successful discovery as
\begin{equation}
s_0(c)
=
\Pr_{x\sim p_0(x\mid c)}
\left[
S(x,c)=1
\right],
\label{eq:csp_success_app}
\end{equation}
where $S$ denotes a chosen discovery criterion. If compositions are sampled from a distribution $r(c)$, successful candidates from composition $c$ occur with probability proportional to
\begin{equation}
r(c)\,s_0(c).
\label{eq:search_success_app}
\end{equation}
Direct DNG therefore allocates discovery effort according to $p_0(c)s_0(c)$. A composition can have high discovery potential under the structural model while remaining rarely observed because $p_0(c)$ is small. Replacing the composition marginal with an adaptive search policy instead gives $\pi_\theta(c)s_0(c)$ and allows search effort to move toward compositions where the frozen CSP model is empirically productive.

The distinction is therefore between reproducing the chemical statistics of the training distribution and allocating future discovery effort. The composition marginal $p_0(c)$ describes which chemistries resemble those already represented in the data. A discovery policy $\pi_\theta(c)$ instead learns which chemistries should be investigated next. The former can provide a useful prior without being the optimal search distribution.

\paragraph{KL-regularized reward tilting.}
Reward-guided fine-tuning provides one way to move beyond the pretrained DNG distribution. It can modify the composition marginal, but can simultaneously modify the conditional structural distribution. We use the standard KL-regularized reward-tilting identity to examine the latter effect, in the form stated by \citet{bergmeister2026reinforce}.

For the remainder of this analysis, $p_0(x\mid c)$ denotes a pretrained conditional structural distribution, $q(x\mid c)$ a reward-adapted endpoint distribution, and $q^*(x\mid c)$ the reward-optimal endpoint distribution under a KL-regularized objective. For a pretrained endpoint distribution and scalar reward, \citet{bergmeister2026reinforce} consider
\[
\max_q
\left\{
\mathbb{E}_{x\sim q(\cdot\mid c)}
\left[
R(x,c)
\right]
-
\beta
D_{\mathrm{KL}}
\left(
q(\cdot\mid c)
\,\|\,
p_0(\cdot\mid c)
\right)
\right\},
\qquad
\beta>0.
\]
The corresponding reward-optimal endpoint distribution is
\begin{equation}
q^*(x\mid c)
=
\frac{
p_0(x\mid c)
\exp\!\left(R(x,c)/\beta\right)
}{
Z(c)
},
\label{eq:reward_tilt}
\end{equation}
with
\[
Z(c)
=
\mathbb{E}_{x\sim p_0(\cdot\mid c)}
\left[
\exp\!\left(R(x,c)/\beta\right)
\right].
\]
This is the Gibbs solution of the KL-regularized objective, written with an explicit regularization coefficient $\beta$ and conditioned on composition, which \citet{bergmeister2026reinforce} derive from the Donsker-Varadhan variational principle.

\paragraph{Application to novelty.}
We first consider the illustrative novelty reward
\begin{equation}
R_{\mathrm{nov}}(x,c)
=
-\alpha\log p_0(x\mid c),
\qquad
\alpha>0,
\end{equation}
which assigns higher reward to structures receiving lower probability under the pretrained structural distribution. Because composition is fixed, this example isolates the structural effect of novelty and does not itself encourage exploration of new compositions. Low conditional probability also does not distinguish a useful new polymorph from an implausible structure that departs from the structural preferences learned during pretraining.

Substitution into Eq.~\ref{eq:reward_tilt} gives
\begin{equation}
q^*(x\mid c)
\propto
p_0(x\mid c)^{\,1-\alpha/\beta}.
\label{eq:novelty_power}
\end{equation}
Increasing novelty pressure relative to KL regularization reduces the exponent $1-\alpha/\beta$, weakening the preference for high-probability structures and flattening the pretrained structural distribution. As $\alpha/\beta$ approaches one, differences inherited from $p_0$ are progressively suppressed. For $\alpha/\beta>1$, the probability ordering can reverse wherever the resulting distribution remains normalizable.

Equation~\ref{eq:novelty_power} is an illustrative specialization of Eq.~\ref{eq:reward_tilt}. The novelty measure used in \textsc{ANCHOR} is not defined through model likelihood. Structural diversity within $p_0(x\mid c)$ remains useful for discovery. The point is that novelty alone does not identify whether an unusual structure is a useful new polymorph or an implausible departure from the pretrained CSP distribution.

\paragraph{General reward objectives.}
The inverse-density example gives a simple closed-form case, while the direction of reward-driven shift can be characterized more generally. Consider the infinitesimal tilt
\begin{equation}
q_\epsilon(x\mid c)
\propto
p_0(x\mid c)
\exp\!\left(
\epsilon R(x,c)
\right).
\end{equation}
The change in expected pretrained log-likelihood at $\epsilon=0$ is
\begin{equation}
\left.
\frac{\mathrm{d}}{\mathrm{d}\epsilon}
\mathbb{E}_{x\sim q_\epsilon(\cdot\mid c)}
\left[
\log p_0(x\mid c)
\right]
\right|_{\epsilon=0}
=
\operatorname{Cov}_{p_0(\cdot\mid c)}
\left(
\log p_0(x\mid c),
R(x,c)
\right).
\label{eq:novelty_covariance}
\end{equation}
A negative covariance therefore means that increasing reward pressure decreases expected pretrained log-likelihood. For the illustrative novelty reward above,
\[
\operatorname{Cov}_{p_0}
\left(
\log p_0,
-\alpha\log p_0
\right)
=
-\alpha\operatorname{Var}_{p_0}
\left[
\log p_0
\right]
\leq 0.
\]

For crystal discovery, the relevant reward component is novelty. A negative association between novelty and pretrained likelihood is plausible because structural novelty is usually measured against collections of known crystals drawn from the same materials space, and often from the same or closely related databases used to train the structural model. Structures farther from this reference set may therefore also lie in regions assigned lower probability by $p_0(x\mid c)$. This association is not guaranteed because a generative model can generalize beyond its training examples.

\paragraph{Semantic credit assignment.}
Reward-guided DNG can partly correct the composition-marginal problem because the reward-adapted joint distribution
\begin{equation}
q(c,x)
=
q(c)\,q(x\mid c)
\end{equation}
is free to move its composition marginal away from the likelihood-trained distribution. The same optimization can also change the conditional structural distribution.

A joint crystal-level objective does not enforce how adaptation is allocated between these factors. A high reward may arise from selecting promising chemistry, from generating a useful structure within that chemistry, or from both. Updating the full DNG generator can therefore modify both the composition distribution and the conditional structural distribution.

This creates a semantic credit-assignment problem distinct from assigning a terminal reward across the steps of a denoising trajectory. Reward obtained after moving into productive chemistry can still change $q(x\mid c)$ even when no change to the structural distribution was required. Structural exploration within $p_0(x\mid c)$ can remain useful without requiring the conditional distribution itself to adapt.

This distinction does not imply that structural adaptation is inherently harmful. A reward that measures structural quality can provide useful information for improving the conditional model. Stability is one example. The concern is that a joint crystal-level objective also permits structural adaptation when reward originates from chemical exploration or from discovery signals that do not imply any corresponding change in structural plausibility.

\paragraph{The reward-optimal composition marginal.}
Applying \eqref{eq:reward_tilt} to the joint rather than the conditional makes the effect on composition explicit. For a crystal-level reward $R(x,c)$ and the pretrained joint $p_0(c,x)=p_0(c)p_0(x\mid c)$,
\begin{equation}
q^*(c,x)
\propto
p_0(c)\,p_0(x\mid c)\exp\!\left(R(x,c)/\beta\right),
\end{equation}
and marginalizing over $x$ gives
\begin{equation}
q^*(c)
\propto
p_0(c)\,
\mathbb{E}_{x\sim p_0(\cdot\mid c)}\!\left[\exp\!\left(R(x,c)/\beta\right)\right]
=
p_0(c)\,Z(c).
\label{eq:optimal_composition_marginal}
\end{equation}
The reward-optimal composition marginal is therefore the pretrained marginal reweighted by a per-composition partition function. For a binary discovery reward $R=\mathbbm{1}[S(x,c)=1]$, $Z(c)=1+(e^{1/\beta}-1)s_0(c)$, which is monotone in $s_0(c)$: \eqref{eq:optimal_composition_marginal} is the soft analogue of the allocation $p_0(c)s_0(c)$ in \eqref{eq:search_success_app}, and recovers the same ordering over compositions. The conditional is tilted in parallel, $q^*(x\mid c)\propto p_0(x\mid c)\exp(R(x,c)/\beta)$, which is exactly the structural adaptation that the frozen construction removes.

\paragraph{The tilt is bounded.}
If the reward is bounded, $R_{\min}\le R(x,c)\le R_{\max}$, then $e^{R_{\min}/\beta}\le Z(c)\le e^{R_{\max}/\beta}$ for every composition, and the normalizer $\sum_{c'}p_0(c')Z(c')$, a convex combination of these values, lies in the same interval. With $\Delta R=R_{\max}-R_{\min}$,
\begin{equation}
e^{-\Delta R/\beta}\;\le\;\frac{q^*(c)}{p_0(c)}\;\le\;e^{\Delta R/\beta}.
\label{eq:bounded_tilt}
\end{equation}
The reward-optimal marginal therefore keeps the support of $p_0(c)$, and a composition to which the pretrained model assigns probability $p_0(c)$ receives at most $e^{\Delta R/\beta}p_0(c)$. KL-regularized fine-tuning can reweight chemistry the pretrained model already proposes, but it cannot reach chemistry to which that model assigns negligible probability unless $\beta$ becomes small, which is also the regime in which the structural tilt of \eqref{eq:reward_tilt} is least constrained. Our DDPO baselines use no KL term, so \eqref{eq:bounded_tilt} does not constrain them, but neither does anything constrain their conditional, which is then free to adapt to the full reward. The reward of \textsc{ANCHOR} is bounded as well, with $r(c)\in[r_{\mathrm{inv}},1+\lambda_c]$ (\eqref{eq:reward}), but its composition policy is not regularized toward $p_0(c)$ and its conditional is frozen, so it escapes the bound without the structural drift.

\paragraph{KL regularization and a frozen CSP model.}
The same distinction appears in the KL penalty. Writing the pretrained and reward-adapted joint endpoint distributions as
\[
p_0(c,x)=p_0(c)p_0(x\mid c)
\qquad\text{and}\qquad
q(c,x)=q(c)q(x\mid c),
\]
the chain rule for KL divergence gives
\begin{equation}
D_{\mathrm{KL}}\!\left(q(c,x)\|p_0(c,x)\right)
=
D_{\mathrm{KL}}\!\left(q(c)\|p_0(c)\right)
+
\mathbb{E}_{c\sim q(c)}
D_{\mathrm{KL}}\!\left(q(x\mid c)\|p_0(x\mid c)\right).
\label{eq:kl_factorization_app}
\end{equation}
A single KL coefficient therefore penalizes changes in both chemical exploration through $q(c)$ and structural adaptation through $q(x\mid c)$. Increasing its strength to protect the conditional structural distribution also restricts movement of the composition distribution away from its pretrained marginal.

Freezing the CSP model imposes the hard constraint
\begin{equation}
q(x\mid c)=p_0(x\mid c),
\label{eq:frozen_conditional}
\end{equation}
which gives
\begin{equation}
\mathbb{E}_{c\sim q(c)}
D_{\mathrm{KL}}\!\left(
q(x\mid c)\|p_0(x\mid c)
\right)
=
0.
\label{eq:zero_structural_kl}
\end{equation}
The conditional structural distribution is then fixed exactly while a separately parameterized composition policy remains free to adapt.

One could instead introduce separate regularization strengths
\begin{equation}
\beta_c
D_{\mathrm{KL}}\!\left(q(c)\|p_0(c)\right)
+
\beta_x
\mathbb{E}_{c\sim q(c)}
D_{\mathrm{KL}}\!\left(q(x\mid c)\|p_0(x\mid c)\right).
\label{eq:factor_specific_kl}
\end{equation}
which makes the distinction explicit. A single joint KL coefficient couples chemical and structural regularization. Freezing the CSP model fixes the second factor directly while leaving chemical adaptation separate.

The frozen construction therefore differs qualitatively from simply increasing the strength of joint KL regularization. Stronger joint regularization can reduce structural drift, but it simultaneously restricts the same compositional adaptation required to move beyond the likelihood-trained DNG marginal. \textsc{ANCHOR} instead removes structural adaptation as a route for optimizing the discovery reward while leaving the composition policy unconstrained by the structural KL term.

\paragraph{Application to history-dependent novelty.}
History-dependent novelty follows a standard idea from reinforcement learning, where exploration bonuses decrease as states become familiar through visitation counts or episodic memory~\citep{bellemare2016unifying,savinov2018episodic}. Applying this principle to crystal discovery makes novelty depend on the state of the search. At search step $t$,
\begin{equation}
R_t(x)
=
R_{\mathrm{nov}}
\left(
x,
\mathcal{D}_{\mathrm{ref}},
\mathcal{H}_t
\right),
\end{equation}
where $\mathcal{D}_{\mathrm{ref}}$ is the fixed novelty reference set and $\mathcal{H}_t$ contains the structures discovered so far.

Under the same endpoint-distribution analysis, Eq.~\ref{eq:reward_tilt} gives
\begin{equation}
q_t^*(x\mid c)
=
\frac{
p_0(x\mid c)
\exp\!\left(R_t(x)/\beta\right)
}{
Z_t(c)
}.
\label{eq:adaptive_target}
\end{equation}
The mathematics follows directly from the same tilting identity. As $\mathcal{H}_t$ grows, a previously discovered structure can receive lower novelty reward and therefore lower relative probability under $q_t^*(x\mid c)$ even though the structure itself, its physical properties, and its probability under $p_0(x\mid c)$ have not changed.

This exposes the distinction between search utility and structural quality particularly clearly. A crystal does not become less structurally plausible because it has already been discovered. It only becomes less useful to sample again. A structural generator optimized directly with history-dependent novelty is therefore asked to change because the search has already encountered a candidate, not because new information has changed the candidate's physical plausibility. The preferred reward-optimal structural distribution consequently evolves with campaign history.

A search policy should respond to this information. The structural prior need not. Campaign history can change which compositions should be queried and how often they should be revisited without requiring a corresponding change in the conditional structural distribution. Stronger KL regularization can suppress this effect when fine-tuning the generator, but again reduces the influence of the discovery reward on the joint model.

\paragraph{Connection to diffusion fine-tuning.}
The analysis above concerns endpoint distributions and does not describe the exact parameter updates of a particular diffusion-RL algorithm. Diffusion models generate samples through reverse dynamics associated with a target distribution and its noisy marginals~\citep{song2020score}. Chemeleon2, for example, regularizes its learned denoising policy against a pretrained reference throughout the generative process~\citep{park2026guiding}. The endpoint analysis isolates the corresponding distribution-level effect of reward optimization. Changing the reward changes the endpoint distribution favored by the regularized objective.

This distinction is important because successful reward fine-tuning does not require structural quality to deteriorate. A joint reward containing stability can improve the structural distribution, and recent reward-guided crystal generators demonstrate that such optimization can substantially improve discovery metrics~\citep{park2026guiding,chen2025accelerating}. The argument here is narrower. Joint fine-tuning permits both $q(c)$ and $q(x\mid c)$ to adapt, while the reward does not generally specify which factor should change. Reward-driven diffusion fine-tuning can also exhibit loss of diversity or sample quality when optimization moves sufficiently far from pretrained behavior~\citep{uehara2024fine}, providing an additional practical reason to control unnecessary structural adaptation. \citet{bergmeister2026reinforce} additionally show that the KL-regularized optimum tilts only the clean-endpoint distribution and leaves the conditional law of noisy states given that endpoint unchanged. Reward fine-tuning of a diffusion model therefore changes which endpoints are likely, not how a fixed endpoint is noised. The separation \textsc{ANCHOR} makes architecturally is the composition-space analogue: the reward changes which compositions are queried, not how a structure is generated once one is fixed.

\paragraph{Implication for \textsc{ANCHOR}.}
\textsc{ANCHOR} instantiates the composition--structure separation directly as
\begin{equation}
p_{\textsc{ANCHOR}}(c,x)
=
\pi_\theta(c)\,p_0(x\mid c),
\label{eq:anchor_ansatz_app}
\end{equation}
where $\pi_\theta(c)$ is an adaptive composition policy and $p_0(x\mid c)$ is a separately pretrained and frozen CSP model. Unlike the conceptual factorization of a conventional DNG distribution, Eq.~\ref{eq:anchor_ansatz_app} is an architectural decomposition. The two factors are represented by separate models and only the composition policy is updated during discovery.

The composition policy can therefore increase the probability of querying regions where the CSP model produces useful candidates even if those regions receive little probability under a likelihood-trained DNG composition marginal. At the same time,
\begin{equation}
p_{\textsc{ANCHOR}}(x\mid c)
=
p_0(x\mid c)
\end{equation}
by construction. This removes reward-driven changes to the conditional structural distribution, including both the semantic credit-assignment effects of joint generator fine-tuning and the non-stationarity introduced by history-dependent novelty.

Freezing the CSP model does not freeze structural exploration. Repeated queries to the same composition can still sample different structures from $p_0(x\mid c)$. \textsc{ANCHOR} therefore changes where and how often the structural model is queried without restricting the structural diversity already represented by the pretrained CSP distribution.

This separation becomes increasingly natural as CSP models improve. In the limiting case of a CSP model that perfectly captures the physically plausible structural distribution for any queried composition, a separately trained DNG model becomes unnecessary for discovery. Discovery reduces to learning where and how often to query the CSP model. A learned DNG composition marginal may remain useful as an initialization or chemical prior, but there is no reason for the discovery distribution itself to reproduce the composition frequencies of the training data or to modify a conditional structural distribution that is already correct.

\newpage
\section{Method and Training Details}
\label{app:method-training}

\subsection{Element Vocabulary and Action Space}
\label{app:elem_vocab_action_space}

\paragraph{Vocabulary.} 
The policy draws elements from the $E = 84$ symbols listed in Table~\ref{tab:elements}. These are the $89$ elements appearing in MP-20, minus the five noble gases (He, Ne, Ar, Kr, Xe). The noble gases are removed because they have no tabulated ICSD oxidation states in pymatgen~\citep{ong2013python}, so any composition containing one of these 5 elements will fail the charge-neutrality half of the validity gate by construction (\appref{app:reward/validity}). 
Keeping these elements in the vocabulary would spend needless rollouts on actions whose reward is known in advance to be $r_{\mathrm{inv}}$, which adds variance to the group baseline (\eqref{eq:advantage}) without adding information. 
\begin{table}[h]
    \centering
    \caption{The $E = 84$ elements available to the policy, grouped by block and listed by ascending atomic number within each row. Removed from MP-20: He, Ne, Ar, Kr, Xe.}
    \label{tab:elements}
    \small
    \begin{tabular}{ll}
        \toprule
        \textbf{Block} & \textbf{Elements} \\
        \midrule
            Period 1--3 & H, Li, Be, B, C, N, O, F, Na, Mg, Al, Si, P, S, Cl \\
            Period 4    & K, Ca, Sc, Ti, V, Cr, Mn, Fe, Co, Ni, Cu, Zn, Ga, Ge, As, Se, Br \\
            Period 5    & Rb, Sr, Y, Zr, Nb, Mo, Tc, Ru, Rh, Pd, Ag, Cd, In, Sn, Sb, Te, I \\
            Period 6    & Cs, Ba, Hf, Ta, W, Re, Os, Ir, Pt, Au, Hg, Tl, Pb, Bi \\
            Lanthanides & La, Ce, Pr, Nd, Pm, Sm, Eu, Gd, Tb, Dy, Ho, Er, Tm, Yb, Lu \\
            Actinides   & Ac, Th, Pa, U, Np, Pu \\
        \bottomrule
    \end{tabular}
\end{table}

\paragraph{Action-space bounds.} 
A composition action (\eqref{eq:action}) uses $k \in \gK = \{2,3,4\}$ distinct elements and a cell size $T \in \{k,\dots,N\}$ with $N = 20$, with at most $12$ atoms of any single element. We exclude $k = 1$ because unary compositions carry no compositional-discovery signal, and cap $k$ at $4$. 


\paragraph{Size of the search space.} 
Enumerating all charge-neutral compositions consistent with the tabulated ICSD states~\citep{ong2013python} gives roughly $4.4 \times 10^{4}$ binary, $1.1 \times 10^{7}$ ternary and $1.1 \times 10^{9}$ quaternary candidates. The policy therefore searches a space of order $10^{9}$ compositions while a training run evaluates on the order of $10^{5}$ of them. This is the regime that makes the exploration mechanisms, described in \secref{sec:method} and ablated in \secref{sec:results}, load-bearing rather than cosmetic.



\subsection{Structure Generation, Relaxation, and the Validity Gate}
\label{app:reward/validity}

\paragraph{Frozen CSP backbone and relaxation.}
Given a reduced formula $f(c)$, the frozen KLDM CSP model~\citep{cornet2025kinetic} samples a structural realization. The CSP model is queried anew for every composition action, so repeated queries of the same reduced formula may produce different structures. The sampled crystal is then relaxed with BFGS under the MatterSim interatomic potential~\citep{yang2024mattersim}, over atomic positions only and with the lattice fixed at the backbone output. Relaxation and all subsequent reward calculations form part of the environment observed by the composition policy. Neither the CSP backbone nor the interatomic potential receives gradient updates. The complete path for a sampled action is therefore
\[
c
\longrightarrow
f(c)
\longrightarrow
x \sim p_0(x\mid f(c))
\longrightarrow
x_{\mathrm{relaxed}}
\longrightarrow
r(c),
\]
and only the mapping from the rollout state to the composition action is learned.

\paragraph{Evaluation order.}
Each relaxed structure is evaluated in a fixed order: \textbf{(i)}~validity gate, \textbf{(ii)}~stability (\eqref{eq:stab-score}, valid only), \textbf{(iii)}~novelty (\eqref{eq:novelty}, valid only), \textbf{(iv)}~within-group diversity (\eqref{eq:diversity}, over all $G$ samples in the rollout, after per-structure scoring).

\paragraph{Validity gate.}
A structure is \emph{valid} only if it passes both of the following checks.
\begin{enumerate}
    \item \textbf{Bond lengths.} No two atoms may sit closer than an element-pair minimum. For a site pair of elements $(a,b)$ at minimum-image distance $d$, the structure is rejected if $d < \theta_{ab}$, with
    \begin{equation}
      \theta_{ab} =
      \begin{cases}
        (1-\tau)\,d^{\min}_{ab}, & (a,b)\ \text{in the reference table},\\[2pt]
        d_{\mathrm{floor}},      & \text{otherwise}.
      \end{cases}
      \label{eq:bond-thresh}
    \end{equation}
    Here $d^{\min}_{ab}$ is the shortest $a$--$b$ separation observed in MP-20, and $\tau = 0.2$ is a fractional tolerance that admits bonds up to $20\%$ below that reference. Pairs never seen in MP-20 fall back to a global floor $d_{\mathrm{floor}} = 1.2$\,\AA, applied without tolerance. A single too-close pair fails the check.
    \item \textbf{Charge neutrality.} The composition must admit a charge-neutral oxidation-state assignment under pymatgen's \texttt{oxi\_state\_guesses}, using ICSD oxidation states~\citep{ong2013python}. The ICSD list only restricts which integer states may be combined; the test itself is a global neutrality search, and any non-empty set of assignments is accepted (unweighted). Per-element oxidation bounds are available in the implementation but disabled in all reported runs.
\end{enumerate}
Invalid structures receive the fixed penalty $r_{\mathrm{inv}}$ (\eqref{eq:reward}) and skip stability and novelty evaluation. They are still embedded and kept in the group when within-group diversity (\eqref{eq:diversity}) is computed: an invalid structure can lower a valid sibling's diversity score, even though its own reward never reflects that term.
Treating validity as a gate rather than as a fifth weighted objective is deliberate: as a weighted term it could be traded away. The magnitude of $r_{\mathrm{inv}}$ still controls how strongly invalid actions are discouraged; we sweep it in \appref{app:stage2-invalid_penalty}.

\subsection{Stability Score}
\label{app:reward/stability}

For valid structures we score thermodynamic viability from the energy of the relaxed cell above the convex hull~\citep{jain2013commentary},
\begin{equation}
  s_{\mathrm{stab}}
  \;=\;
  1 - \mathrm{clip}(E_{\mathrm{hull}},\, 0,\, 1).
  \label{eq:stab-score}
\end{equation}
$E_{\mathrm{hull}}$ is smaller for more stable crystals, so the score inverts it and places stability on the same higher-is-better $[0,1]$ scale as novelty and diversity. Thus $s_{\mathrm{stab}}=1$ on the hull and falls to $0$ for $E_{\mathrm{hull}}\geq1$~eV/atom. 
Sharing one bounded scale between the objectives is what makes the Dirichlet weights $\vw$ meaningful: the scalarization $\vm^{\top}\vw$ is then a convex combination of comparable terms, so no objective can dominate through magnitude alone and the group-relative advantages stay on a common scale.
%
%
%
Given the relaxed structure, \eqref{eq:stab-score} is deterministic; this section is the path from that cell to $E_{\mathrm{hull}}$.
\paragraph{From a generated cell to $E_{\mathrm{hull}}$.} 
Each valid structure passes through four steps, all using the MatterSim potential of \appref{sec:app-setup}.
\begin{enumerate}
  \item \textbf{Relaxation.} The atomic positions are relaxed with the ASE BFGS optimizer~\citep{ase-paper} to a force tolerance $f_{\max}=0.02$~eV/\AA, capped at a step budget of 100. The lattice is held fixed at the backbone output. Relaxations that reach the cap are not discarded, and the potential energy at the final step is taken as the structure's energy $E_p$. At this budget, most relaxations of ANCHOR structures are truncated (Appendix~\ref{app:relax-budget}), so the stability term is a ranking signal within a group rather than a converged $E_{\text{hull}}$.
  \item \textbf{MP2020-compatible entry.} The relaxed energy $E_{\mathrm{p}}$ and the cell's composition are wrapped as a computed entry and passed through \texttt{MaterialsProject2020Compatibility}~\citep{ong2013python}. This places the MatterSim energy on the same corrected scale as the reference entries, so structure and hull are directly comparable.
  \item \textbf{Reference convex hull.} For the structure's chemical system we gather the reference entries of that system and all of its subsystems from MatterGen's Alex-MP reference dataset~\citep{zeni2025generative}, which combines Materials Project and Alexandria~\citep{schmidt2024improving} entries, and build a pymatgen \texttt{PhaseDiagram}. The reference entries carry the same MP2020 compatibility correction. Note that this is a different and far larger corpus than the MP-20 index used for novelty (\appref{app:reward/novelty}). Phase diagrams are cached per chemical system and reused across the run. 
  \item \textbf{Energy above convex hull.} $E_{\mathrm{hull}}$ (eV/atom) signifies the distance  of the structure's $E_{\mathrm{p}}$ above the reference convex hull. We retain its sign, so a structure on or below the hull is not clipped to zero before \eqref{eq:stab-score}.
\end{enumerate}
%




\subsection{Continuous Adaptive Novelty and the Count Bonus}
\label{app:reward/novelty}

Our novelty reward, which we refer to throughout as continuous adaptive novelty (CAN), makes two design choices, each isolated by an ablation in \appref{app:explore_ablation}. 
It is \emph{continuous}: instead of a binary novel/not-novel flag, every structure receives a graded score in $[0,1]$ from a kNN--RBF comparison in descriptor space~(\eqref{eq:knn-novelty}--\eqref{eq:novelty}). And it is \emph{adaptive}: the reference is not the frozen MP-20 corpus alone but a growing index of the agent's own discoveries, so rediscovering a structure the policy already found is penalized. 
This section specifies the pieces that implement both; the RDF descriptor, the frozen MP-20 reference index and the online agent-history index. Novelty is computed only for valid relaxed structures (\appref{app:reward/validity}).

\paragraph{RDF embedding.}
Given a relaxed structure $x$, let $d_{ab}$ denote the minimum-image distance between sites $a$ and $b$. We collect all pairwise distances with $0 < d_{ab} \le r_{\max}$, histogram them into $n_{\text{bins}}$ equal-width bins on $(0, r_{\max}]$, and $\ell_1$-normalize, writing $h_b$ for the number of pairs that fall in bin $b$:
\begin{equation}
  z_b \;=\; \frac{h_b}{\sum_{b'} h_{b'}},
  \qquad
  \vz(x) = (z_1, \dots, z_{n_{\text{bins}}}) \in \R^{n_{\text{bins}}}.
\end{equation}
Normalization reduces sensitivity to unit-cell size. The same $\vz$ is reused for within-group structural diversity (\eqref{eq:diversity}). 

\paragraph{MP-20 reference index.}

Before training, every MP-20 structure is bucketed by reduced formula $f$ and converted to $\vz(x)$. 
We write $Z^{\mathrm{ref}}$ for the embeddings of the formula currently being scored, and $\mathcal{F}_{\mathrm{ref}}$ for the reduced formulas that have at least one such embedding. 
At runtime the index is loaded from a prebuilt cache; it is never updated during RL.

\paragraph{kNN--RBF scoring.}
Novelty comparisons are \emph{within-formula}: a candidate $\vz$ is scored only against reference embeddings sharing its reduced formula $f$. This isolates structural novelty within a composition family; exploration \emph{across} families is handled instead by the composition policy and the count bonus.
In \eqref{eq:knn-novelty}, $\mathrm{kNN}_{k_{nn}}(\vz, Z)$ is the set of indices of the $\min(k_{nn}, |Z|)$ embeddings in $Z$ with smallest $\ell_2$ distance to $\vz$, so $\gJ$ is truncated when a formula has fewer than $k_{nn}$ references and the average is over that (possibly smaller) set. 
The RBF kernel converts distances in RDF space into graded similarity, and averaging over the neighbourhood reduces sensitivity to a single nearest reference while remaining a local measure of structural similarity.

\paragraph{Bandwidth $\sigma$.}
For the MP-20 branch we use a precomputed per-formula bandwidth $\sigma_f$: the median pairwise distance among embeddings in $Z^{\mathrm{ref}}$ (median heuristic). For the agent-history branch, $\sigma$ is computed on-the-fly from the current $Z^{\mathrm{hist}}$. Both paths then apply a global floor $\sigma \leftarrow \max(\sigma, \sigma_{\mathrm{floor}})$.
The floor is the 10th percentile of leave-one-out, within-formula $k$-NN distances on MP-20 itself (how close known polymorphs of the same formula already sit in RDF space), computed once offline. Without it, a family with two near-identical references would drive $\sigma$ to zero and the RBF would collapse to a $\delta$: novelty $0$ on an exact match and $1$ otherwise.

\paragraph{Agent-history index.}
The two indices ask different questions. The fixed index asks whether a structure differs from known data; the growing index asks whether it differs from what the search itself has already found. During training the agent maintains $Z^{\mathrm{hist}}$, a FIFO buffer of RDF embeddings keyed by reduced formula. After each novelty evaluation, $\vz(x)$ is appended, so a candidate is always scored against the history as it stood before that candidate was generated and never against itself; when the buffer exceeds a per-formula cap $C_{\mathrm{hist}}$ the oldest entry is evicted. 


\paragraph{Scoring procedure.}
Algorithmically, novelty evaluation follows the rule in~\eqref{eq:novelty}:
\begin{enumerate}
  \item Compute $\vz(x)$ and look up $Z^{\mathrm{ref}}$ and $Z^{\mathrm{hist}}$.
  \item If $f \in \mathcal{F}_{\mathrm{ref}}$: score against MP-20; if
    $|Z^{\mathrm{hist}}| > 0$, also score against agent history and take the
    minimum (anti-rediscovery for known chemistries).
  \item If $f \notin \mathcal{F}_{\mathrm{ref}}$ and $|Z^{\mathrm{hist}}| = 0$:
    return the bootstrap default $\nu_0 = 0.75$.
  \item If $f \notin \mathcal{F}_{\mathrm{ref}}$ and $|Z^{\mathrm{hist}}| > 0$:
    score by kNN against $Z^{\mathrm{hist}}$.
  \item Append $\vz(x)$ to $Z^{\mathrm{hist}}$.
\end{enumerate}
The bootstrap $\nu_0 = 0.75$ is set below the ceiling of $1$ on purpose. It is a placeholder returned before any same-formula reference exists, not a claim that the structure is maximally novel; leaving headroom above it means that once references for that formula accumulate, a genuinely distinct structure can be scored \emph{higher} than the first-seen bootstrap by the kNN--RBF metric. A value of $1$ would saturate the term on the first sample and flatten within-formula novelty thereafter.


\paragraph{Hyperparameters.}
Table~\ref{tab:novelty-hparams} summarizes the novelty configuration used in the main runs.
\begin{table}[h]
    \centering
    \caption{Novelty hyperparameters (main runs).}
    \label{tab:novelty-hparams}
    \begin{tabular}{lcl}
    \toprule
    \textbf{Symbol / parameter} & \textbf{Value} & \textbf{Role} \\
    \midrule
        $r_{\max}$ & 7\,\AA & RDF distance cutoff \\
        $n_{\text{bins}}$ & 80 & RDF histogram resolution \\
        $k_{\mathrm{cfg}}$ & 10 & kNN neighbourhood size \\
        $\nu_0$ & 0.75 & Bootstrap for first off-MP-20 formula \\
        $C_{\mathrm{hist}}$ & 20 & FIFO cap per reduced formula \\
        $\sigma_{\mathrm{floor}}$ & 10th pct.\ calibration & Minimum RBF bandwidth \\
    \bottomrule
    \end{tabular}
\end{table}

\paragraph{Count Bonus and the Absence of a Uniqueness Term}
The count bonus in \eqref{eq:reward} adds
\[
\frac{\lambda_c}{\sqrt{\max(n_f,\,1)}}
\]
to the reward of every valid action, where $n_f$ is the number of times reduced formula $f$ has been generated as a valid structure so far in the current run, including the current group. 
The inverse-square-root dependence gives a large incentive for early visits to a formula and progressively reduces the bonus as that formula becomes familiar.

We deliberately do not add a fifth, explicit uniqueness term. The count bonus only sees how often a formula has been \emph{proposed}; whether the crystal itself is new is the job of CAN, so a separate uniqueness objective would charge the same repeat twice. The two mechanisms also act at different levels: the count bonus discourages repeatedly allocating search effort to the same reduced formula, while CAN discourages rediscovery of the same or closely related structures \emph{within} a formula family. 
%
Both also persist for the whole run, whereas uniqueness is conventionally measured as repetition within a single generated group~\citep{zeni2025generative,betala2025lemat}.


\subsection{Within-group Diversity Kernels}
\label{app:reward/diversity}
A group that contains only one structural or chemical motif provides little useful ranking signal for a group-relative update. We therefore reward candidates that differ from the other $G-1$ samples in the same rollout,
\begin{equation}
  d_{\bullet}(x_i)
  =
  \mathrm{clip} \Bigl(
      1 - \tfrac{1}{G-1} \sum_{j \neq i} K_{\bullet}(x_i, x_j),
      \; 0,\; 1
    \Bigr),
  \qquad
  \bullet \in \{\mathrm{struct}, \mathrm{comp}\},
  \label{eq:diversity}
\end{equation}
where $K_{\mathrm{struct}}$ is an RBF kernel on the RDF embeddings $\vz$ and $K_{\mathrm{comp}}$ is the Jaccard similarity of chemical-role sets. The clip is a numerical guard, since both kernels already return similarities in $[0,1]$, which places the two terms on the same scale as the other objectives.
%
%
%
%
%
%
Writing $\mathrm{roles}_i$ for the set of chemical families present in action $c_i$,
\begin{equation*}
  K_{\mathrm{comp}}(x_i, x_j)
  \;=\;
  \frac{\lvert \mathrm{roles}_i \cap \mathrm{roles}_j \rvert}{\lvert \mathrm{roles}_i \cup \mathrm{roles}_j \rvert}.
\end{equation*}
These families partition the vocabulary of \appref{app:elem_vocab_action_space} into $11$ classes; alkali, alkaline earth, early transition metal, late transition metal, lanthanide, actinide, p-block metal/metalloid, chalcogen, halogen, pnictogen, and a residual class.
The partition is deliberately coarse enough that a $k \in \{2,3,4\}$ action maps to $2$--$4$ families, which keeps this similarity discriminative at every admissible $k$. Comparing role sets rather than raw element sets means two actions that differ only by substituting Na for K are treated as compositionally similar, which is the behaviour we want from a term whose job is to prevent a group from collapsing onto one chemical motif.

Both terms are computed once per rollout, after every sample in the group has been individually scored for stability and novelty. Unlike novelty (\eqref{eq:novelty}), they compare a structure only against its $G{-}1$ siblings and carry no memory across steps. The diversity objectives therefore serve to limit within-rollout redundancy rather than the adaptive novelty's long-horizon role to prevent rediscovery.





\subsection{Policy Parametrization and Optimization Details}
\label{app:policy}


\paragraph{Composition action space and policy parameterization}



Throughout this section $\gE$ is the element vocabulary of size $E = |\gE|$, $N$ the maximum number of atoms per cell, and $\gK = \{2,3,4\}$ the admissible number of distinct elements; their values are given in \appref{app:elem_vocab_action_space} and all symbols in \appref{app:notation}. The composition action $c = (k, \ve, T, \rho)$ of \eqref{eq:action} deliberately carries no structural degrees of freedom: once the reduced formula $f(c)$ is fixed, atomic coordinates and lattice parameters are generated entirely by the frozen CSP backbone, so the reinforcement-learning problem acts on the chemical query supplied to that model.

The policy factorizes the composition action autoregressively in the order in which its components are sampled,
\begin{equation}
  \pi_\theta(c \mid \vs)
  \;=\;
  \underbrace{p_k(k \mid \vs)}_{\text{cardinality}}
  \;\prod_{i=1}^{k}
  \underbrace{p_e(e_i \mid e_{<i}, \vs)}_{\text{elements}}
  \;\underbrace{p_T(T \mid \ve, \vs)}_{\text{cell size}}
  \;\underbrace{p_\rho(\rho \mid T, \ve, \vs)}_{\text{stoichiometry}} .
  \label{eq:policy-factor}
\end{equation}
Each factor is implemented as a masked categorical head, with the corresponding multinomial construction used for $\rho$. The cell-size head chooses $T$ before the stoichiometry is written, so the policy knows how many atoms must be distributed across the selected elements. The stoichiometry mask ensures that every selected element appears at least once.


The rollout state $\vs$ carries both the objective weighting $\vw$ and a size-invariant summary of the element sub-group $\gS$ active for the rollout; its concrete form is given in \eqref{eq:state}. Element draws are restricted to $\gS$ by masking, so $p_e(e_i\mid e_{<i},\vs)=0$ for every $e_i\notin\gS$ (\eqref{eq:head-e}). The element vocabulary, maximum cell size, categorical masks, and resulting size of the discrete action space are fixed before training.

\paragraph{Element sub-group sampling.} 
Each rollout restricts element choice to a random sub-group $\gS \subseteq \gE$, drawn by first sampling its size and then the sub-group uniformly among all sets of that size,
\begin{equation}
  n_{\gS} \sim \mathrm{Uniform}\{\,k_{\max}, \dots, E\,\},
  \qquad
  \gS \sim \mathrm{Uniform}\bigl(\{\, \gS' \subseteq \gE : |\gS'| = n_{\gS} \,\}\bigr),
  \label{eq:subgroup}
\end{equation}
after which all element draws in \eqref{eq:policy-factor} are masked to $\gS$. The lower bound $k_{\max}=4$ guarantees that every sub-group can support the largest admissible number of distinct elements.

\paragraph{Per-head distributions.} 
Each factor of the policy in \eqref{eq:policy-factor} is a small MLP head conditioned on the state vector $\vs$ (\eqref{eq:state}). Logits outside an admissible set $A$ are set to $-\infty$ before the softmax; we write this $\mathrm{mask}(g, A)$.
\begin{align}
  k
    &\sim \mathrm{Categorical} \bigl(\softmax\, g_k(\vs)\bigr),
    \label{eq:head-k}\\
  e_i \mid e_{<i}
    &\sim \mathrm{Categorical} \bigl(\softmax\,
      \mathrm{mask}(g_e(\vs, e_{<i}),\, \gE_i)\bigr),
    \qquad
    \gE_i = \gS \setminus \{e_{<i}\},
    \label{eq:head-e}\\
  T
    &\sim \mathrm{Categorical} \bigl(\softmax\,
      \mathrm{mask}(g_T(\vs,\ve),\, \{k,\dots,N\})\bigr),
    \label{eq:head-T}\\
  \tilde{\rho} &\sim \mathrm{Multinomial} \bigl(T-k,\; \softmax\, g_\rho(\vs,\ve)\bigr), 
  \qquad \qquad
    \rho = \1_k + \tilde{\rho}.
    \label{eq:head-c}
\end{align}
These four draws are the components of a composition action $c$ in \eqref{eq:action}, in that order. Elements are drawn without replacement from the active sub-group $\gS$; $T \in \{k,\dots,N\}$ is sampled as its own categorical; and \eqref{eq:head-c} distributes the $T-k$ atoms over the $k$ chosen elements. Only $T-k$ atoms are to be distributed since we pre-distribute one atom for each element; ensuring each sampled element appears at least once in the composition.

\paragraph{State vector.} 
The head inputs concatenate a size-invariant summary of the active sub-group $\gS$ with the objective weights $\vw$,
\begin{equation}
  \vs \;=\; \bigl[\, \mW_{\!f}\,\bar{\vf}_{\gS}\; ;\; \vw \,\bigr],
  \qquad
  \bar{\vf}_{\gS} = \frac{1}{|\gS|}\sum_{e \in \gS} \vf_e,
  \label{eq:state}
\end{equation}
where $\mW_{\!f}$ is a learned linear projection and $\vf_e \in \R^{38}$ is a fixed (non-trained) descriptor of element $e$,
\begin{equation}
  \begin{aligned}
    \vf_e = \bigl[\,
      &\mathrm{onehot}(\mathrm{period}_e);\;
      \mathrm{onehot}(\mathrm{group}_e);\;
      \mathrm{onehot}(\mathrm{block}_e);\\
      &\tfrac{\phi_s}{2},\, \tfrac{\phi_p}{6},\, \tfrac{\phi_d}{10},\, \tfrac{\phi_f}{14};\;
      \tilde{\chi}_e,\, \tilde{r}_e,\, \tilde{o}^{\min}_e,\, \tilde{o}^{\max}_e,\, \tilde{o}^{\mathrm{mean}}_e
    \,\bigr],
  \end{aligned}
  \label{eq:element-feat}
\end{equation}
which concatenates one-hot period ($\R^{7}$), group ($\R^{18}$) and block ($\R^{4}$) encodings, the capacity-normalized valence-shell occupancies $(\phi_s, \phi_p, \phi_d, \phi_f)$, and the min--max-normalized Pauling electronegativity $\tilde{\chi}_e$, atomic radius $\tilde{r}_e$, and common-oxidation-state min, max and mean. The choice of period, group, block and valence-shell occupancies as the element descriptor follows the subatomic token of Crystalite~\citep{veljkovic2026crystalite}, though we keep these as one-hot and normalized channels rather than their PCA-compressed continuous token; the five scalar channels and the mean-pooling over $\gS$ in \eqref{eq:state} are ours.

\paragraph{Log-probability and entropy.} 
The log-probability of an action and the policy entropy each decompose as a sum over the four heads,
\begin{align}
  \log \pi_\theta(c \mid \vs)
    &= \log p_k + \textstyle\sum_{i=1}^{k}\log p_e^{(i)} + \log p_T + \log p_\rho, \label{eq:logp}\\
  H[\pi_\theta](\vs)
    &= H_k + \textstyle\sum_{i=1}^{k}H_e^{(i)} + H_T + H_\rho,
    \qquad
    H_\rho = (T-k)\,H\!\bigl(\softmax\, g_\rho\bigr). \label{eq:entropy}
\end{align}
The stoichiometry head draws the $T-k$ surplus atoms from a single categorical $\softmax\, g_\rho$ (\eqref{eq:head-c}). Its log-probability (\eqref{eq:logp}) is the exact multinomial log-probability, and its entropy contribution is the summed entropy of those $T-k$ independent trials, $H_\rho = (T-k)\,H(\softmax\, g_\rho)$, a closed-form upper bound on the exact multinomial entropy. 


\paragraph{Multi-objective rollout conditioning.}
Each rollout samples objective weights $\vw \sim \mathrm{Dirichlet}(\bm{\alpha})$ together with an element sub-group $\gS$ (\eqref{eq:subgroup}). Both are held fixed across all $G$ actions of the rollout and are supplied to the policy through the state $\vs$ (\eqref{eq:state}).
Sampling $\vw$ avoids committing to one scalarization in advance, which would otherwise be a weighting to tune. Because $\vw$ enters the conditioning state, a single set of parameters covers the whole weight family rather than one policy per scalarization. All evaluations fix $\vw$ to the Dirichlet mean.
Sampling $\gS$ restricts each rollout to a random chemical subspace, forcing the policy to keep practicing chemistries that could otherwise disappear from the sampled distribution as training becomes more exploitative.
Sharing both variables within a group matters for the update: every action entering a group-relative comparison is then evaluated under the same objective and the same available element set.


\paragraph{Group-relative advantages.}
Rewards are converted into advantages using the statistics of the sampled group as the baseline, so no learned value function is required~\citep{shao2024deepseekmath},
\begin{equation}
  A_i
  \;=\;
  \frac{r_i - \bar{r}}
       {\max\!\bigl(\mathrm{std}(r),\,\eps_A\bigr)},
  \qquad
  \bar{r}
  =
  \frac{1}{G}\sum_{j=1}^{G}r_j.
  \label{eq:advantage}
\end{equation}
The advantage measures how well an action performs relative to the alternative compositions sampled under the same rollout context. Sharing $\vw$ and $\gS$ within the group is what makes this meaningful: it prevents the normalization from comparing candidates evaluated under different objective trade-offs or different chemical constraints.

\paragraph{Clipped policy update.}
The policy is updated over $K$ epochs using the clipped PPO surrogate~\citep{schulman2017proximal}
\begin{equation}
  \Ls^{\mathrm{clip}}(\theta)
  =
  \E_i\!\left[
      \min\!\bigl(
        \eta_i A_i,\;
        \mathrm{clip}(\eta_i,1-\eps,1+\eps)A_i
      \bigr)
    \right],
  \qquad
  \eta_i
  =
  \frac{\pi_\theta(c_i\mid\vs)}
       {\pi_{\theta_{\mathrm{old}}}(c_i\mid\vs)}.
  \label{eq:ppo}
\end{equation}
Only the composition policy is updated. No gradient is propagated through the frozen CSP model (unless stated otherwise), the relaxation procedure, or the reward evaluator.


\paragraph{Full objective.} 
The loss we minimize is the negative clipped surrogate (\eqref{eq:ppo}) with an entropy bonus:
\begin{equation}
  \Ls(\theta)
  = -\,\Ls^{\mathrm{clip}}(\theta)
    \;-\; \beta_{H}\,\E_{i}\!\bigl[H[\pi_\theta]\bigr].
  \label{eq:loss}
\end{equation}

\paragraph{Entropy schedule.} 
The coefficient $\beta_{H}$ follows a warmup/plateau/cosine schedule, modulated by a proportional controller that nudges the normalized entropy $H[\pi_\theta]/H_{\max}$ toward a target fraction of its maximum. We take $H_{\max}$ as the largest per-sample entropy the chain \eqref{eq:entropy} can attain, with every head at its uniform maximum,
\begin{equation}
  H_{\max}
  = \log|\gK|
  + \max_{k \in \gK}\Bigl[
      \textstyle\sum_{i=0}^{k-1}\log(E-i)
      \;+\; \log(N-k+1)
      \;+\; (N-k)\log k
    \Bigr],
  \label{eq:hmax}
\end{equation}
where $\log|\gK|$ bounds the cardinality head and the three bracketed terms bound the $k$ without-replacement element draws, the cell-size head $H_T$, and the count head $H_c$ (at $T=N$). The element bound uses the full vocabulary $E$ rather than a per-rollout sub-group $\gS$, so $H_{\max}$ is a single constant for the whole run and the controller targets the same absolute entropy scale at every step, whichever sub-group was sampled. 
The schedule is keyed to cumulative structures generated (with $G$ per step) rather than to optimizer steps, so runs with different group sizes $G$ see the same entropy pressure at the same sample budget; this is what makes the comparison across $G$ in \appref{app:stage1} a fair one. 

\subsection{Training Algorithm}
\label{app:GRPO_alg}

Algorithm~\ref{alg:anchor} gives the full ANCHOR training procedure. Each optimization step has three phases. The \emph{rollout} phase draws one context (a preference vector $\vw$ and an element sub-group $\gS$) and samples $G$ compositions from the current policy under it. Each composition is decoded by the frozen backbone $\gG$ and relaxed. The \emph{scoring} phase computes the reward of each structure and converts the group's rewards into group-relative advantages. The \emph{update} phase performs $K$ clipped policy-gradient epochs on the stored rollout. Only the update phase changes $\theta$; no gradient passes through $\gG$, the relaxer, or the reward.

\begin{algorithm}[h]
\caption{ANCHOR: group-relative policy optimization over compositions,
with a frozen CSP backbone.}
\label{alg:anchor}
\begin{algorithmic}[1]
\Require frozen backbone $\gG$ and relaxer; frozen MP-20 index $Z^{\mathrm{ref}}$;
         hyperparameters (Table~\ref{tab:hparams})
\State $\theta \gets$ initial policy parameters; \quad
       $Z^{\mathrm{hist}} \gets \emptyset$; \quad
       $n_f \gets 0$ for all formulas $f$
\For{$\mathrm{step} = 1, 2, \dots$}
  \State $\theta_{\mathrm{old}} \gets \theta$
  \Statex \hspace{\algorithmicindent}\emph{Rollout (no gradients)}
  \State $\vw \sim \mathrm{Dirichlet}(\bm{\alpha})$; \quad
         $\gS \gets$ \Call{SampleSubgroup}{$\gE$}; \quad
         $\vs \gets [\,\mW_{\!f}\bar{\vf}_{\gS}\,;\,\vw\,]$
         \Comment{(\plaineqref{eq:subgroup}, \plaineqref{eq:state})}
  \For{$i = 1$ \textbf{to} $G$}
    \State $c_i \sim \pi_{\theta_{\mathrm{old}}}(\cdot \mid \vs)$; \quad
           $x_i \gets$ \Call{Relax}{$\gG(f(c_i))$}
           \Comment{(\plaineqref{eq:action}, \plaineqref{eq:policy-factor})}
  \EndFor
  \Statex \hspace{\algorithmicindent}\emph{Scoring}
  \State $(r_1,\dots,r_G) \gets$ \Call{ScoreGroup}{$\{c_i\}, \{x_i\}, \vw$}
  \State $A_i \gets \bigl(r_i - \bar{r}\bigr)\big/\max\bigl(\mathrm{std}(r), \eps_A\bigr)$
         \quad for $i = 1,\dots,G$
         \Comment{(\plaineqref{eq:advantage})}
  \Statex \hspace{\algorithmicindent}\emph{Update}
  \For{$\mathrm{epoch} = 1$ \textbf{to} $K$}
    \State $\Ls \gets
            -\,\Ls^{\mathrm{clip}}(\theta)
            \;-\; \beta_H\, \E_{i}\bigl[H[\pi_\theta(\cdot \mid \vs)]\bigr]$
           \Comment{(\plaineqref{eq:ppo}, \plaineqref{eq:loss})}
    \State $\theta \gets$ \Call{AdamStep}{$\theta,\, \nabla_\theta \Ls$}
           \Comment{gradients reach $\pi_\theta$ only}
  \EndFor
\EndFor
\State \Return trained policy $\pi_\theta$
\Statex
\Procedure{ScoreGroup}{$\{c_i\}_{i=1}^{G},\ \{x_i\}_{i=1}^{G},\ \vw$}
  \For{$i = 1$ \textbf{to} $G$} \Comment{in sampling order}
    \If{$x_i$ fails the validity gate} \Comment{(\appref{app:reward/validity})}
      \State \textbf{continue}
    \EndIf
    \State $s_{\mathrm{stab},i} \gets 1 - \mathrm{clip}\bigl(E_{\mathrm{hull}}(x_i),\, 0,\, 1\bigr)$
    \Comment{(\plaineqref{eq:stab-score})}
    \State $n_{f(c_i)} \gets n_{f(c_i)} + 1$
    \State $\nu_i \gets \nu(x_i)$; \quad
           append $\vz(x_i)$ to $Z^{\mathrm{hist}}$
           \Comment{(\plaineqref{eq:novelty})}
  \EndFor
  \State $\bigl(d_{\mathrm{struct},i},\, d_{\mathrm{comp},i}\bigr)_{i=1}^{G} \gets$
         diversity over all $G$ structures, valid or not
         \Comment{(\plaineqref{eq:diversity})}
  \For{$i = 1$ \textbf{to} $G$} \Comment{scalarize}
    \If{$x_i$ valid}
      \State $\vm_i \gets \bigl(s_{\mathrm{stab},i},\, \nu_i,\, d_{\mathrm{struct},i},\, d_{\mathrm{comp},i}\bigr)$
      \State $r_i \gets \vm_i^{\!\top}\vw + \lambda_c\big/\sqrt{n_{f(c_i)}}$
             \Comment{(\plaineqref{eq:reward})}
    \Else
      \State $r_i \gets r_{\mathrm{inv}}$
    \EndIf
  \EndFor
  \State \Return $(r_1, \dots, r_G)$
\EndProcedure
\end{algorithmic}
\end{algorithm}

\subsection{Training Setup and Hyperparameters}
\label{sec:app-setup}

\paragraph{Backbone and training potential.} 
Every policy trained with KLDM uses a single KLDM CSP checkpoint as the frozen structure-prediction backbone, sampled with a 50-step Euler--Maruyama solver, one structure per formula. 
Generated cells are relaxed under BFGS with the training potential MatterSim-1M, over atomic positions only and for at most 100 steps (\appref{app:reward/stability}).

Two MatterSim variants appear in the paper and it matters which is which: the \emph{training potential}, which supplies the stability term of the reward inside the training loop, and the \emph{evaluation potential}, which scores final structures after training (\appref{app:eval_protocol}). The training potential is a base-model-selection variable; the evaluation potential is fixed to MatterSim-5M for every internal table.
LeMat-GenBench and MatterGen evaluation submissions are scored by their respective default pipelines, not by our MatterSim-5M (\appref{app:eval-mattergen} and \ref{app:eval-lemat}).

\paragraph{Hyperparameters.} 
Table~\ref{tab:hparams} lists every value needed to reproduce a main model run. Entries marked ``swept'' are varied in the corresponding appendix experiment and kept at the listed default elsewhere.
\begin{table}[h]
\centering
\caption{Hyperparameters for main runs.}
\label{tab:hparams}
\small
\begin{tabular}{llp{0.42\linewidth}}
\toprule
\textbf{Parameter} & \textbf{Value} & \textbf{Note} \\
\midrule
\multicolumn{3}{l}{\emph{Action space}} \\
$E$ & 84 & Element vocabulary (\appref{app:elem_vocab_action_space}) \\
$\gK$ & $\{2,3,4\}$ & Distinct elements per action \\
$N$ & 20 & Max atoms per cell \\
max atoms / element & 12 & \\
\addlinespace
\multicolumn{3}{l}{\emph{Policy network}} \\
state dim & 32 & Projected sub-group summary \\
hidden dim & 256 & Per-head MLP width \\
element embedding & 32 & \\
\addlinespace
\multicolumn{3}{l}{\emph{GRPO}} \\
$G$ & 32 & Group size; swept in \appref{app:stage1} \\
$K$ & 2 & Update epochs per step \\
$\eps$ & 0.2 & PPO clip \\
$\eps_A$ & 0.05 & Advantage std floor (\plaineqref{eq:advantage}) \\
optimizer & Adam, learning rate $10^{-4}$ & \\
budget & $10^{5}$ structures & Entropy-schedule horizon; see below \\
\addlinespace
\multicolumn{3}{l}{\emph{Entropy schedule}} \\
$\beta_H$ base / min / max & $7.5\!\times\!10^{-3}$ / $2\!\times\!10^{-3}$ / $2.5\!\times\!10^{-2}$ & \\
target fraction & 0.85 & Of $H_{\max}$ \\
deadband / gain & 0.03 / 3.0 & Controller \\
warmup / warm start & 0.02 / 0.5 & Fraction of horizon; fraction of base at step 0 \\
decay start & 1.0 & No cosine decay within the budget \\
\addlinespace
\multicolumn{3}{l}{\emph{Reward}} \\
$\bm{\alpha}$ (Dirichlet) & $(1, 4, 1, 1)$ & Order: stability, novelty, $d_{\mathrm{struct}}$, $d_{\mathrm{comp}}$ \\
$\lambda_c$ & 1.0 & Count bonus; switched on/off in \appref{app:explore_ablation} \\
$r_{\mathrm{inv}}$ & $-0.2$ & $-0.2$ in base-model selection; $\{-0.1, -0.5\}$ in the invalid-penalty sweep (\appref{app:stage2-invalid_penalty}) \\
\addlinespace
\multicolumn{3}{l}{\emph{Evaluator}} \\
bond-length tolerance & 0.2 & Fractional \\
fallback min distance & 1.2\,\AA & Unlisted element pairs \\
oxidation states & ICSD & pymatgen oxidation-state mode (\appref{app:reward/validity}) \\
\bottomrule
\end{tabular}
\end{table}

\paragraph{Training budget.}
Because the group size $G$ is one of the variables we compare, wall-clock and optimizer steps are not comparable across runs; the number of structures generated is. 
The entropy schedule is keyed to a horizon of $10^{5}$ structures (\appref{app:policy}). Jobs were stopped by a common seven-day wall-clock, so the number actually consumed is $G\times$~(final step) and is not exactly $10^{5}$. Table~\ref{tab:run-budget} is that count for every trained policy in the paper. Collapsed cells consumed more, not less, because repeating small cells makes each BFGS relaxation cheap (\appref{app:explore_ablation}).

\begin{table}[h]
  \centering
  \caption{Structures generated by each trained policy, $G\times$~(final checkpoint step). All jobs are a single A100 and a seven-day wall. The entropy-schedule horizon is $10^{5}$; this table is what was actually consumed. The selected base model is $G{=}32$ / MatterSim-1M.}
  \label{tab:run-budget}
  \small
  \begin{tabular}{llrrr}
    \toprule
    Experiment & Configuration & $G$ & Step & Structures \\
    \midrule
    Base-model sweep
      & $G{=}16$ / 1M & 16 & 7950 & $127{,}200$ \\
      & $G{=}16$ / 5M & 16 & 7800 & $124{,}800$ \\
      & $G{=}32$ / 1M & 32 & 3850 & $123{,}200$ \\
      & $G{=}32$ / 5M & 32 & 3850 & $123{,}200$ \\
      & $G{=}64$ / 1M & 64 & 2000 & $128{,}000$ \\
      & $G{=}64$ / 5M & 64 & 1750 & $112{,}000$ \\
    \addlinespace
    Invalid penalty
      & $r_{\mathrm{inv}}=-0.1$ & 32 & 3950 & $126{,}400$ \\
      & $r_{\mathrm{inv}}=-0.2$ & 32 & 3850 & $123{,}200$ \\
      & $r_{\mathrm{inv}}=-0.5$ & 32 & 3850 & $123{,}200$ \\
    \addlinespace
    Exploration factorial
      & full method & 32 & 3850 & $123{,}200$ \\
      & -- count bonus & 32 & 4100 & $131{,}200$ \\
      & -- sub-group mask & 32 & 3200 & $102{,}400$ \\
      & -- adaptive nov. & 32 & 4000 & $128{,}000$ \\
      & adaptive nov.\ only & 32 & 3300 & $105{,}600$ \\
      & count bonus only & 32 & 3250 & $104{,}000$ \\
      & sub-group mask only & 32 & 7400 & $236{,}800$ \\
      & none & 32 & 20250 & $648{,}000$ \\
      & binary novelty & 32 & 4050 & $129{,}600$ \\
    \addlinespace
    Validity gate
      & KLDM, LeMat-GenBench gate & 32 & 6750 & $216{,}000$ \\
    \addlinespace
    Trainable KLDM-CSP
      & \textsc{ANCHOR} shared rewards & 32 & 3800 & $121{,}600$ \\
      & Stability only & 32 & 3850 & $123{,}200$ \\
    \bottomrule
  \end{tabular}
\end{table}

\paragraph{Compute.}
Each reinforcement-learning run occupies a single NVIDIA A100-40GB under a seven-day wall, so one run is $168$ GPU-hours. 
Every such run hit that wall, which is why runs are compared here by structures generated (Table~\ref{tab:run-budget}). 
Within that wall the selected configuration generates $\sim$$123\mathrm{k}$ structures and most other cells $102$--$131\mathrm{k}$. 
The collapsed and the LeMat-GenBench gate cells generate more structures because repeated small cells make each BFGS cheap. 
The cost is backbone sampling plus BFGS of every generated cell; the policy update is a few small MLP heads and is negligible beside that.

Counting unique training jobs, Table~\ref{tab:run-budget} holds $19$ reinforcement-learning runs: six base-model cells, two further invalid-penalty cells, eight further exploration cells (the selected $G{=}32$ / MatterSim-1M cell is also the $r_{\mathrm{inv}}{=}-0.2$ run and the full method), one validity-gate cell, and two trainable KLDM-CSP runs, or $19\times 168 = 3{,}192$ GPU-hours. 

Likelihood training sits outside that loop and adds about $350$ GPU-hours. KLDM releases CSP weights only, with no DNG checkpoint, so the KLDM-DNG baseline is one we trained ourselves ($46$). 
On top of that come four CSP fine-tunes, which continue training KLDM-CSP and Crystalite-CSP on MP-20 alone and on MP-20 augmented with \textsc{ANCHOR} structures ($44$); four DNG reward fine-tunes, KLDM-DNG and Crystalite-DNG under a mean and a balanced weighting ($4\times 48 = 192$); and generating the \textsc{ANCHOR} corpus that the augmented arms train on ($61$).

Evaluating three seeds of $N_{\mathrm{eval}}{=}1000$ for every main-text row adds about $540$. The single-seed appendix protocols add about $80$ A100-hours, plus about $70$ RTX~3090-24GB GPU-hours for LeMat-GenBench. The total is roughly $4{,}200$ A100 GPU-hours, excluding unreported exploratory runs.

\newpage
\section{Evaluation Protocols}
\label{app:evaluation-protocols}



\subsection{Internal Evaluation Protocol}
\label{app:eval_protocol}

\paragraph{Two relaxation protocols.}
Every internal number in this paper uses MatterSim-5M and BFGS for at most 400 steps. The protocols differ only in whether the lattice is relaxed. All internal numbers in the main text (Tables~\ref{tab:explore-ablation}--\ref{tab:internal}) relax positions and lattice. The lattice-fixed appendix tables (Tables~\ref{tab:app-internal}, \ref{tab:app-lemat-gate}, \ref{tab:stage1-eval},\ref{tab:stage1-late-window}, \ref{tab:invalid-penalty}, \ref{tab:app-explore-fixed} and~\ref{tab:kldm-ladder}) 
hold the lattice fixed at the backbone output, and their captions are marked \emph{Lattice fixed}. Neither is the training reward, which uses MatterSim-1M and 100 steps with the lattice fixed (\secref{app:reward/validity} and \ref{app:reward/stability}). The same checkpoint therefore has different values in the main text and in this appendix: the selected policy reaches 38.5\% MSUN with the lattice fixed and 47.6\% with it free. Tables~\ref{tab:explore-ablation} and~\ref{tab:app-explore-fixed} give both. The MatterGen and LeMat-GenBench tables in \appref{app:eval-mattergen} and \appref{app:eval-lemat} use those pipelines' own relaxation and are unaffected.  Statistics logged during training measure a policy along the trajectory it happened to take, which makes them path-dependent and unsuitable for comparing final policies. Every headline number in this paper is therefore produced after training, from fresh samples. We load a checkpoint, sample new actions, generate and relax the corresponding structures, and score them with a potential that is the same for every policy being compared.

\paragraph{Sampling.} 
Each policy generates $N_{\mathrm{eval}} = 1000$ structures per sampling seed. The lattice-free tables of the main text report mean and standard deviation over three sampling seeds, and the MatterGen and LeMat-GenBench appendix tables use seed~0. The lattice-fixed appendix tables use seed~0 alone and serve as a check on the protocol. Actions are sampled stochastically rather than by \texttt{argmax}, since a deterministic policy would collapse to one composition and the quantity of interest is the distribution the policy induces. The objective weights are fixed to the mean of the training Dirichlet, $\vw = \E[\mathrm{Dirichlet}(\bm{\alpha})]$, and the sub-group mask is disabled so the policy may use the full $84$-element vocabulary. 

\paragraph{Backbone transfer.}
Unless stated otherwise, sampled compositions are realized by the same frozen KLDM backbone used during training. In the backbone-transfer rows of \secref{sec:results} the policy checkpoint and the composition set it produces are held fixed and only the structure model is replaced, querying a different frozen CSP model through the same composition-to-structure interface. Neither component is fine-tuned. In the internal tables, these rows are scored by the same MatterSim-5M protocol as every other row. Under MatterGen and LeMat-GenBench they are scored by those pipelines.

\paragraph{Evaluation potential.} 
Structures are relaxed with MatterSim-5M under BFGS for at most $400$ steps, and $E_{\mathrm{hull}}$ is computed from the resulting energies against the Alex-MP hull (\appref{app:reward/stability}). 
We distinguish two roles for the machine-learned interatomic potential throughout the paper: the \emph{training potential} supplies the stability term of the reward during a run, whereas the \emph{evaluation potential} used here scores the final structures. The evaluation potential is fixed to MatterSim-5M for every policy, including policies trained against the smaller 1M variant. Without this, a policy trained against the weaker potential would be graded by a weaker judge than one trained against the stronger, and the base-model comparison in \appref{app:stage1} would confound model quality with grader leniency. The relaxation budget is larger than the budget used in training because stability is being reported here rather than used as a learning signal.
Main-text internal tables additionally relax the lattice. The internal appendix tables reported below hold it fixed (\appref{sec:app-cellrelax}). These choices define the internal tables. 
Structures submitted to MatterGen's evaluation pipeline or to LeMat-GenBench are scored by those pipelines (\appref{app:eval-mattergen}, \appref{app:eval-lemat}).

\paragraph{Metric definitions.} 
We report the following, all as percentages of the $N_{\mathrm{eval}}$ generated structures:
\begin{itemize}
  \item \textbf{Valid}: passes the gate of \appref{app:reward/validity}.
  \item \textbf{Unique}: distinct reduced formulas, $|\{f(a_i)\}|/N$. This is deliberately composition-level rather than structure-level and therefore does not depend on the choice of unit-cell representation. Structural novelty within a formula is handled separately by the continuous CAN score.
  \item \textbf{Novel}: valid and $\nu(x) > 0.5$. The reported rates of ANCHOR variants are insensitive to this threshold (\appref{app:stage1}).
    \item \textbf{Stable}: valid and $E_{\mathrm{hull}} \le 0$.
  \item \textbf{Meta}: valid and $E_{\mathrm{hull}} \le 0.1$~eV/atom.
  \item \textbf{SUN}: stable $\wedge$ unique $\wedge$ novel.
    \item \textbf{MSUN}: meta $\wedge$ unique $\wedge$ novel.
  \item \textbf{MP-20 hits}: number of structures whose reduced formula occurs in MP-20.
\end{itemize}
The validity gate and the novelty score are applied to the relaxed structure under both relaxation protocols.

\paragraph{Novelty at evaluation time.}
At evaluation, we switch off the adaptive part of CAN. The agent-history index is neither scored against nor updated, so novelty is the graded kNN-RBF score against the frozen MP-20 index alone. Keeping the history live would usually change little, because $1000$ samples from such a large action space rarely repeat. It matters for policies that collapse, however. On the ablation arm without any exploration mechanism, which repeats a handful of formulas, the same structures are $99.9\%$ novel against MP-20 alone but only $0.9\%$ novel once the search history is included. A growing reference is a training signal, so a metric computed against it would depend on draw order and would charge a second time for repetition that Unique already reports. Across the $5{,}737$ valid structures of the base-model selection grid, $5{,}731$ ($99.9\%$) received the bootstrap $\nu_0 = 0.75$ because their reduced formula is absent from MP-20; only $6$ were scored by the kNN--RBF machinery. Novel\% in our tables is therefore almost always a statement that the formula does not occur in MP-20. The graded score and the growing index are exercised during training (\appref{app:reward/novelty}) and isolated by the ablation (\appref{app:explore_ablation}); they are not what Novel\% at model inference reports.

\subsubsection{Shared rows and reading conventions}
\label{sec:app-eval-rows}

The internal, MatterGen, and LeMat-GenBench tables in this appendix score the \emph{same} set of rows with different evaluators, so any row can be read across pipelines.

\paragraph{Reading the rows.}
\textsc{ANCHOR} $\to$ \emph{X} is a single composition policy, trained against a frozen KLDM-CSP backbone and queried at inference through generator \emph{X}. The \emph{trained with LeMat validity} block instead changes the validity gate entering the reward during GRPO, with the same KLDM-CSP backbone, and is not a post-hoc swap. \emph{Trainable KLDM-CSP} lets GRPO update the backbone as well, under two reward variants: the shared \textsc{ANCHOR} reward and the stability term alone. \emph{DNG} models choose composition and structure themselves with no policy. FT marks DDPO reward fine-tuning of the generator, at either Dirichlet-mean objective weights (mean, novelty-heavy) or equal stability-novelty weights (bal.). \emph{FT} $\to$ \emph{CSP} rows query the frozen CSP model of the same family with the fine-tuned model's compositions, so they differ from the \emph{FT} row only in the structure generator.

\paragraph{Reading the numbers.}
All entries are percentages of $N$ unless stated otherwise. A \texttt{"-"} is a missing or non-final run, never a zero, and is never ranked. Best value in each column is \textbf{bold}, second-best \underline{underlined}.

\subsubsection{Internal results}
\label{app:eval-internal} 
Table~\ref{tab:app-internal} applies the protocol above to every row.
%
\begin{table}[t]
\centering
\footnotesize
\setlength{\tabcolsep}{2.6pt}
\caption{\latfix Internal evaluation pipeline (MatterSim-5M, BFGS-400). Valid = bonds + ICSD oxidation. Unique = distinct formulas / $N$. Novel = $\mathrm{novelty\_score}>0.5$. Stable / Meta: $E_{\mathrm{hull}}\le 0$ / $0.1$. The last column counts valid samples whose reduced formula occurs in the frozen MP-20 reference index (not ranked). Every \textsc{ANCHOR} row is $0$ in the MP-20 column, the DNG rows are not: the novelty reward moves the policy off the reference set, whereas a finetuned generator re-emits it. Table~\ref{tab:internal} gives the lattice-free values of the headline rows; main-text tables relax the lattice.}
\label{tab:app-internal}
\begin{tabular}{@{}lrrrrrrrrr@{}}
\toprule
Configuration & $N$ & Valid & Unique & Novel & Meta & Stable & MSUN & SUN & MP-20 \\
\midrule
\multicolumn{10}{@{}l}{\textbf{\textsc{ANCHOR}, inference-time backbone}} \\
\addlinespace[1pt]
\textsc{ANCHOR} $\to$ KLDM-CSP & 1000 & 97.9 & \textbf{100.0} & 97.9 & \textbf{38.5} & \textbf{18.1} & \textbf{38.5} & \textbf{18.1} & 0 \\
\textsc{ANCHOR} $\to$ Crystalite-CSP & 1000 & 94.1 & \textbf{100.0} & 94.1 & 37.0 & 13.7 & 37.0 & 13.7 & 0 \\
\textsc{ANCHOR} $\to$ OMatG-CSP & 998 & \textbf{98.2} & \textbf{100.0} & \textbf{98.2} & \underline{37.9} & \underline{14.9} & \underline{37.9} & \underline{14.9} & 0 \\
\addlinespace[3pt]
\multicolumn{10}{@{}l}{\textbf{\textsc{ANCHOR} trained with LeMat validity}} \\
\addlinespace[1pt]
KLDM backbone & 1000 & 5.6 & \textbf{100.0} & 5.6 & 0.7 & 0.0 & 0.7 & 0.0 & 1 \\
\addlinespace[3pt]
\multicolumn{10}{@{}l}{\textbf{Trainable KLDM-CSP}} \\
\addlinespace[1pt]
ANCHOR reward & 1000 & \underline{98.1} & \textbf{100.0} & \underline{98.1} & 15.0 & 4.2 & 15.0 & 4.2 & 0 \\
Stability only & 1000 & \underline{98.1} & \textbf{100.0} & \underline{98.1} & 19.4 & 7.4 & 19.4 & 7.4 & 0 \\
\addlinespace[3pt]
\multicolumn{10}{@{}l}{\textbf{DNG}} \\
\addlinespace[1pt]
Crystalite-DNG & 1000 & 49.0 & 98.7 & 33.9 & 22.5 & 1.5 & 11.4 & 0.6 & 163 \\
Crystalite-DNG $\to$ CSP & 1000 & 50.5 & 98.7 & 34.9 & 24.5 & 2.3 & 12.4 & 1.0 & 165 \\
Crystalite-DNG FT (mean) & 1000 & 96.4 & 40.1 & 32.3 & 37.2 & 0.9 & 8.0 & 0.4 & 617 \\
Crystalite-DNG FT (bal.) & 1000 & 96.4 & 50.0 & 37.4 & 36.9 & 0.6 & 10.3 & 0.2 & 594 \\
KLDM-DNG & 1000 & 44.7 & 98.9 & 34.5 & 11.4 & 0.9 & 5.6 & 0.5 & 124 \\
KLDM-DNG $\to$ CSP & 1000 & 43.7 & 98.9 & 32.6 & 12.4 & 0.9 & 5.1 & 0.4 & 123 \\
KLDM-DNG FT (mean) & 1000 & 53.9 & 99.4 & 45.0 & 11.1 & 0.3 & 6.3 & 0.1 & 107 \\
KLDM-DNG FT (bal.) & 1000 & 50.0 & 99.3 & 41.4 & 10.5 & 0.6 & 4.7 & 0.3 & 100 \\
\bottomrule
\end{tabular}
\end{table}  
The lattice-fixed evaluator preserves the same separation as the main-text cell-relaxed results. \textsc{ANCHOR}'s composition policy transfers across all three frozen CSP backbones with $37.0$-$38.5\%$ MSUN, whereas allowing the KLDM-CSP backbone to receive the reward lowers MSUN to $15.0\%$ under the shared reward and $19.4\%$ under stability alone. The DNG fine-tunes show the opposite failure mode: Crystalite validity rises to $96.4\%$, but formula uniqueness falls to $40.1$-$50.0\%$ and $594$-$617$ samples return to MP-20 formulas, leaving SUN at $0.2$-$0.4\%$. The conclusions therefore do not depend on lattice relaxation: freezing the structural prior preserves the part of the generator that already works, while reward adaptation of the joint generator can raise validity without improving discovery.

\subsubsection{Sensitivity to the validity gate}
\label{app:eval-lemat-gate}
Validity is the definition our protocol and the LeMat-GenBench leaderboard protocol disagree on most (\appref{app:eval-lemat}), and it gates every downstream count. Table~\ref{tab:app-lemat-gate} isolates that one choice: the same samples, relaxer, hull and novelty index as Table~\ref{tab:app-internal}, with the LeMat-GenBench validity gate substituted for ours. Differences between the two tables are attributable to the gate alone, not to the energy model or the reference hull.
%
\begin{table}[t]
\centering
\footnotesize
\setlength{\tabcolsep}{2.6pt}
\caption{\latfix Internal evaluation pipeline with the LeMat-GenBench validity gate (charge + distance + plausibility) substituted for our own. Relaxer, hull, novelty indices and column definitions are unchanged from Table~\ref{tab:app-internal}; the alternative gate changes which chemistries are admitted rather than acting as a uniformly looser filter. The \emph{trained with LeMat validity} block was trained against this gate, so its Valid and Novel entries are graded by their own training signal and are not comparable across blocks. Table~\ref{tab:internal} gives the lattice-free values of the headline rows. Tables in the main text relax the lattice.}
\label{tab:app-lemat-gate}
\begin{tabular}{@{}lrrrrrrrrr@{}}
\toprule
Configuration & $N$ & Valid & Unique & Novel & Meta & Stable & MSUN & SUN & MP-20 \\
\midrule
\multicolumn{10}{@{}l}{\textbf{\textsc{ANCHOR}, inference-time backbone}} \\
\addlinespace[1pt]
\textsc{ANCHOR} $\to$ KLDM-CSP & 1000 & 72.9 & \textbf{100.0} & 72.9 & 25.0 & \textbf{9.4} & 25.0 & \textbf{9.4} & 0 \\
\textsc{ANCHOR} $\to$ Crystalite-CSP & 1000 & 72.7 & \textbf{100.0} & 72.7 & 26.0 & \underline{8.4} & 26.0 & \underline{8.4} & 0 \\
\textsc{ANCHOR} $\to$ OMatG-CSP & 998 & 72.8 & \textbf{100.0} & 72.8 & 26.2 & 7.9 & \underline{26.2} & 7.9 & 0 \\
\addlinespace[3pt]
\multicolumn{10}{@{}l}{\textbf{\textsc{ANCHOR} trained with LeMat validity}} \\
\addlinespace[1pt]
KLDM backbone & 1000 & \underline{98.3} & \textbf{100.0} & \underline{97.5} & 9.2 & 0.0 & 8.7 & 0.0 & 10 \\
\addlinespace[3pt]
\multicolumn{10}{@{}l}{\textbf{Trainable KLDM-CSP}} \\
\addlinespace[1pt]
ANCHOR reward & 1000 & 86.5 & \textbf{100.0} & 86.5 & 17.7 & 4.8 & 17.7 & 4.8 & 0 \\
Stability only & 1000 & 81.5 & \textbf{100.0} & 81.5 & 16.1 & 4.6 & 16.1 & 4.6 & 0 \\
\addlinespace[3pt]
\multicolumn{10}{@{}l}{\textbf{DNG}} \\
\addlinespace[1pt]
Crystalite-DNG & 1000 & 93.2 & 98.7 & 64.4 & \underline{43.6} & 2.0 & 21.6 & 0.8 & 311 \\
Crystalite-DNG $\to$ CSP & 1000 & 94.4 & 98.7 & 64.8 & \textbf{47.3} & 3.0 & 23.6 & 1.2 & 311 \\
Crystalite-DNG FT (mean) & 1000 & 96.8 & 40.1 & 31.7 & 37.5 & 1.1 & 8.2 & 0.5 & 625 \\
Crystalite-DNG FT (bal.) & 1000 & 96.4 & 50.0 & 36.8 & 37.8 & 0.8 & 11.0 & 0.3 & 596 \\
KLDM-DNG & 1000 & 93.2 & 98.9 & 73.9 & 26.0 & 1.1 & 13.5 & 0.5 & 232 \\
KLDM-DNG $\to$ CSP & 1000 & 90.4 & 98.9 & 70.1 & 24.9 & 1.0 & 10.5 & 0.4 & 231 \\
KLDM-DNG FT (mean) & 1000 & 91.7 & 99.4 & 75.9 & 19.0 & 0.6 & 9.9 & 0.3 & 189 \\
KLDM-DNG FT (bal.) & 1000 & 91.0 & 99.3 & 75.7 & 19.3 & 1.0 & 9.2 & 0.7 & 177 \\
\bottomrule
\end{tabular}
\end{table}

This substitution is not a uniformly looser or stricter filter. Relative to Table~\ref{tab:app-internal}, \textsc{ANCHOR}$\to$KLDM-CSP validity falls from $97.9\%$ to $72.9\%$, while Crystalite-DNG rises from $49.0\%$ to $93.2\%$; the gate changes which chemistry is admitted rather than applying a common severity shift. The downstream metrics move accordingly: Crystalite-DNG$\to$CSP rises from $12.4\%$ to $23.6\%$ MSUN, while \textsc{ANCHOR}$\to$KLDM-CSP falls from $38.5\%$ to $25.0\%$. The policy trained directly against the LeMat gate reaches $98.3\%$ validity under that same gate but $0\%$ SUN under the unchanged MatterSim hull, separating optimisation of the validity rule from discovery of on-hull structures.

\subsubsection{Sensitivity to lattice relaxation}
\label{sec:app-cellrelax}
\label{app:eval-cellrelax}
The main text protocol relaxes the lattice together with the atomic positions, following the cell-relaxing convention of the MatterGen pipeline (\appref{app:eval-mattergen}). The training reward instead relaxes positions only, under MatterSim-1M and a $100$-step budget (\appref{app:reward/stability}). Tables~\ref{tab:app-internal} and~\ref{tab:app-lemat-gate} hold the lattice fixed and leave everything else unchanged (MatterSim-5M, BFGS, $400$ steps), so they are not the training reward, which also differs in potential and budget. Holding the lattice fixed raises $E_{\mathrm{hull}}$ and lowers every stability-derived column, because a lattice taken directly from the backbone leaves structures strained. We report the free-cell numbers as the headline for that reason, and because free-cell relaxation is not the signal the policies were optimised against. Tables~\ref{tab:internal} and~\ref{tab:app-internal} place the two protocols side by side.
\paragraph{Conclusions across protocols.}
Freeing the lattice raises every stability-derived number but leaves the conclusions of the main text unchanged. In the exploration ablation (Tables~\ref{tab:explore-ablation} and~\ref{tab:app-explore-fixed}), the full method has the highest SUN under both protocols ($18.1\%$ fixed, $22.1\%$ free) and the collapsed cells are last under both. The full method and adaptive novelty alone swap order on MSUN ($38.5\%$ against $37.1\%$ fixed, $47.6\%$ against $52.0\%$ free), while their SUN order holds under both. In Table~\ref{tab:internal} and its lattice-fixed counterpart Table~\ref{tab:app-internal}, the frozen backbone outperforms both reward-trained backbones and the stability-only reward outperforms the shared reward under both protocols ($38.5 > 19.4 > 15.0\%$ MSUN fixed, $47.6 > 29.5 > 22.6\%$ free), every reward fine-tune lowers the on-hull fraction of its DNG model, and replacing the composition source moves MSUN far more than replacing the structure generator. The one comparison that changes sign is the MSUN of the KLDM-DNG fine-tunes, slightly below or above the pretrained model with the lattice fixed ($4.7\%$ and $6.3\%$ against $5.6\%$) and slightly above it with the lattice free ($12.0\%$ and $11.4\%$ against $11.1\%$). The main text therefore bases its fine-tuning claims on the on-hull fraction, which falls under both.

\subsubsection{Relaxation budget}
\label{app:relax-budget}
The training relaxation is chosen for throughput, since the reward relaxes $G$ structures per policy step and relaxation is essentially its whole cost. We measure the trade-off on the four structure sets relaxed under both protocols: $2{,}820$ paired structures, identical CIFs, MatterSim-5M. The evaluator records the BFGS step count and wall-clock time of every structure, so the $100$-step budget can be read exactly from the $400$-step runs. A relaxation that converged within $100$ steps under the $400$-step cap would also have converged under a $100$-step cap, and its time is rescaled as the per-step cost times the truncated step count. These rates are measured under the evaluation potential rather than MatterSim-1M, so they indicate the relative cost of the two protocols rather than the exact convergence of the training runs.

\begin{table}[h]
\centering
\caption{Relaxation convergence and cost at the training ($100$-step) and evaluation ($400$-step) budgets, on $2{,}820$ structures relaxed under both protocols with MatterSim-5M. $100$-step values are read from the $400$-step runs (see text). Training uses positions only at $100$ steps; the main-text evaluation frees the lattice at $400$ steps. ``--'': timing unavailable for that set.}
\label{tab:relax-budget}
\small
\begin{tabular}{llrrrr}
\toprule
 & & \multicolumn{2}{c}{100-step budget} & \multicolumn{2}{c}{400-step budget} \\
\cmidrule(lr){3-4}\cmidrule(lr){5-6}
Structures & Relaxation & Conv.\ (\%) & s/struct & Conv.\ (\%) & s/struct \\
\midrule
\textsc{ANCHOR}$\to$KLDM-CSP       & positions & 16.4 & --   &  98.8 & --    \\
                                   & cell free &  6.5 & 7.18 &  99.4 & 14.00 \\
\textsc{ANCHOR}$\to$Crystalite-CSP & positions & 27.7 & 5.47 &  99.5 &  8.00 \\
                                   & cell free &  9.4 & 7.08 &  99.5 & 12.86 \\
Crystalite-DNG                     & positions & 90.2 & 2.10 & 100.0 &  2.28 \\
                                   & cell free & 82.5 & 3.63 & 100.0 &  4.27 \\
KLDM-DNG                           & positions & 68.4 & 3.58 &  99.8 &  4.55 \\
                                   & cell free & 57.0 & 5.18 &  99.8 &  7.25 \\
\bottomrule
\end{tabular}
\end{table}

Pooled over the four sets, freeing the lattice costs $1.28\times$ the force calls at the $400$-step budget and $1.6$--$1.9\times$ the wall clock, more than the step ratio because each cell-free step also updates the nine lattice degrees of freedom and evaluates the stress. At the $100$-step training budget, freeing the lattice lowers convergence from $16.4\%$ to $6.5\%$ on \textsc{ANCHOR}$\to$KLDM structures and from $27.7\%$ to $9.4\%$ on \textsc{ANCHOR}$\to$Crystalite. Restoring convergence requires the $400$-step budget, at which $99.4$--$99.5\%$ of cell-free \textsc{ANCHOR} relaxations converge. On \textsc{ANCHOR}$\to$Crystalite structures, where both timings are available, that costs $2.4\times$ the per-structure wall clock of the positions-only relaxation used in training ($12.86$\,s against $5.47$\,s). Because the multiplier applies to the dominant cost of every policy step, a cell-free reward at matched convergence would have reduced the number of structures generated within the seven-day wall (Table~\ref{tab:run-budget}) by roughly the same factor.

The positions-only reward is itself truncated: at $100$ steps, only $16.4\%$ of \textsc{ANCHOR}$\to$KLDM relaxations converge. \textsc{ANCHOR} structures converge far less often than DNG structures under either protocol ($68.4$--$90.2\%$ positions-only at $100$ steps for DNG), consistent with their larger relaxation displacement in Table~\ref{tab:app-mattergen} (RMSD $1.27$-$1.40$\,\AA\ against $0.31$--$0.56$\,\AA). The stability term of the training reward is therefore an energy along a truncated relaxation, used to rank candidates within a group. The reported numbers use the evaluation relaxation, at which at least $98.8\%$ of structures converge.

\subsection{MatterGen Evaluation Pipeline}
\label{app:eval-mattergen}
\label{app:mattergen-validity}

\begin{table}[t]
\centering
\footnotesize
\setlength{\tabcolsep}{2.6pt}
\caption{MatterGen evaluation pipeline (unmodified), which supplies its own relaxer, reference hull and novelty definition. MSUN is unique $\wedge$ novel $\wedge$ $E_{\mathrm{hull}}\le 0.1$, which the pipeline reports as S.U.N.; SUN uses $E_{\mathrm{hull}}\le 0$. Meta is the stability rate at $E_{\mathrm{hull}}\le 0.1$; Stable is the raw stability rate at $E_{\mathrm{hull}}\le 0$, unfiltered by uniqueness or novelty. $\langle E_{\mathrm{hull}}\rangle$ (eV/atom) and RMSD (\AA, against the pre-relaxation cell): lower is better. $N$ is the number of structures the pipeline scored, below $1000$ where it dropped structures of its own accord. We cannot report RMSD for Crystalite-DNG $\to$ CSP for its particular structures: the official pipeline fails while computing it. The other columns on that row use the same relaxed cells. Rows are defined in \appref{app:eval_protocol}. The pipeline relaxes positions and lattice (FIRE with \texttt{ExpCellFilter}). Entries are from sampling seed~0, and Table~\ref{tab:mattergen_dng} reports the three-seed means.}
\label{tab:app-mattergen}
\begin{tabular}{@{}lrrrrrrrrr@{}}
\toprule
Configuration & $N$ & Unique & Novel & Meta & Stable & MSUN & SUN & $\langle E_{\mathrm{h}}\rangle$ & RMSD \\
\midrule
\multicolumn{10}{@{}l}{\textbf{\textsc{ANCHOR}, inference-time backbone}} \\
\addlinespace[1pt]
\textsc{ANCHOR} $\to$ KLDM-CSP & 993 & \textbf{100.0} & \textbf{100.0} & 42.3 & 17.1 & 42.3 & \underline{17.1} & \textbf{0.133} & 1.40 \\
\textsc{ANCHOR} $\to$ Crystalite-CSP & 993 & \textbf{100.0} & \textbf{100.0} & 42.8 & 14.5 & \underline{42.8} & 14.5 & 0.138 & 1.27 \\
\textsc{ANCHOR} $\to$ OMatG-CSP & 991 & \textbf{100.0} & \textbf{100.0} & 43.2 & 17.3 & \textbf{43.2} & \textbf{17.3} & \underline{0.134} & 1.28 \\
\addlinespace[3pt]
\multicolumn{10}{@{}l}{\textbf{\textsc{ANCHOR} trained with LeMat validity}} \\
\addlinespace[1pt]
KLDM backbone & 998 & \textbf{100.0} & 97.2 & 22.0 & 0.3 & 19.4 & 0.0 & 0.189 & 0.60 \\
\addlinespace[3pt]
\multicolumn{10}{@{}l}{\textbf{Trainable KLDM-CSP}} \\
\addlinespace[1pt]
ANCHOR reward & 997 & \textbf{100.0} & \textbf{100.0} & 18.3 & 5.0 & 18.3 & 5.0 & 0.202 & 1.37 \\
Stability only & 995 & \textbf{100.0} & \underline{99.9} & 23.8 & 8.6 & 23.7 & 8.6 & 0.181 & 1.27 \\
\addlinespace[3pt]
\multicolumn{10}{@{}l}{\textbf{DNG}} \\
\addlinespace[1pt]
Crystalite-DNG & 922 & 99.1 & 63.7 & \underline{54.8} & 7.2 & 22.9 & 2.3 & 0.195 & \underline{0.31} \\
Crystalite-DNG $\to$ CSP & 922 & 99.1 & 62.8 & \textbf{57.3} & 8.0 & 23.6 & 2.8 & 0.187 & {--} \\
Crystalite-DNG FT (mean) & 989 & 60.8 & 64.5 & 52.5 & 4.8 & 17.2 & 1.9 & 0.158 & 0.45 \\
Crystalite-DNG FT (bal.) & 989 & 67.9 & 63.9 & 51.6 & 4.7 & 19.0 & 1.9 & 0.184 & 0.49 \\
KLDM-DNG & 943 & 99.8 & 75.9 & 39.8 & 6.3 & 18.6 & 2.7 & 0.213 & 0.56 \\
KLDM-DNG $\to$ CSP & 943 & 99.8 & 73.7 & 41.0 & 6.5 & 17.8 & 2.5 & 0.201 & 0.59 \\
KLDM-DNG FT (mean) & 943 & 99.8 & 78.5 & 33.6 & 3.4 & 15.2 & 1.2 & 0.239 & 0.70 \\
KLDM-DNG FT (bal.) & 948 & 99.8 & 80.0 & 32.0 & 4.2 & 15.4 & 2.1 & 0.223 & 0.69 \\
\bottomrule
\end{tabular}
\end{table}  
MatterGen gives the same qualitative decomposition under a different evaluator. The Crystalite-DNG rows have higher raw metastability than \textsc{ANCHOR} ($54.8$--$57.3\%$ Meta versus $42.3$--$43.2\%$), but only $22.9$--$23.6\%$ survives the unique-and-novel conjunction. For \textsc{ANCHOR}, Unique and Novel are both $100\%$, so MSUN is numerically the raw metastable rate: its advantage comes from allocating compositions to regions where the frozen CSP succeeds, not from a higher raw metastable fraction. Reward fine-tuning the DNGs again fails to convert adaptation into discovery, lowering both Stable and MSUN, while replacing only DNG structure generation with the corresponding frozen CSP changes MSUN by at most $0.8$ points.  \begin{table}[t]
\centering
\footnotesize
\setlength{\tabcolsep}{3pt}
\caption{Representative published DNG results on MP-20 under MatterGen's evaluation pipeline, rounded to one decimal, with the frozen-backbone \textsc{ANCHOR} rows of Table~\ref{tab:mattergen_dng} for reference (mean$_{\pm\mathrm{std}}$ over three sampling seeds). MSUN is the quantity MatterGen and the cited works report as SUN. Table~\ref{tab:mattergen_dng} keeps the strongest variant of each model family. Best in each column is bold, second-best underlined. \textsuperscript{a}\cite{cornet2025kinetic}, \textsuperscript{b}\cite{hollmer2025open}, \textsuperscript{c}\cite{veljkovic2026crystalite}, \textsuperscript{d}\cite{zhang2026crystalrepa}.}
\label{tab:app-mattergen-baselines}
\begin{tabular*}{\linewidth}{@{\extracolsep{\fill}}lcccccc@{}}
\toprule
Model & Unique & Novel & Meta & $\overline{E}_h$ & RMSD & MSUN \\
\midrule
\multicolumn{7}{@{}l}{\textbf{Published DNG baselines}} \\[2pt]
MatterGen-MP\textsuperscript{a}       & --   & --   & 47.1 & 0.20 & \underline{0.15} & 25.8 \\
DiffCSP\textsuperscript{a}            & --   & --   & 41.3 & 0.19 & 0.41 & 20.1 \\
KLDM-\(x_0\)(C)\textsuperscript{a}    & --   & --   & 38.6 & 0.27 & 0.37 & 16.7 \\
KLDM-\(x_0\)(C-AB)\textsuperscript{a} & --   & --   & 49.8 & 0.19 & 0.30 & 17.9 \\
KLDM-\(x_0\)(D)\textsuperscript{a}    & --   & --   & 59.2 & 0.16 & 0.28 & 18.5 \\
FlowMM\textsuperscript{b}             & --   & --   & 37.5 & 0.25 & 0.65 & 13.9 \\
OMatG-SBD\textsuperscript{b}          & --   & --   & --   & --   & --   & 22.1 \\
ADiT\textsuperscript{c}               & 89.6 & 43.2 & \textbf{70.0} & 0.15 & 0.49 & 17.2 \\
Crystalite\textsuperscript{c}         & 94.7 & 56.6 & 64.5 & 0.15 & 0.27 & 24.3 \\
CrystalFlow\textsuperscript{d}        & 99.6 & \underline{73.1} & 41.2 & 0.20 & 0.60 & 17.2 \\
CrystalFlow+REPA\textsuperscript{d}   & \underline{99.9} & 69.0 & 55.0 & 0.17 & 0.42 & 20.0 \\
DiffCSP+REPA\textsuperscript{d}       & 99.5 & 72.0 & 56.5 & 0.15 & 0.32 & 24.8 \\
MatterGen+REPA\textsuperscript{d}     & 99.7 & 63.5 & \underline{66.2} & \textbf{0.12} & \textbf{0.11} & 29.2 \\
\midrule
\multicolumn{7}{@{}l}{\textbf{\textsc{ANCHOR} + frozen CSP (ours)}} \\[2pt]
Trained on KLDM            & \ms{\textbf{100.0}}{0.0} & \ms{\textbf{100.0}}{0.0} & \ms{41.3}{1.0} & \ms{\underline{0.14}}{0.01} & \ms{1.39}{0.01} & \ms{\underline{41.3}}{1.0} \\
\quad$\to$ Crystalite      & \ms{\textbf{100.0}}{0.0} & \ms{\textbf{100.0}}{0.0} & \ms{41.1}{2.1} & \ms{\underline{0.14}}{0.01} & \ms{1.27}{0.02} & \ms{41.1}{2.1} \\
\quad$\to$ OMatG           & \ms{\textbf{100.0}}{0.0} & \ms{\textbf{100.0}}{0.0} & \ms{43.1}{0.9} & \ms{\underline{0.14}}{0.01} & \ms{1.27}{0.01} & \ms{\textbf{43.1}}{0.9} \\
\bottomrule
\end{tabular*}
\end{table}

\paragraph{What the pipeline measures.}
MatterGen treats validity as two independent quantities rather than one~\citep{zeni2025generative}. \emph{Composition validity} is a SMACT test~\citep{davies2019smact}: the reduced formula must admit a charge-neutral assignment of SMACT oxidation states that also respects the Pauling electronegativity ordering, with unary cells and all-metal alloys exempt from that check. \emph{Structure validity} is far weaker, requiring only a minimum pair distance of $0.5$\,\AA\ and a cell volume above $0.1$\,\AA$^3$. Neither quantity coincides with LeMat Valid (\appref{app:eval-lemat}) or with our own gate (\appref{app:reward/validity}).

\paragraph{Its columns are not nested.}
Unique, Novel and S.U.N.\ (MSUN in our tables) do not filter through either validity column. Unique and Novel are \texttt{StructureMatcher} flags~\citep{ong2013python} evaluated on every scored cell, and S.U.N.\ is unique $\wedge$ novel $\wedge$ $E_{\mathrm{hull}}\le 0.1$~eV/atom. A model can therefore report Unique and Novel near $100\%$ alongside a much lower composition validity, an ordering our own columns do not admit. Table~\ref{tab:app-mattergen} uses the pipeline's definitions throughout, in its default configuration: MatterSim-v1-1M, FIRE with \texttt{ExpCellFilter}, \texttt{DefaultDisorderedStructureMatcher}, and the Alex-MP \texttt{reference\_MP2020correction} hull. Table~\ref{tab:app-mattergen-baselines} lists every published DNG result on this pipeline. The main text keeps the strongest variant of each model family.

\paragraph{Element coverage.}
Our $84$-element vocabulary is broader than the element coverage of that reference hull, and a cell cannot be placed on the hull unless a terminal entry exists for every element it contains. A single unsupported element voids the energy columns for the entire submitted batch, so such cells are removed beforehand, screened against the same reference object the pipeline is then given. This is why $N$ falls below $1000$ in Table~\ref{tab:app-mattergen}; the surviving cells are scored exactly as the pipeline defines them. The same discard precedes published results on this pipeline~\citep{cornet2025kinetic}.

\subsection{LeMat-GenBench Evaluation}
\label{app:eval-lemat}

\begin{table}[t]
\centering
\footnotesize
\setlength{\tabcolsep}{2.6pt}
\caption{LeMat-GenBench evaluation against the Bulk ensemble hull, NequIP pre-relaxation unless noted. All rates are fractions of the submitted $N$, with validity, uniqueness and novelty applied in that order. $\langle E_{\mathrm{hull}}\rangle$ (eV/atom): lower is better. The \emph{trained with LeMat validity} block optimised a LeMat validity filter during training, so its Valid, Unique and Novel entries are not comparable across blocks. Rows are defined in \appref{app:eval_protocol}. }
\label{tab:app-lemat-bulk}
\begin{tabular}{@{}lrrrrrrrrr@{}}
\toprule
Configuration & $N$ & Valid & Unique & Novel & Meta & Stable & $\langle E_{\mathrm{h}}\rangle$ & MSUN & SUN \\
\midrule
\multicolumn{10}{@{}l}{\textbf{\textsc{ANCHOR}, inference-time backbone}} \\
\addlinespace[1pt]
\textsc{ANCHOR} $\to$ KLDM-CSP & 1000 & 73.0 & 73.0 & 73.0 & 2.9 & 0.0 & 0.20 & 2.9 & 0.0 \\
\textsc{ANCHOR} $\to$ Crystalite-CSP & 1000 & 73.0 & 73.0 & 73.0 & 3.1 & 0.0 & 0.20 & 3.1 & 0.0 \\
\textsc{ANCHOR} $\to$ OMatG-CSP & 998 & 72.9 & 72.9 & 72.9 & 3.1 & 0.0 & 0.20 & 3.1 & 0.0 \\
\addlinespace[3pt]
\multicolumn{10}{@{}l}{\textbf{\textsc{ANCHOR} trained with LeMat validity}} \\
\addlinespace[1pt]
KLDM backbone & 1000 & \underline{99.4} & \underline{99.4} & \underline{96.2} & 28.1 & 0.1 & 0.17 & \underline{26.0} & 0.0 \\
\addlinespace[3pt]
\multicolumn{10}{@{}l}{\textbf{Trainable KLDM-CSP}} \\
\addlinespace[1pt]
ANCHOR reward & 1000 & 86.5 & 86.5 & 86.5 & 6.9 & 0.0 & 0.19 & 6.9 & 0.0 \\
Stability only & 1000 & 81.5 & 81.5 & 81.4 & 5.3 & 0.0 & 0.20 & 5.2 & 0.0 \\
\addlinespace[3pt]
\multicolumn{10}{@{}l}{\textbf{DNG}} \\
\addlinespace[1pt]
Crystalite-DNG & 1000 & 95.6 & 94.9 & 57.5 & \underline{45.4} & 8.9 & \textbf{0.11} & 19.6 & 1.4 \\
Crystalite-DNG $\to$ CSP & 1000 & 95.6 & 95.0 & 55.0 & \textbf{48.4} & 9.7 & \textbf{0.11} & 20.4 & \underline{1.5} \\
Crystalite-DNG FT (mean) & 1000 & 97.8 & 59.3 & 53.8 & 45.3 & \underline{10.5} & 0.16 & 15.8 & 1.2 \\
Crystalite-DNG FT (bal.) & 1000 & 97.0 & 66.4 & 57.1 & 41.9 & \textbf{14.3} & 0.17 & 17.5 & \textbf{2.3} \\
KLDM-DNG & 999 & 95.4 & 95.3 & 68.1 & 35.0 & 5.7 & 0.16 & 16.8 & 0.6 \\
KLDM-DNG $\to$ CSP & 999 & 95.4 & 95.1 & 67.1 & 35.1 & 6.9 & \underline{0.15} & 17.4 & 0.5 \\
KLDM-DNG FT (mean) & 997 & 93.6 & 93.5 & 69.6 & 28.7 & 4.4 & 0.18 & 13.2 & 0.6 \\
KLDM-DNG FT (bal.) & 999 & 92.8 & 92.7 & 69.8 & 27.5 & 3.8 & 0.18 & 13.4 & 0.2 \\
\bottomrule
\end{tabular}
\end{table}

\paragraph{Bulk-hull results.}
Against the Bulk hull, all three frozen-backbone \textsc{ANCHOR} variants retain roughly $73\%$ Valid, Unique, and Novel samples but yield $0\%$ SUN and only $2.9$--$3.1\%$ MSUN. The Crystalite-DNG variants retain $1.4$--$2.3\%$ SUN and $15.8$--$20.4\%$ MSUN. Training against the LeMat validity gate raises MSUN to $26.0\%$ while still producing $0\%$ SUN, so a high metastable count under this evaluator need not imply any on-hull yield. Because the Bulk hull is not the reference used by the published Table~11 leaderboard, we treat this table as evaluator sensitivity and use the MP-20 hull below for direct comparison.

\begin{table}[t]
\centering
\footnotesize
\setlength{\tabcolsep}{2.6pt}
\caption{LeMat-GenBench evaluation against the MP-20 ensemble hull, following the published Table~11 protocol so the rows can be read against that leaderboard. Columns as in Table~\ref{tab:app-lemat-bulk}; only the reference hull changes. The same caveat applies to the \emph{trained with LeMat validity} block.}
\label{tab:app-lemat-mp20}
\begin{tabular}{@{}lrrrrrrrrr@{}}
\toprule
Configuration & $N$ & Valid & Unique & Novel & Meta & Stable & $\langle E_{\mathrm{h}}\rangle$ & MSUN & SUN \\
\midrule
\multicolumn{10}{@{}l}{\textbf{\textsc{ANCHOR}, inference-time backbone}} \\
\addlinespace[1pt]
\textsc{ANCHOR} $\to$ KLDM-CSP & 1000 & 73.0 & 73.0 & 73.0 & 8.2 & 0.4 & 0.170 & 7.8 & 0.4 \\
\textsc{ANCHOR} $\to$ Crystalite-CSP & 1000 & 73.0 & 73.0 & 73.0 & 8.6 & 0.8 & 0.172 & 7.8 & 0.8 \\
\textsc{ANCHOR} $\to$ OMatG-CSP & 998 & 72.9 & 72.9 & 72.9 & 9.1 & 0.6 & 0.170 & 8.5 & 0.6 \\
\addlinespace[3pt]
\multicolumn{10}{@{}l}{\textbf{\textsc{ANCHOR} trained with LeMat validity}} \\
\addlinespace[1pt]
KLDM backbone & 1000 & 99.4 & \underline{99.4} & \underline{98.9} & 34.0 & 1.1 & 0.154 & \underline{32.5} & 1.0 \\
\addlinespace[3pt]
\multicolumn{10}{@{}l}{\textbf{Trainable KLDM-CSP}} \\
\addlinespace[1pt]
ANCHOR reward & 1000 & 86.5 & 86.5 & 86.5 & 12.9 & 0.4 & 0.169 & 12.5 & 0.4 \\
Stability only & 1000 & 81.5 & 81.5 & 81.5 & 11.4 & 0.5 & 0.172 & 10.9 & 0.5 \\
\addlinespace[3pt]
\multicolumn{10}{@{}l}{\textbf{DNG}} \\
\addlinespace[1pt]
Crystalite-DNG & 1000 & 95.6 & 94.9 & 73.8 & \underline{61.3} & \underline{19.8} & \underline{0.083} & 30.0 & \underline{9.8} \\
Crystalite-DNG $\to$ CSP & 1000 & 95.6 & 95.0 & 70.7 & \textbf{64.4} & \textbf{21.6} & \textbf{0.076} & 29.0 & \textbf{10.9} \\
Crystalite-DNG FT (mean) & 1000 & 97.8 & 59.3 & 67.0 & 57.6 & 13.9 & 0.150 & 19.1 & 2.7 \\
Crystalite-DNG FT (bal.) & 1000 & 97.0 & 66.4 & 65.6 & 58.6 & 18.7 & 0.160 & 20.8 & 3.9 \\
KLDM-DNG & 999 & 95.4 & 95.3 & 80.3 & 45.8 & 11.7 & 0.133 & 26.2 & 4.5 \\
KLDM-DNG $\to$ CSP & 999 & 95.4 & 95.1 & 79.3 & 48.2 & 13.9 & 0.120 & 26.0 & 6.2 \\
KLDM-DNG FT (mean) & 997 & 93.6 & 93.5 & 83.0 & 38.3 & 9.9 & 0.153 & 23.0 & 5.1 \\
KLDM-DNG FT (bal.) & 999 & 92.8 & 92.7 & 82.5 & 37.2 & 9.0 & 0.157 & 23.0 & 4.1 \\
\bottomrule
\end{tabular}
\end{table}

\paragraph{MP-20-hull results.}
Changing only the reference hull leaves Valid, Unique, and Novel unchanged but changes the stability picture substantially. \textsc{ANCHOR}$\to$KLDM-CSP moves from $2.9\%$ to $7.8\%$ MSUN and from $0\%$ to $0.4\%$ SUN, while Crystalite-DNG moves from $19.6\%$ to $30.0\%$ MSUN and from $1.4\%$ to $9.8\%$ SUN. The reference hull alone therefore accounts for a large shift in absolute stability rates within the same LeMat pipeline. Under the published MP-20 protocol, the DNG rows remain much stronger on-hull than \textsc{ANCHOR} trained under our internal evaluator ($9.8$--$10.9\%$ versus $0.4$--$0.8\%$ SUN), while the LeMat-validity-trained policy reaches $32.5\%$ MSUN but only $1.0\%$ SUN. This is why we keep MSUN and on-hull SUN separate when comparing evaluators.
\paragraph{What the pipeline measures.}
Structures submitted to LeMat-GenBench~\citep{betala2025lemat} are scored entirely by that pipeline, which defines every column differently from our internal tables. A structure is valid only if it passes three independent checks, and only valid structures are eligible for the downstream counts. Charge neutrality is a cascade rather than a single SMACT screen: compositions with a high SMACT metallicity score are exempted and pass, and the rest are tried against bond-valence sums, then \texttt{BVAnalyzer} decoration, then a frequency-weighted guess using LeMat's ICSD oxidation-state mapping, and finally all tabulated states. This cascade is the dominant filter, accounting for essentially all of the structures the benchmark rejects in our submissions. The minimum-interatomic-distance check sets its threshold per atom pair from the two atomic radii rather than as one global cutoff. The plausibility check screens bulk and cell quantities --- density, atoms per unit volume, lattice lengths and angles, and space-group assignment --- rather than local coordination environments.

\paragraph{Where it diverges from our protocol.}
The pipeline deduplicates by structure rather than by reduced formula, so two polymorphs of one formula count twice there and once for us. It computes $E_{\mathrm{hull}}$ as the mean of a three-potential ensemble (ORB-v3, MACE-MP, UMA) against its own reference hull, where ours comes from a single MatterSim-5M relaxation, and its novelty is a structure-level match against that same reference corpus, where ours during training is the graded CAN score of \appref{app:reward/novelty}, computed per reduced formula against the frozen MP-20 index and against the history the training run itself accumulates. Two consequences matter for Valid\%. Our gate accepts any pymatgen ICSD assignment whereas theirs can reject compositions for which \texttt{oxi\_state\_guesses} still returns a solution, so the two columns measure different charge searches, not ``neutrality versus no neutrality,'' and should not be equated; Table~\ref{tab:app-lemat-gate} isolates that substitution inside our own pipeline. And the metallicity exemption leaves the charge check silent on compositions the heuristic treats as metallic, a potentially hackable trait under RL that could bias training toward the exempt set rather than toward charge-neutral formulas.

\paragraph{Reference hull.}
The benchmark admits two reference hulls and they are not interchangeable, so both are reported: the LeMat-Bulk hull~\citep{siron2025lemat} in Table~\ref{tab:app-lemat-bulk} and the MP-20 hull in Table~\ref{tab:app-lemat-mp20}. 

\begin{figure}[t!]
    \centering
    \includegraphics[width=0.99\linewidth]{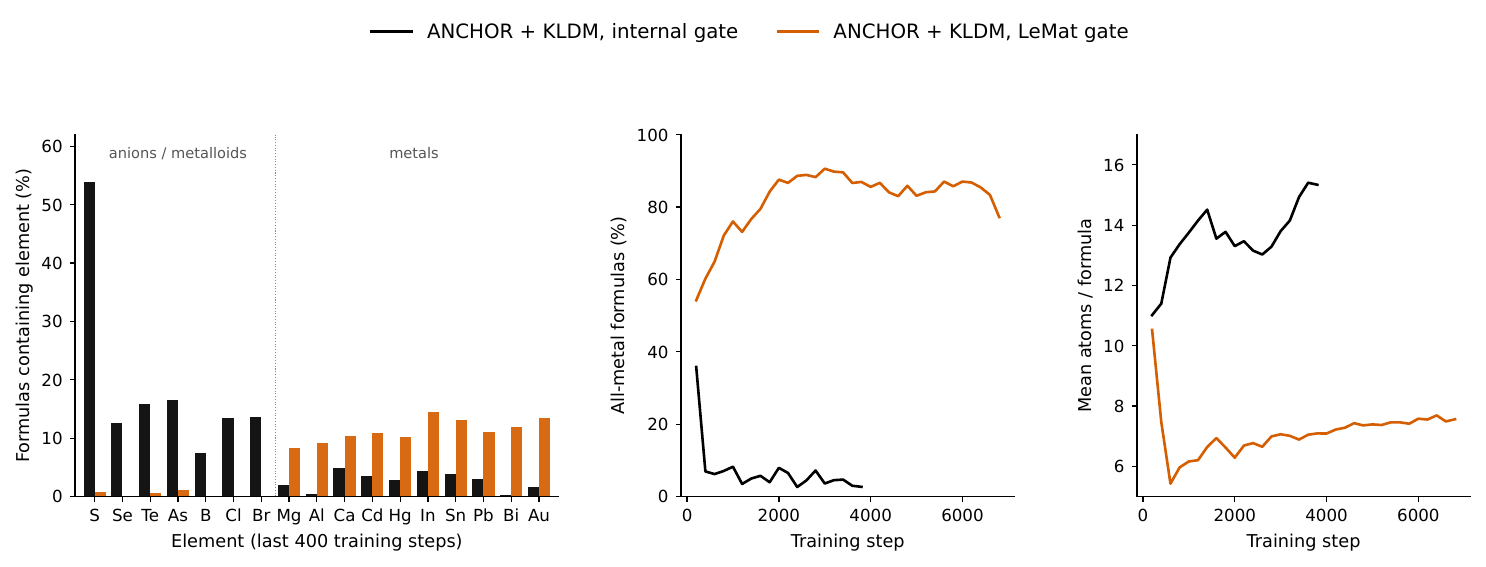}
    \caption{\textbf{Training gate sensitivity.}
Policies trained with KLDM-CSP against our validity gate and against the LeMat-GenBench gate. Left: elements appearing in formulas over the last 400 training steps. Middle: fraction of all-metal compositions during training. Right: mean atoms per formula.}
    \label{fig:element_sampling}
\end{figure}

\subsubsection{Training against the LeMat validity gate}
\label{app:gates}
\begin{table*}[t]
\centering
\scriptsize
\renewcommand{\arraystretch}{0.95}
\setlength{\tabcolsep}{4pt}
\caption{\textbf{The gate and the threshold decide the ranking.} Identical structures scored three
ways. \emph{Internal} and \emph{LeMat gate} differ only in the validity definition; relaxer (MatterSim-5M, BFGS-400, lattice free), hull and novelty index are held fixed (Table~\ref{tab:internal}).
\emph{LeMat-GenBench} is that benchmark's own pipeline against its MP-20 ensemble hull, so its
columns follow the published leaderboard. SUN is on-hull throughout, MSUN is
$E_{\mathrm{hull}}\le0.1$. Single sampling seed (seed~0), so the Internal-gate columns differ slightly from the three-seed means of Table~\ref{tab:internal}. Policies in the second block were trained against the LeMat validity
filter.}
\label{tab:gates}
\begin{tabular*}{\textwidth}{@{\extracolsep{\fill}}lccc ccc ccc@{}}
\toprule
& \multicolumn{3}{c}{Internal gate} & \multicolumn{3}{c}{LeMat gate, our pipeline} & \multicolumn{3}{c}{LeMat-GenBench} \\
\cmidrule(lr){2-4}\cmidrule(lr){5-7}\cmidrule(lr){8-10}
Configuration & Valid & SUN & MSUN & Valid & SUN & MSUN & Valid & SUN & MSUN \\
\midrule
\multicolumn{10}{@{}l}{\textbf{ANCHOR trained against the internal validity gate}} \\
\textsc{ANCHOR} $\rightarrow$ KLDM-CSP & 97.9 & 23.2 & 49.9 & 72.9 & 13.6 & 34.1 & 73.0 & 0.4 & 7.8 \\
\midrule
\multicolumn{10}{@{}l}{\textbf{ANCHOR trained against the LeMat validity gate}} \\
KLDM backbone              & 5.6 & 0.0 & 2.1 & 98.3 & 0.1 & 28.1 & 99.4 & 1.0 & 32.5 \\
\midrule
\multicolumn{10}{@{}l}{\textbf{DNG reference}} \\
Crystalite-DNG             & 49.0 & 1.4 & 14.8 & 93.2 & 1.7 & 27.1 & 95.6 & 9.8 & 30.0 \\
\bottomrule
\end{tabular*}
\end{table*}

The validity gate in \textsc{ANCHOR}'s reward determines what the policy learns. Table~\ref{tab:gates} trains the same architecture, with the same KLDM-CSP backbone, against the LeMat-GenBench validity gate and scores the result in three ways. Using the same gate during training and evaluation gives an essentially perfect policy, at $98.3\%$ validity. Under our validity gate, however, only $5.6\%$ of the same structures survive. The mechanism is documented in the benchmark: its charge cascade exempts compositions classified as metallic by the SMACT metallicity heuristic~\citep{betala2025lemat}, which assigns every atom an oxidation state of zero rather than testing neutrality. The LeMat-trained policy is already about $54\%$ all-metal within its first $200$ steps and saturates near $85\%$ by step $2000$, ending at $80.0\%$ over the last $400$ steps against $3.1\%$ for the policy trained with our gate (Figure~\ref{fig:element_sampling}). Anions largely disappear. Sulfur, present in $53.9\%$ of formulas under our gate, appears in $0.8\%$ under the LeMat gate, and the mean cell shrinks from $15.2$ to $7.6$ atoms per formula.

The stability threshold is exploitable in the same way. The LeMat-trained policy reaches $28.1\%$ MSUN under the LeMat gate in our pipeline and $32.5\%$ on LeMat-GenBench itself, above every DNG model, while producing almost no on-hull structures under either of our evaluators. Metallic alloys near the hull therefore satisfy every criterion the benchmark applies except on-hull stability, which is why we report SUN at $E_{\mathrm{hull}}\le0$ as the headline and treat MSUN as secondary.

The dependence also cuts against \textsc{ANCHOR}. On LeMat-GenBench, Crystalite-DNG reaches $9.8\%$ SUN and $30.0\%$ MSUN against the MP-20 hull, while \textsc{ANCHOR}, trained against a different gate, relaxer and hull in our pipeline, reaches $0.4\%$ and $7.8\%$ at $73.0\%$ validity. Training against the LeMat gate recovers metastability and beats every DNG model, but still loses on on-hull SUN. Which model leads therefore depends on the gate and threshold a benchmark uses, including our own. Full results against both LeMat-GenBench reference hulls are in Tables~\ref{tab:app-lemat-bulk} and~\ref{tab:app-lemat-mp20}.

\newpage
\section{Additional Experiments and Baselines}
\label{app:additional-experiments}

\subsection{Base-model Selection}
\label{app:stage1}

We sweep group size  $G \in \{16, 32, 64\}$ against MatterSim calculators $\{1\mathrm{M},5\mathrm{M}\}$ under a fixed model setup including most significantly the exploration mechanisms; CAN, element sub-grouping and count bonus. Each job ran until a 7-day wall-clock limit with a structure-keyed entropy schedule without decay.

\paragraph{Reported Metrics.}
We report validity and uniqueness as the fraction of all generated samples whose formula is unique within the generated set i.e. not conditioned on validity. 
%
%
We additionally report SUN and MSUN rates (SUN = Valid $\wedge$ Unique $\wedge$ Novel $\wedge$ ($E_{\mathrm{hull}} \le 0$); MSUN = Valid $\wedge$ Unique $\wedge$ Novel $\wedge$ ($E_{\mathrm{hull}} \le 0.1$ eV/atom)).
Novel uses our continuous score with a threshold of $\nu(x) > 0.5$, the midpoint of its $[0,1]$ range. The reported rates are insensitive to where this cut is placed. A candidate whose reduced formula is absent from MP-20 receives the bootstrap value $\nu_0 = 0.75$, and $5{,}731$ of the $5{,}737$ valid structures in this grid did so; only $6$ were scored by the kNN--RBF comparison at all. Any threshold below $\nu_0$ therefore moves Novel\% by at most those $6$ structures ($0.1\%$), and the choice of $0.5$ over any other value in that range does not affect a single number we report. \appref{app:eval_protocol} gives the full breakdown.


\paragraph{Selection protocol.}
We select the base configuration from the post-training evaluation of each final checkpoint (Table~\ref{tab:stage1-eval}): $N{=}1000$ samples, seed~0, full element vocabulary, frozen KLDM prior, and a \emph{shared} MatterSim-5M evaluation potential (BFGS, 400 steps, lattice held fixed) for every structure. This selects $G{=}32$ with MatterSim-1M. Selection therefore used the lattice-fixed protocol, and we did not re-select under the lattice-free protocol that Section~\ref{sec:results} reports. \appref{sec:app-cellrelax} shows that the conclusions drawn from the two protocols agree. 

Table~\ref{tab:stage1-late-window} reports late-window metrics over the final $10\mathrm{k}$ training structures and ranks the same cell first, but it is not independent corroboration: each run is scored with its own training-time MatterSim potential, so the 1M and 5M rows are graded by different potentials and grader leniency is confounded with one of the two axes under comparison. We include it as a descriptive summary of where each run ended and rest the selection on the shared-potential evaluation alone.
Subsequent invalid-penalty ablations use this base model setup, with this experiment run providing the
$\mathrm{invalid\_penalty}=-0.2$ reference model for the investigation in \appref{app:stage2-invalid_penalty}.
%
%
%
%
\begin{table}[ht]
  \centering
  \caption{\latfix Post-training evaluation.
  Unique\% is computed over all structures.
    $\mathrm{SUN}$: $E_{\mathrm{hull}}\le 0$;
  $\mathrm{MSUN}$: $E_{\mathrm{hull}}\le 0.1\,\mathrm{eV/atom}$.}
  \label{tab:stage1-eval}
  \begin{tabular}{rlcccc}
    \toprule
    Rank & Config & Valid\% & Unique\% & SUN\% & MSUN\% \\
    \midrule
    1 & \textbf{G32 / 1M} & \textbf{97.9} & 100.0 & \textbf{18.1} & \textbf{38.5}\\
    2 & G16 / 5M         & 94.3 & 100.0 & \underline{8.6}           & \underline{20.8} \\
    3 & G16 / 1M         & 93.6 & 100.0 & 6.0         & 13.5 \\
    4 & G64 / 1M         & \underline{96.5} & 100.0 & 3.5            & 13.4 \\
    5 & G32 / 5M         & 96.0 & 100.0 & 3.9           & 11.5 \\
    6 & G64 / 5M         & 95.4 & 100.0 & 1.5          & 8.5 \\
    \bottomrule
  \end{tabular}
\end{table}

\begin{table}[h]
  \centering
  \caption{Late-window comparison on the final $10\mathrm{k}$
  training structures. Unique\% is computed over all structures.
  $\mathrm{SUN}$/$\mathrm{MSUN}$ use training-time MatterSim scores
  (stability ${=}1.0$ / ${\ge}0.9$ as proxies for
  $E_{\mathrm{hull}}\le 0$ / ${\le}0.1$).}
  \label{tab:stage1-late-window}
  \begin{tabular}{rlcccc}
    \toprule
    Rank & Config & Valid\% & Unique\% & SUN\% & MSUN\% \\
    \midrule
    1 & \textbf{G32 / 1M} & \underline{87.0} & 99.8 & \textbf{7.2}  & \textbf{22.1} \\
    2 & G16 / 5M         & \textbf{88.0} & 99.7 & \underline{3.8}          & \underline{11.7}  \\
    3 & G16 / 1M         & 85.1 & 99.8 & 3.3           & 11.2  \\
    4 & G64 / 5M         & 85.2 & 99.8 & 1.2          & 7.5  \\
    5 & G32 / 5M         & 84.8 & 99.8 & 2.1          & 7.5  \\
    6 & G64 / 1M         & 84.9 & 99.8 & 1.4          & 7.0  \\
    \bottomrule
  \end{tabular}
\end{table}



\subsection{Invalid-structure Penalty Sweep}
\label{app:stage2-invalid_penalty}

Fixing the group size and training potential (\appref{app:stage1}) leaves another important reward parameter undetermined: the penalty $r_{\mathrm{inv}}$ assigned to a structure that fails the validity gate. It is tempting to read this as a knob on the validity rate, and we test that reading below, but $r_{\mathrm{inv}}$ is better understood as the exchange rate between the gate and the discovery objectives, and it enters the update through three distinct channels.

\emph{It sets the dynamic range of the group-relative advantage.}
Valid structures score in $[0,1]$ by construction (\appref{app:reward/validity}), while an invalid one scores $r_{\mathrm{inv}} < 0$, so a group containing both spans a reward range wider than $1$. Because the advantage is standardised within the group (\eqref{eq:advantage}), a more negative $r_{\mathrm{inv}}$ lowers the group mean but also inflates the group standard deviation, and the second effect \emph{compresses} the normalised differences among the valid members. Enlarging the penalty does not simply add pressure against invalidity; it spends the group's available signal on a valid-versus-invalid contrast that the policy has largely already resolved, leaving less to separate a good valid structure from a mediocre one. This is the channel we care about most, because it acts on discovery rather than on validity.

\emph{It is the only weight-invariant term in the reward.} 
An invalid structure is assigned $r_{\mathrm{inv}}$ in every objective coordinate, so its scalarised reward is $r_{\mathrm{inv}} \sum_j w_j = r_{\mathrm{inv}}$ for any $\vw$ on the simplex. The penalty is therefore the one component of \eqref{eq:reward} that does not move with the sampled trade-off: it is a common floor shared by every policy in the weight-conditioned family, and any distortion it introduces is introduced everywhere at once rather than in one corner of the trade-off surface.

\emph{It prices exploration risk.} 
The mechanisms of \appref{app:explore_ablation} deliberately push the policy into unfamiliar chemistries, where the invalid rate is higher than in the well-covered regions it would otherwise exploit. $r_{\mathrm{inv}}$ is what that risk costs, so a penalty set too high has a plausible failure mode that is invisible in the validity rate: a policy that retreats to safe, familiar composition space and looks well behaved precisely because it has stopped exploring.

We therefore sweep $r_{\mathrm{inv}} \in \{-0.1, -0.5\}$ on the selected base model, holding everything else fixed; the selected base model supplies the $r_{\mathrm{inv}} = -0.2$ reference. All three arms are independent runs from the same frozen configuration under the same $7$-day wall-clock, and they are budget-matched to within $3\%$ ($123$--$126\mathrm{k}$ generated structures), so they can be compared directly. The outcome, together with \appref{app:stage1}, defines the final configuration used everywhere else in the paper. Results of the sweep are in Table \ref{tab:invalid-penalty}.
\begin{table}[t]
  \centering
  \caption{\latfix Invalid-penalty sweep on the selected base model
  ($G{=}32$, MatterSim-1M), evaluated post-training under
  \appref{app:eval_protocol} ($N{=}1000$, seed~0).
  \emph{Bond} and \emph{Charge} are the pass rates of the two components of the
  validity gate (\appref{app:reward/validity}); \emph{Valid} is the conjunction.}
  \label{tab:invalid-penalty}
  \begin{tabular}{lcccccc}
    \toprule
    $r_{\mathrm{inv}}$ & Valid\% & Bond\% & Charge\% & Unique\% & SUN\% & MSUN\% \\
    \midrule
    $-0.1$           & 96.6 & 96.7 & \textbf{99.9} & 100.0 & 8.7   & 22.0 \\
    $-0.2$ (selected) & \textbf{97.9} & \textbf{98.4} & 99.5 & 100.0 & \textbf{18.1} & \textbf{38.5} \\
    $-0.5$           & \underline{97.4} & \underline{97.6} & \underline{99.8} & 100.0 & \underline{12.0} & \underline{28.4} \\
    \bottomrule
  \end{tabular}
\end{table}

\paragraph{Validity is insensitive to the penalty.}
Over a five-fold range of $r_{\mathrm{inv}}$ the validity rate moves by $1.3$ percentage points ($96.6$--$97.9\%$), and not monotonically. At this point in training the gate is already satisfied by $\sim$97\% of samples whatever the penalty, so the marginal structure that $r_{\mathrm{inv}}$ could discourage is rare enough that its contribution to the objective is small: what sets the validity rate is the frozen backbone and the gate itself, not the price attached to failing it. This also bounds the first channel above, since a small and near-constant invalid fraction leaves the group standard deviation dominated by the spread \emph{among} valid structures.
The gate's two components are not equally binding: charge neutrality passes at $99.5$--$99.9\%$ while bond-length plausibility passes at $96.7$--$98.4\%$, so essentially all of our validity failures are geometric. 

\paragraph{No cell becomes conservative.}
The failure mode that Valid\% would hide, a policy buying its $97\%$ by retreating into safe, familiar chemistry, does not appear at any of the swept penalties: compositional uniqueness is $100.0\%$ in all three arms. The measure is sensitive enough to have caught it, since removing a genuinely load-bearing exploration mechanism collapses uniqueness to $1.0\%$ (\appref{app:explore_ablation}). Over $[-0.5, -0.1]$ the penalty fixes the ordering of valid above invalid without changing how much the policy explores.
$\mathrm{SUN}$ and $\mathrm{MSUN}$ move more than validity or uniqueness across the sweep ($8.7$--$18.1\%$ and $22.0$--$38.5\%$), peaking at the default $-0.2$. We use $-0.2$ on that basis but do not read three single-seed points as an interior optimum.




\subsection{Exploration-mechanism Ablations}
\label{app:explore_ablation}
\begin{table}[t] 
\centering 
\footnotesize
\renewcommand{\arraystretch}{0.95}
\setlength{\tabcolsep}{2.6pt}
\caption{\latfix Exploration mechanism ablation, evaluated post-training ($N_{\text{eval}}=1000$, full chemical space, shared MatterSim-5M evaluation potential, BFGS relaxation of positions only for at most 400 steps, see \appref{app:eval_protocol}). \\
$\bullet$ marks that the mechanism is enabled, -- means disabled.
\emph{Adapt.}\ = novelty scored against the growing agent-history index or frozen MP-20 index; \emph{bin.}\ = the binary Chemeleon2 flag of \appref{app:explore_ablation}.  \emph{CB} = count bonus, \emph{SG} = element sub-group mask. Runs were matched on wall clock. 
} 
\label{tab:app-explore-fixed} 
\begin{tabular}{lcccccccccc}
\toprule
Configuration & Adapt. & CB & SG & Valid (\%) & Unique (\%) & Stable (\%) & Meta (\%) & MSUN (\%) & SUN (\%) \\
\midrule
\textbf{Full method} & $\bullet$ & $\bullet$ & $\bullet$ & 97.9 & \textbf{100.0} & 18.1 & 38.5 & \textbf{38.5} & \textbf{18.1} \\
\midrule
\quad-- count bonus & $\bullet$ & -- & $\bullet$ & 96.3 & \textbf{100.0} & 8.7 & 24.6 & 24.6 & 8.7 \\
\quad-- sub-group mask & $\bullet$ & $\bullet$ & -- & 96.4 & \textbf{100.0} & 7.6 & 27.9 & 27.9 & 7.6 \\
\quad-- adaptive nov. & -- & $\bullet$ & $\bullet$ & 96.9 & \underline{99.9} & 8.4 & 26.4 & 26.4 & 8.4 \\
\midrule
adaptive nov.\ only & $\bullet$ & -- & -- & \underline{98.1} & \textbf{100.0} & 12.3 & 37.2 & \underline{37.1} & \underline{12.3} \\
count bonus only & -- & $\bullet$ & -- & 96.4 & \textbf{100.0} & 7.3 & 28.3 & 28.3 & 7.3 \\
sub-group mask only & -- & -- & $\bullet$ & \textbf{99.8} & 6.3 & \underline{19.3} & \underline{46.0} & 2.0 & 0.6 \\
none & -- & -- & -- & \textbf{99.8} & 1.0 & \textbf{60.0} & \textbf{68.7} & 0.6 & 0.6 \\
\midrule
binary novelty & bin. & $\bullet$ & $\bullet$ & 97.2 & \underline{99.9} & 5.7 & 16.6 & 16.6 & 5.7 \\
\bottomrule
\end{tabular}
\end{table}

  Table~\ref{tab:app-explore-fixed} repeats Table~\ref{tab:explore-ablation} with the lattice held fixed, and the numbers below refer to it. It varies the three exploration mechanisms as a full $2^{3}$ factorial rather than one ablation at a time. Because all three act on the same failure mode, the question is not only whether each helps but whether any one of them is \emph{sufficient}, and a leave-one-out design alone cannot separate redundancy from complementarity. 
All cells share the configuration of Table~\ref{tab:hparams} fixed by the two selection studies ($G{=}32$ with MatterSim-1M, \appref{app:stage1}; $r_{\mathrm{inv}} = -0.2$, \appref{app:stage2-invalid_penalty}) and are evaluated under \appref{app:eval_protocol} with the lattice held fixed. Every cell is a mechanism switched on or off, and nothing here re-selects a hyperparameter. Disabling the adaptive axis of CAN means novelty is scored by the same continuous kNN-RBF metric against the frozen MP-20 index alone. The novelty term is never removed from the reward. All cells are single-seed, so we read the ordering of widely separated rows and not differences of a few points.

\paragraph{Reading the factorial.}
The collapse of the all-off cell is the reward being optimized, not a failure to optimize it. That cell attains the highest per-structure stability in the whole study ($68.7\%$ of samples at $E_{\mathrm{hull}}\le0.1$ against $38.5\%$ for the full method, and $60.0\%$ at $E_{\mathrm{hull}}\le0$ against $18.1\%$) while proposing only $10$ distinct reduced formulas in $1000$ samples. The policy locates a composition the potential scores favorably and re-proposes it, which is exactly the degenerate solution a stability-weighted objective admits when nothing prices repetition. Per-structure stability is therefore not a safe headline metric for a generative policy; it has to be read together with a diversity measure.

The two halves of the factorial rank the mechanisms differently, and the disagreement is the point. In isolation, adaptive novelty is clearly the most valuable ($\mathrm{MSUN} = 37.1\%$, against $28.3\%$ for the count bonus and $2.0\%$ for the mask alone). The leave-one-out costs are instead tightly clustered ($24.6$--$27.9\%$, each roughly a third of $\mathrm{MSUN}$) and put the count bonus first. That is what redundancy looks like: with two of the three still active, the survivors partly cover for the one removed. Either of the two depth mechanisms alone restores full compositional diversity ($100.0\%$ unique), so their anti-repetition role is genuinely interchangeable in kind, though not in quality.

One caveat runs in our favor. The runs were stopped by a common wall clock rather than a common structure budget, and the collapsed cells consumed \emph{more} structures than the full method ($648\mathrm{k}$ for the all-off cell and $237\mathrm{k}$ for sub-group-only, against $102$--$131\mathrm{k}$ elsewhere), because repeatedly proposing the same small cells makes each relaxation cheap (Table \ref{tab:run-budget}). Those cells had more training and still produced two orders of magnitude fewer distinct formulas.

\paragraph{Adaptive versus fixed-reference novelty}
The control disables the agent-history index so that novelty is scored against the frozen MP-20 corpus alone, with everything else held fixed. Beyond the post-training row, a single full-method run shows the adaptive branch is exercised as intended. We instrument the selected base model (\appref{app:stage1}) by recording, for every scored structure, which branch of \eqref{eq:novelty} produced its score. Over the run ($123{,}968$ generated structures, $101{,}237$ valid) the agent produced $97{,}599$ distinct reduced formulas, a formula-repeat rate of $3.6\%$. Every repeat is routed through an anti-rediscovery branch rather than silently collecting the bootstrap value $\nu_0$: the $3{,}638$ repeated formulas split into $3{,}550$ scored against agent history and $88$ against both indices.
Splitting the run into fifths shows the reference moving. Collisions with MP-20 fall from $346$ in the first fifth to $3$ in the last, so the policy leaves the corpus the backbone was trained on. More tellingly, collisions with the agent's \emph{own} history fall from $1{,}078$ to $635$ over the same span, even as $Z^{\mathrm{hist}}$ grows by orders of magnitude. At a fixed sampling distribution a larger history can only raise the collision probability, so a falling count means the sampling distribution itself moved away from already-visited chemistries, the behavior the $\min$ over history in \eqref{eq:novelty} is meant to induce. These counts are descriptive and come from one run: they show the mechanism is active, not that it \emph{causes} the downstream gains. The fixed-reference row is what isolates the effect.

\paragraph{Continuous versus binary novelty}
This ablation replaces our graded novelty score with a static, binary one and changes nothing else. For each generated structure we compute a flag $v \in \{0,1\}$ from two same-formula tests:
%
\begin{itemize}
  \item \textbf{Novel} if no reference of the \emph{same reduced formula}
    matches the candidate under pymatgen's \texttt{StructureMatcher}
    (tolerances $l_{\mathrm{tol}}=0.2$, $s_{\mathrm{tol}}=0.3$,
    $\theta_{\mathrm{tol}}=5^{\circ}$). The reference is our frozen MP-20 index
    ($45{,}229$ structures, $37{,}217$ reduced formulas).
  \item \textbf{Unique} if the candidate is the first member of its
    \texttt{StructureMatcher} group within the current batch of generated
    structures.
\end{itemize}
The policy is trained on $v$ alone: the novelty entry of every valid structure is overwritten with $1$ when $v$ holds and $0$ otherwise, replacing \eqref{eq:knn-novelty}. Our continuous RDF novelty and the agent-history index still run, but only as logged diagnostics. Binary novel/known tests of this kind are the standard evaluation convention in crystal generation~\citep{zeni2025generative,betala2025lemat}, and these two flags are the ones used in Chemeleon2~\citep{park2026guiding}; rather than design our own, we adopt them here, with uniqueness evaluated within one GRPO group of $G=32$. Chemeleon2 combines the flags differently: its ''creativity'' reward is $1$ when a candidate is both unique and novel, $0$ when it is neither, and the Chebyshev distance between average-minimum-distance (AMD) descriptors~\citep{widdowson2022average}, a continuous isometry invariant of periodic crystals, only when the two disagree. We log that reward but do not train on it, so this cell is a binary variant of \emph{our} novelty term rather than a reproduction of their objective.

\paragraph{Why a binary signal saturates in our setting.}
The tests are defined relative to structures of the same reduced formula in MP-20. Our composition policy is explicitly rewarded for proposing formulas that lie outside MP-20, and once it does so the same-formula reference list is empty, the tests cannot fail, and $v = 1$ identically. Being constant across a GRPO group, the flag then contributes nothing to the group-relative advantage (\eqref{eq:advantage}): the novelty term cancels and the policy is trained on validity and the count bonus alone. The resulting loss is therefore one of \emph{gradient} rather than the collapse of the all-off cell (validity and uniqueness stay at $97.2\%$ and $99.9\%$) yet the cell still scores below every configuration that retains a continuous novelty signal, including single-mechanism cells with \emph{fewer} active mechanisms. What the comparison isolates is the value of a graded signal to a group-relative update, in a search whose objective is to leave the reference corpus; it is not a statement about binary novelty tests as evaluation metrics.

\paragraph{Count bonus and sub-group mask.}
The informative cell for the count bonus is its isolation row rather than its removal. On its own it holds compositional uniqueness at $100.0\%$, so it is sufficient to prevent the all-off collapse, but it converts that diversity into far fewer stable discoveries than adaptive novelty does ($\mathrm{MSUN}$ $28.3\%$ against $37.1\%$). The count bonus prices \emph{repetition of a formula}; adaptive novelty prices \emph{proximity to what has already been made}, which additionally pushes the policy away from near-duplicates that a formula counter cannot see.
The sub-group mask is the only one of the three that cannot stand alone: by itself it leaves uniqueness at $6.3\%$, barely better than the $1.0\%$ of the all-off cell, and its removal is the cheapest of the three ($38.5 \to 27.9\%$). Both are the expected signature of a breadth mechanism. Restricting the element pool per rollout changes which region of the periodic table a group explores, but nothing in it penalises proposing the same composition twice inside that region, so it redistributes exploration without bounding repetition. It is valuable in combination and inert in isolation.

\subsection{Learned Composition Preferences}
\label{app:composition_ladder}
To isolate whether ANCHOR simply learns elementary compositional validity, we perform an experiment keeping the complete generation, relaxation, and scoring fixed and compare uniformly random formulas, uniformly sampled charge-neutral formulas, and the trained ANCHOR policy (Table~\ref{tab:kldm-ladder}). Random formulas are drawn uniformly from the untrained action space of \appref{app:elem_vocab_action_space}, and enumerated formulas uniformly from its charge-neutral enumeration. Both are queried through the same frozen KLDM-CSP backbone and scored by the same evaluator as the policy, so only the formula source changes. With the lattice held fixed (\appref{sec:app-cellrelax}), charge-neutral enumeration raises Valid from $9.5\%$ to $70.1\%$, but MSUN increases only from $0.4\%$ to $3.1\%$. ANCHOR reaches $97.9\%$ Valid and $38.5\%$ MSUN. All three composition sources remain $100\%$ unique, showing that the improvement cannot be explained by collapse or by learning charge neutrality alone. The policy learns to allocate queries to compositions for which the frozen CSP model produces substantially more successful discovery candidates.
\begin{table}[h!]
  \centering
  \caption{\latfix Effect of composition selection with the structural pipeline held fixed. All rows use the same frozen KLDM-CSP backbone and post-training evaluation protocol ($N{=}1000$, seed~0, full chemical space, shared MatterSim-5M, BFGS${=}400$). Unique\% is distinct reduced formulas over all $N$. }
  \label{tab:kldm-ladder}
  \begin{tabular}{lcccc}
    \toprule
    Composition source & Valid\% & Unique\% & SUN\% & MSUN\% \\
    \midrule
    KLDM + random formulas
      & 9.5 & 100.0 & 0.2 & 0.4 \\
    KLDM + enumerated charge-neutral
      & 70.1 & 100.0 & 0.8 & 3.1 \\
    ANCHOR (selected policy)
      & \textbf{97.9} & 100.0 & \textbf{18.1} & \textbf{38.5} \\
    \bottomrule
  \end{tabular}
\end{table}

\subsection{Reward Fine-tuning Baselines}
\label{app:reward-ft}

Two families of baseline optimize the same reward as \textsc{ANCHOR} \eqref{eq:reward} but place the gradient on a structure model instead of, or in addition to, a composition policy. Together they measure what reward fine-tuning of the generative model buys on its own, and so isolate the contribution of keeping the backbone frozen and searching over compositions.

\paragraph{DDPO fine-tuning of DNG models.}
We fine-tune the released Crystalite-DNG checkpoint and a KLDM-DNG model retrained on MP-20 with DDPO~\citep{black2024training}, treating each denoising step as an action and assigning the terminal reward of \eqref{eq:reward} to every step of the trajectory with no discounting. The implementation is the clipped surrogate of \eqref{eq:ppo} averaged over denoising steps rather than over a group of compositions,
\begin{equation}
  \Ls^{\mathrm{clip}}(\theta)
  = \E\Bigl[
      \tfrac{1}{|\gT|}\textstyle\sum_{t \in \gT}
      \min\bigl(\eta_t A,\; \mathrm{clip}(\eta_t, 1-\eps, 1+\eps)\,A\bigr)
    \Bigr],
  \qquad
  \eta_t = \frac{p_\theta(x_{t-1} \mid x_t)}{p_{\theta_{\mathrm{old}}}(x_{t-1} \mid x_t)},
  \label{eq:ddpo}
\end{equation}
where $A$ is the sample's terminal reward standardized over the sampling batch, shared by every step of its trajectory, and $\gT$ is the set of scored steps. For Crystalite-DNG $\gT$ is restricted to steps whose churn noise lies in $[0.2, 2.0]$, where that sampler's stochasticity makes the step log-probability informative; for KLDM-DNG all EM steps are scored.

Both runs take a single inner epoch per batch, which is DDPO's score-function form: the update is $-A\,\nabla_\theta \log p_\theta(x_{t-1} \mid x_t)$ averaged over the scored steps, with the clip at $\eps = 0.05$ and a trust-region stop on the mean $|\log \eta_t|$ retained for configurations that take more epochs. Reward fine-tuning costs uniqueness on Crystalite-DNG, which falls from $98.8\%$ to $40.5\%$ (\emph{mean}) and $49.5\%$ (\emph{bal.}) in our pipeline (Table~\ref{tab:internal}), so part of the reward there is collected by re-emitting structures, while KLDM-DNG stays above $99\%$.

The reward uses the same validity gate, stability term, adaptive novelty index, both diversity terms and count bonus as \textsc{ANCHOR}, with $r_{\mathrm{inv}} = -0.2$ and $\lambda_c = 1$, evaluated by the same scorer. Two details differ from the evaluation protocol of \appref{app:eval_protocol} and are properties of training, not of the reported metrics: relaxation during fine-tuning is BFGS for at most $100$ steps against $400$ at evaluation, and the training potential is MatterSim-1M against MatterSim-5M at evaluation. Objective weights are fixed rather than resampled per rollout, either at the mean of the training Dirichlet (\emph{mean}, $\vw = (0.143, 0.571, 0.143, 0.143)$) or with stability lifted to novelty's level (\emph{bal.}, $\vw = (0.4, 0.4, 0.1, 0.1)$), both in the order $(s_{\mathrm{stab}}, \nu, d_{\mathrm{struct}}, d_{\mathrm{comp}})$. Hyperparameters are in Table~\ref{tab:ddpo-hparams}.
Both reported rows sample from a pinned snapshot taken after about $4.1$k episodes, roughly $66$k structures generated, about half the budget of an \textsc{ANCHOR} run (Table~\ref{tab:run-budget}). By then Crystalite-DNG had already doubled its validity and lost more than half its uniqueness, so further steps would extend the same shift rather than reverse it.

\begin{table}[h]
\centering
\small
\caption{DDPO fine-tuning of the two DNG models. One Adam step per episode, taken after accumulating gradients over the scored denoising steps of the batch.}
\label{tab:ddpo-hparams}
\begin{tabular}{lcc}
\toprule
& Crystalite-DNG & KLDM-DNG \\
\midrule
Initialization & released checkpoint & MP-20 retrain (below) \\
Learning rate & $5\times10^{-6}$ & $5\times10^{-6}$ \\
Batch size (structures / episode) & 16 & 16 \\
Denoising steps per sample & 150 & 50 \\
Scored steps $\gT$ & churn $\in [0.2, 2.0]$ & all \\
Inner epochs per batch & 1 & 1 \\
Clip $\eps$ & 0.05 & 0.05 \\
Checkpoint & rolling, every 50 episodes & rolling, every 50 episodes \\
\bottomrule
\end{tabular}
\end{table}

\paragraph{Why DDPO rather than GRPO on the generator.}
In this setting, the two use the same advantage estimator, so the choice is not a substantive one. GRPO replaces a learned value baseline with the mean reward of a group of samples drawn under a shared context. For \textsc{ANCHOR}'s composition policy that context is a preference vector and an element sub-group, and the group is the $G$ compositions sampled under it (\eqref{eq:advantage}). Our DNG baselines are sampled unconditionally, so the only group available is the sampling batch, and the group-relative advantage becomes exactly the batch-standardized terminal reward $A$ of \eqref{eq:ddpo}. \citet{park2026guiding} instead group rollouts by a conditioning input such as the number of atoms. What separates these baselines from Chemeleon2 is therefore not the estimator, which differs only in how the group is formed, but the treatment of drift from the pretrained denoiser, the axis this comparison is designed around, and the reward, since Chemeleon2 scores novelty against a fixed reference rather than a search history. Chemeleon2 restrains drift with a KL penalty, the DNG baselines leave it unrestrained, and \textsc{ANCHOR} removes it exactly by freezing the backbone (\eqref{eq:zero_structural_kl}).

\paragraph{MP-20 retraining of KLDM-DNG.}
The KLDM checkpoint used elsewhere in this paper is a CSP model, so the DNG baseline is a separate KLDM-DNG model we train from scratch on MP-20 with KLDM's own \texttt{dng} recipe, on the same split as the CSP backbone. Fine-tuning initializes from its final checkpoint.

\paragraph{Reward-trained CSP backbone.}
In the \emph{reward-trained KLDM-CSP} rows, GRPO updates the backbone alongside the composition policy. All $14.9$M parameters of the KLDM network are unfrozen; there are no adapters. The gradient does not flow through the sampling chain: rollouts are drawn under \texttt{no\_grad}, and the stored denoising trajectory is then replayed to give the backbone the same clipped per-step objective as \eqref{eq:ddpo}, so the backbone is trained by DDPO nested inside the GRPO update of the policy. The backbone is held fixed for the first $50$ policy steps, then updated with learning rate $5\times10^{-5}$ against $10^{-4}$ for the policy, and clip $\eps = 0.05$ against $0.2$. Unlike the DNG runs above, these use $K = 2$ epochs per batch, so on the second epoch the ratio departs from one and both the clip and a trust-region guard are live: an epoch whose mean $|\log \eta_t|$ exceeds $0.05$ is discarded rather than applied. There is still no divergence penalty to the pretrained weights in the loss.

With the \emph{shared} reward the backbone receives the full reward of \eqref{eq:reward}. With the \emph{stability-only} reward it receives $s_{\mathrm{stab}}$ alone while the composition policy still receives the full reward, which tests whether the backbone should be asked to chase novelty and diversity that only the composition policy can act on. The two arms are otherwise identical: their configurations differ by this one setting and the run name.

\subsection{Training the CSP Backbone on \textsc{ANCHOR} Structures}
\label{app:csp-finetune} 
We fine-tune each CSP backbone with its own training objective on MP-20 together with $27{,}136$ structures from a single \textsc{ANCHOR} training run, the base model selected in \appref{app:stage1}. None of the added \textsc{ANCHOR} structures occur in the original training set of either backbone. A run does not persist geometry, only a log of the compositions it proposed and their scores, so the corpus is built in two stages: we sample logged candidates, then replay each composition through the frozen backbone to obtain a cell. What is preserved exactly is the composition distribution the trained policy produced, not the individual cells seen during training. Each backbone is fine-tuned on KLDMs rendering of that composition list unless otherwise stated.

Of the run's $123{,}968$ logged candidates we remove $22{,}731$ that fail the validity gate and $3{,}224$ duplicates, then draw $27{,}136$ of the remaining $98{,}013$ uniformly at random without replacement. We apply no stability or novelty threshold and no window over training steps, so the sample represents the run as a whole rather than its converged phase. Validity rises from $68.6\%$ in the first quartile of the run to $87.3\%$ in the last, so this choice dilutes the sample relative to drawing from late steps only.

The sample size equals that of the MP-20 training split, so each \emph{$+$\,\textsc{ANCHOR}} arm trains on $54{,}272$ structures against its control's $27{,}136$. Each \emph{$+$\,\textsc{ANCHOR}} arm is paired with a \emph{ctl} arm trained on pristine MP-20 for the same number of optimizer steps at the same batch size, which equalizes compute and total structure exposures and leaves the corpus as the only difference between them. 
Matching steps rather than epochs means the augmented arm sees each individual structure about half as often as its control; that is the intended trade, since the question is whether a fixed budget is better spent on more diverse data or on more repetitions of less of it. The budget is matched within a backbone, not across the two, which have different published recipes.

\emph{FT} resumes from the released checkpoint for a further ${\sim}150$k steps, reaching global step $849{,}600$ for KLDM, and \emph{from scratch} retrains for the same $699.6$k (KLDM) or $1.4$M (Crystalite) steps as the released model. Two further arms isolate artifacts of how the corpus was built rather than properties of the chemistry in it. \emph{Unrelaxed cells} trains on the \textsc{ANCHOR} cells as generated, before MatterSim relaxation, which tests whether a change is an artifact of the relaxer. \emph{Crystalite-rendered} keeps the \textsc{ANCHOR} compositions but realises them with Crystalite-CSP instead of KLDM-CSP, so that KLDM is not fine-tuned on its own output; it is the control against self-distillation. 
At each of three sampling seeds, every CSP checkpoint is evaluated on the same $1000$ held-out \textsc{ANCHOR} compositions, none of which are part of the $27{,}136$-composition fine-tuning corpus. Table~\ref{tab:app-ft} reports every arm.

\begin{table}[t]
\centering
\footnotesize
\setlength{\tabcolsep}{6pt}
\caption{Training the CSP backbone on \textsc{ANCHOR} structures, under the MatterGen evaluation protocol. \textsc{ANCHOR} contributed $27{,}136$ structures from one of its training runs; each \emph{$+$\,\textsc{ANCHOR}} arm adds them to MP-20 and each \emph{ctl} arm is the step-matched control on pristine MP-20, so the pair isolates the corpus from the extra optimisation. Finetuning resumes from the released checkpoint; \emph{from scratch} retrains it for the same $699.6$k (KLDM) or $1.4$M (Crystalite) steps. Every row is scored on the same $1000$ \textsc{ANCHOR} compositions as the inference-time backbone rows of \appref{app:eval_protocol}; only the CSP checkpoint changes. Entries are from sampling seed~0; Tables~\ref{tab:internal} and~\ref{tab:mattergen_dng} report seed-averaged values for the FT arms. Composition validity and uniqueness are fixed by the shared formula list and are reported per protocol in \appref{app:eval_protocol}.}
\label{tab:app-ft}
\begin{tabular}{@{}lrr@{}}
\toprule
Configuration & MSUN & SUN \\
\midrule
\multicolumn{3}{@{}l}{\textbf{KLDM-CSP}} \\
\addlinespace[1pt]
Released checkpoint & 42.3 & 17.1 \\
FT: MP-20 only (ctl) & 41.9 & 17.1 \\
FT: MP-20 $+$ \textsc{ANCHOR} & 41.6 & 16.5 \\
FT: MP-20 $+$ \textsc{ANCHOR}, unrelaxed cells & 42.5 & 16.7 \\
FT: MP-20 $+$ Crystalite-rendered \textsc{ANCHOR} & 43.9 & 18.9 \\
From scratch: MP-20 only (ctl) & 42.2 & 17.9 \\
From scratch: MP-20 $+$ \textsc{ANCHOR} & 42.5 & 17.1 \\
\addlinespace[3pt]
\multicolumn{3}{@{}l}{\textbf{Crystalite-CSP}} \\
\addlinespace[1pt]
Released checkpoint & 42.8 & 14.5 \\
FT: MP-20 only (ctl) & 41.9 & 14.4 \\
FT: MP-20 $+$ \textsc{ANCHOR} & \textbf{49.8} & \textbf{24.5} \\
From scratch: MP-20 only (ctl) & 37.6 & 13.4 \\
From scratch: MP-20 $+$ \textsc{ANCHOR} & \underline{47.8} & \underline{21.0} \\
\bottomrule
\end{tabular}
\end{table}

The corpus helps Crystalite-CSP and does not help KLDM-CSP. Against its step-matched control, Crystalite-CSP gains $10.1$ points of SUN when finetuned ($14.4 \to 24.5$) and $7.6$ when retrained from scratch ($13.4 \to 21.0$), which makes the finetuned arm the strongest CSP checkpoint we report. 
On KLDM-CSP the same corpus is flat to slightly negative ($17.1 \to 16.5$ finetuned, $17.9 \to 17.1$ from scratch). The two diagnostic arms bear on why. Holding the compositions fixed and rendering them with Crystalite instead of KLDM moves KLDM-CSP from below its control to above it ($16.5 \to 18.9$), whereas removing the relaxation step changes almost nothing ($16.5 \to 16.7$): the geometries matter, and which model produced them matters more than whether they were relaxed. The one arm that fails to improve is also the only one trained on cells from the backbone with which it was trained, which is consistent with a backbone gaining little from its own output, though the two backbones differ enough in starting quality that we do not read the size of the Crystalite gain as attributable to that alone.

\end{document}